\documentclass[pdftex,twocolumn]{aastex63}

\usepackage{amsmath,amssymb}

\submitjournal{ApJ}

\shorttitle{Disk-Jet Structure and Flux Separation in Polarization Images}
\shortauthors{Tsunetoe et al.}

\begin{document}

\title{Investigating the Disk-Jet Structure in M87\\
through Flux Separation in the Linear and Circular Polarization Images
}

\correspondingauthor{Yuh Tsunetoe}
\email{tsunetoe@kusastro.kyoto-u.ac.jp}

\author[0000-0003-0213-7628]{Yuh Tsunetoe}
\affiliation{Department of Astronomy, Kyoto University \\
Kitashirakawa-Oiwake-cho, Sakyo-ku, Kyoto-shi \\
Kyoto, 606-8502, Japan}


\author{Shin Mineshige}
\affiliation{Department of Astronomy, Kyoto University \\
Kitashirakawa-Oiwake-cho, Sakyo-ku, Kyoto-shi \\
Kyoto, 606-8502, Japan}

\author[0000-0001-8527-0496]{Tomohisa Kawashima}
\affiliation{Institute for Cosmic Ray Research, University of Tokyo \\
5-1-5 Kashiwanoha, Kashiwa-shi \\
Chiba, 277-8582, Japan}

\author[0000-0002-2309-3639]{Ken Ohsuga}
\affiliation{Center for Computational Sciences, University of Tsukuba \\
1-1-1 Tennodai, Tsukuba-shi \\
Ibaraki, 305-8577, Japan}

\author[0000-0002-9475-4254]{Kazunori Akiyama}
\affiliation{Massachusetts Institute of Technology, Haystack Observatory \\
99 Millstone Road \\
MA 01886, USA}
\affiliation{Black Hole Initiative, Harvard University \\
20 Garden street, Cambridge \\
MA 02138, USA}
\affiliation{National Astronomical Observatory of Japan \\
2-21-1 Osawa, Mitaka-shi \\
Tokyo, 181-8588, Japan}

\author[0000-0003-0114-5378]{Hiroyuki R. Takahashi}
\affiliation{Department of Natural Sciences, Faculty of Arts and Sciences, Komazawa University \\
1-23-1 Komazawa, Setagaya-ku \\
Tokyo, 154-8525, Japan}



\begin{abstract}

For testing different electron temperature ($T_{\rm e}$) prescriptions in general relativistic magnetohydrodynamics (GRMHD) simulations through observations, we propose to utilize linear polarization (LP) and circular polarization (CP) images.
We calculate the polarization images based on a semi-Magnetically Arrested Disk (MAD)  GRMHD model for various $T_{\rm e}$ parameters, bearing M87 in mind.
We find an LP-CP separation in the images of the low-$T_{\rm e}$ disk cases at 230~GHz; namely, the LP flux mainly originates from downstream of the jet and the CP flux comes from the counter-side jet, while the total intensity is maximum at the jet base. 
This can be understood as follows: although the LP flux is generated through synchrotron emission widely around the black hole, most of the LP flux from the jet base does not reach the observer, since it undergoes Faraday rotation ($\propto T_{\rm e}^{-2}$) when passing through the outer cold disk and is thus depolarized. 
Hence, only the LP flux from the downstream (not passing the cold dense plasmas) can survive.
Meanwhile, the CP flux is generated from the LP flux by Faraday conversion ($\propto T_{\rm e}$) in the inner hot region. 
Stronger CP flux is thus observed from the counter-side jet. 
Moreover, the LP-CP separation is more enhanced at a lower frequency such as 86~GHz but is rather weak at 43~GHz, since the media in the latter case is optically thick for synchrotron self-absorption so that all the fluxes should come from the photosphere. 
The same is true for cases with higher mass accretion rates and/or larger inclination angles.


\end{abstract}



\section{Introduction} \label{sec:intro}

{
Active galactic nuclei (AGN) are known to produce energetic phenomena such as intense radiation and powerful outflows \citep{1989ApJ...347...29S}, and are thought to be driven by a  central supermassive black hole (SMBH) onto which matter accretes  \citep{1969Natur.223..690L,1984ARA&A..22..471R,1999MNRAS.303L...1B,2008bhad.book.....K}. 
A small subset of AGN produce plasma jets which accelerate to relativistic speeds and are highly collimated \citep{1979ApJ...232...34B,1984ARA&A..22..319B,1997ARA&A..35..607Z}. 
In theoretical studies of these AGN jets, magnetic fields are commonly believed to play an  important role in extracting the rotational energy from the black hole and/or accretion flow and thus powering the plasma jets \citep{1977MNRAS.179..433B,1982MNRAS.199..883B}.}

{It is well known that the M87 in the Virgo cluster is a low luminosity AGN (LLAGN; \cite{1997ApJS..112..315H}) with a jet aligned closely to our line of signt with an inclination angle to the jet axis, $i \sim 160^\circ$ ($20^\circ$) (e.g., \cite{2016A&A...595A..54M,2018ApJ...855..128W}).
The jet of M87 has been observed at various length scales over a wide ranges of wavelengths  \citep{1989ApJ...340..698O,1994ApJ...435L..27F,1995ApJ...447..582B,1997ApJ...489..579M,2002ApJ...564..683M,2003ApJ...582..133D,2006Sci...314.1424A,2011ApJ...729..119G,2012ApJ...746..151A}. 
In particular, observations of the jet base with high angular resolution provided by very long baseline interferometry (VLBI) have provided observational evidences of the persistent acceleration and collimation  \citep{1999Natur.401..891J,2007ApJ...660..200L,2007ApJ...668L..27K,2011Natur.477..185H,2012ApJ...745L..28A,2013ApJ...775...70H,2013ApJ...775..118N,2014ApJ...781L...2A,2014ApJ...786....5K,2016A&A...595A..54M,2018ApJ...868..146N,2018A&A...616A.188K,2019ApJ...887..147P}.
In this context, the first ever image of the shadow of the SMBH in M87 by the Event Horizon Telescope (EHT) provides us with a unique opportunity to study, for the first time, the connection between a powerful relativistic jets and the central engine \citep{2012Sci...338..355D,2014ApJ...788..120L,2015ApJ...807..150A,2016ApJ...829...11C,2017ApJ...838....1A,2017AJ....153..159A,2019ApJ...875L...1E,2019ApJ...875L...5E}.}

{The synchrotron emission, emitted within this inner region at radio wavelengths, is also known to have a polarization component which reflects the strength and orientation of the surrounding magnetic fields. 
VLBI observations also point to the existence of ordered magnetic field structure both through  linear polarization (LP) images \citep{2016ApJ...817..131H,2018ApJ...855..128W,2020A&A...637L...6K} and analyses of Faraday rotation measure (RM) and electric vector position angle (EVPA) orientation  \citep{1990ApJ...362..449O,2002ApJ...566L...9Z,2003ApJ...589..126Z,2004ApJ...612..749Z,2016ApJ...823...86A,2019ApJ...871..257P}. }

{Further, the EHT collaboration recently published LP images of M87* in 2017, which exhibited polarization angles in a nearly azimuthal pattern over a region of the asymmetric ring. In addition, day-to-day variation evidence for the temporal evolution of the polarization in this inner region over one week \citep{2021ApJ...910L..12E}. 
They also found low circular polarization fraction of the M87 core ($< 0.3\%$) from ALMA-only 230~GHz observations \citep{2021ApJ...910L..12E,2021arXiv210502272G}. }

{To connect these observations to theoretical models of M87*, researchers have calculated radiative transfer in the Kerr and Schwartzschild metrics (general relativistic radiative transfer; GRRT) based on (semi-)analytical models \citep{2009ApJ...697.1164B,2019ApJ...878...27K,2020MNRAS.493.5606J,2021ApJ...909..168K} or calculation models such as those produced by general relativistic magnetohydrodynamics  \citep[GRMHD;][]{1999ApJ...522..727K,2003ApJ...589..444G,2005MNRAS.359..801K,2006ApJ...641..626N,2011MNRAS.418L..79T,2012MNRAS.423.3083M} simulations.
These models have produced synthetic images which re-produce many of the observed macroscopic features of the synchrotron emission from this inner region  \citep{2012MNRAS.421.1517D,2016A&A...586A..38M,2019MNRAS.486.2873C,2019A&A...632A...2D}.}

{In order to extract useful information on the physical processes we need to specify the regions producing the emission, however, this is a difficult work. 
Since the light rays from near the black hole are bent and lensed by the gravity, the emissions from the jet in the funnel region and from the equatorial accretion disk, which is considered to be radiatively inefficient accretion flow \citep[RIAF;][]{1995ApJ...452..710N,2008bhad.book.....K,2014ARA&A..52..529Y} in LLAGNs, are degenerated into a ring-like image. 
In addition, it is an unresolved issue how to determine the proton-electron coupling in the jet-disk structure in theoretical models.}

{In this regard, the polarization components can provide powerful tools to verify the jet-disk structure, because they carry out the information regarding the plasma properties not only in the emitting plasma, but also in the intervening plasma through the Faraday effects, rotation and conversion. 
\citet{2017MNRAS.468.2214M} presented linearly polarized images and the RMs through GRRT calculations based on GRMHD models with M87* in mind. Their best-fit, jet-dominated (low electron-temperature disk) model gave consistent values of the LP fraction and RM with observations \citep{2014ApJ...783L..33K}.
\citet{2020MNRAS.498.5468R} also showed resolved RM images, which gave strong, spatial and temporal variabilities. 
These two studies demonstrated that the LP vectors originated from the counter (receding) jet are scrambled and depolarized by Faraday rotation in the disk, while those from the foreground (approaching) jet can survive from the Faraday depolarization and thus become dominant on the LP maps. 
The EHT collaboration compared the observed polarization structure with predictions from theoretical models, and attributed low polarization fraction in the image to Faraday rotation internal to the emission region \citep{2021ApJ...910L..13E}. 
Further, the MAD \citep[magnetically arrested disk;][] {2003PASJ...55L..69N,2011MNRAS.418L..79T} models are favored over the SANE \citep[standard and normal evolution;][] {2012MNRAS.426.3241N,2013MNRAS.436.3856S} models in their GRMHD model evaluation.}

{Meanwhile, we suggested in our previous works that the circular polarization (CP) can be amplified by Faraday conversion \citep{1977ApJ...214..522J} from the LP in hot and dense plasma near the black hole, up to the extent comparable with the LP \citep{2020PASJ...72...32T,2020arXiv201205243T}. We there introduced an amplification process of the CP through combination of Faraday conversion and rotation (the rotation-induced conversion), which produces the CP components with signs imprinting the magnetic fields configuration. 
\citet{2021arXiv210300267M} also showed the CP images enhanced by Faraday rotation and conversion. 
\citet{2021MNRAS.tmp.1263R} introduced a CP conversion process through twist of the magnetic field along the line-of-sight on event horizon scales, in addition to the rotation-induced conversion. 
These processes of the rotation-induced and field-twist Faraday conversions, as well as intrinsic CP component of synchrotron emission \citep{1968ApJ...154..499L,1977ApJ...214..522J,1988ApJ...332..678J}, have also been introduced and discussed in the context of CP detection in quasars  \citep{1982ApJ...263..595H,2003Ap&SS.288..143W,2003A&A...401..499E,2008MNRAS.384.1003G,2009ApJ...696..328H}. }

{In this way, studies have established that the unified interpretation of the LP and CP is essential for understanding of the magnetic field structure and plasma properties near the black hole \citep{2017ApJ...837..180G,2020ApJ...896...30A,2021arXiv210105327E}. }

{Along this line, we analyze and quantify the relationship among the polarization components on theoretical polarization images of M87* using correlation functions, focusing on the radiative processes in the jet-disk structure. 
\citet{2018MNRAS.478.1875J} calculated the autocorrelation length of each ($x$- and $y$-) component of LP vectors on the image of Sgr A*, 
and claimed that the correlation length (rather than the LP fraction) provides a reliable indicator of Faraday rotation depth. 
We take a slightly different approach; namely, we calculate auto- and cross- correlations among the total, LP and CP intensities on ray-traced images obtained from GRRT calculation through GRMHD models, in an attempt to understand their relations to the Faraday rotation, the Faraday conversion, the synchrotron self-absorption (SSA), and the underlying plasma  properties in the jet-disk structure in M87. }

{This paper is organized as follows:
We outline the methodology for computing theoretical polarization images in section \ref{sec:method}. 
We show our resultant images (\ref{subsec:figures}), correlation analyses (\ref{subsec:correlations}), and an example of ``LP-CP separation'', a separation along the jet direction between LP and CP intensity distributions (\ref{subsec:Faraday}).
We examine the separation for various electron-temperature parameters in the disk in subsecion \ref{subsec:Rhigh}. 
The dependence on frequency is presented in \ref{subsec:mf}. 
Other possibilities for the inclination angle and models with different mass accretion rates from M87* are discussed in \ref{subsec:inclination} and \ref{subsec:Mdot}, respectively. 
We compare the results with existing observations and discuss prospects for future observations in \ref{subsec:observation}. 
Section \ref{sec:conclusion} presents our conclusion.}

\begin{figure*}[ht!]
\plotone{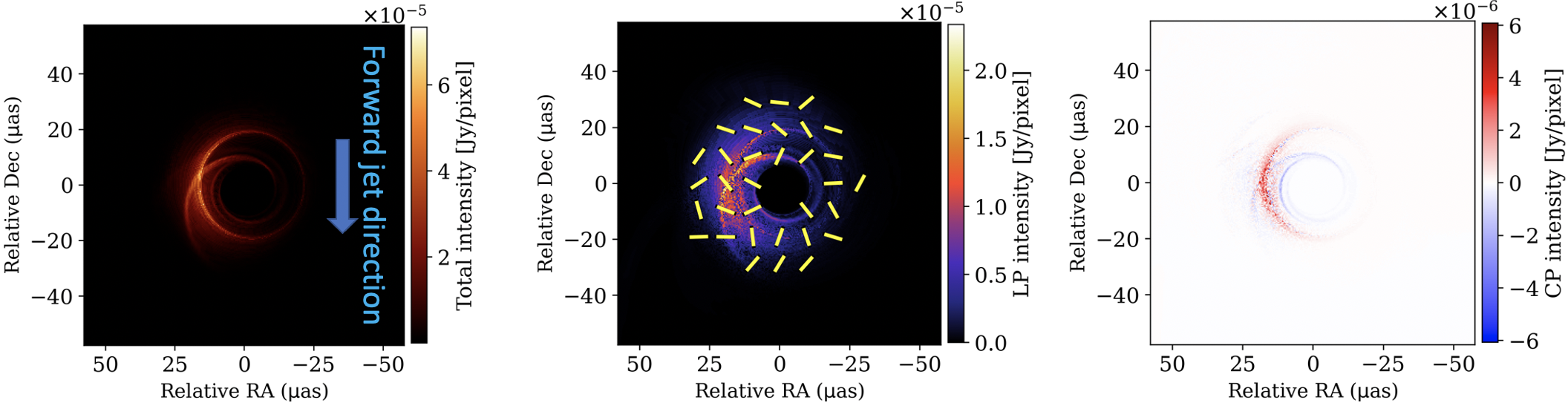}
\caption{Polarimetric images at 230~GHz obtained by the radiative transfer calculation for our fiducial model with an inclination angle of $i = 160^\circ$.
Left: total intensity (Stokes $I$) image, which consists of the photon ring feature and dim  jet components. 
Center: linear polarization (LP) map with color contour of the LP intensity (Stokes $\sqrt{Q^2+U^2}$) and overplotted LP vectors in EVPA (electric vector position angle). The LP vectors are scrambled by the Faraday rotation after the synchrotron emission and show a disordered pattern.
Right: circular polarization (CP) image with color contour of the CP intensity with sign (Stokes $V$). The CP components around the photon ring are amplified by the Faraday conversion process in hot region near the black hole. 
The black hole is located in the center of the images. 
The spin axis of the black hole points upwards in the images, and the approaching jet extends downwards (as shown by a blue arrow in the left image) though it is dim on the images. The images consist of $600\times 600$ pixels.
\label{fig:raw}}
\end{figure*}

\section{Method} \label{sec:method}

\subsection{GRMHD model and proton-electron coupling}\label{subsec:GRMHD}

We performed three-dimensional GRMHD simulation of RIAF around a Kerr black hole (BH) with a dimensionless spin parameter of the black hole $a=0.9375$ by using GR-Radiation-MHD code \texttt{UWABAMI} \citep{Takahashi_2016, Takahashi_2018}, where the radiative effects are turned off \citep{Kawashima_2021} in order to avoid any inconsistency possibly caused by its current one-temperature-fluid approximation.
The modified Kerr-Schild coordinate \citep[e.g.,][]{Gammie_2004} is employed in the simulation. 
The inner- and outer-outflow boundaries are located at $0.96 r_{\rm H} \simeq 1.29 r_{\rm g}$ and $3.33 \times 10^3 r_{\rm g}$, respectively, where $r_{\rm g} \equiv GM_\bullet/c^2$ is the gravitational radius, $G$ is the gravitational constant, $M_\bullet$, black hole mass, $c$ is the speed of light, and $r_{\rm H} (= (1  + \sqrt{1 - a^2})r_{\rm g} \simeq 1.35 r_{\rm g})$ is the outer horizon of the black hole. 
The simulation domain is devited into $r\times \theta \times \phi = 200 \times 128 \times 64$ meshes.

Initially, we set an isentropic hydroequilibrium torus rotating around the Kerr BH \citep{Fishbone_Moncrief_1976} with the single-loop magnetic field configuration, which is embeded in a hot, static, uniform, and non-magnetized ambient gas.
The position of the inner edge and the pressure maximum of the torus are set at $r = 20 r_{\rm g}$ and $r = 33 r_{\rm g}$ on the equatorial plane, respectively.
The specific heat ratio is assumed to be $\gamma_{\rm heat} = 13/9$.
We use a snapshot of the simulation data at $t = 9 \times 10^3 r_{\rm g}/c$, at which the accretion flow in a quasi-steady state after the sufficient mass supply from the initial torus via the growth of the magneto-rotational instability (MRI) \citep{Balbus_Hawley_1991}. 
We calculate the MRI quality factor $Q$-factors, the numbers of the cells across a wavelength of the fastest-growing MRI mode in each direction (\cite{2011ApJ...738...84H}), and obtain $(Q_r, Q_\theta, Q_\phi) = (3.23, 3.97, 11.0)$ in the zero-angular momentum observer frame, averaging over $r \lesssim 20r_{\rm g}$ and $60^\circ \lesssim \theta \lesssim 120^\circ$. (see subsection \ref{subsec:future} for discussion about resolution of the MRI modes.)

GRMHD models are often categorized into two major groups, the MAD and SANE, which are divided by their strength of the dimensionless magnetic flux near the event horizon $\phi \equiv \Phi_{\rm BH} / \sqrt{\dot{M}r_{\rm g}c^2}$, where $\Phi_{\rm BH} = (1/2)\int_\theta \int_\varphi |B^r|{\rm d}A_{\theta \varphi}$. 
The MADs, which typically show the saturation of $\phi \gtrsim 50$ (in Gaussian units), are characterized by the strong, dynamically important magnetic flux near the black hole, while the SANEs ($\phi \le 5$) have the weak magnetic flux.
Our GRMHD simulation shows $\phi \approx 18$, so that the magnitude of $\phi$ is between the typical values of MAD and SANE and this state is sometimes refered to as semi-MAD.

Since the GRMHD simulation only gives temperature for protons, we have to determine electron temperature by post-process to calculate synchrotron radiation transfer. 
As in \citet{2020PASJ...72...32T,2020arXiv201205243T} and previous works including \citet{2019ApJ...875L...5E,2021ApJ...910L..13E}, we implement a relation equation between proton and electron temperature with the plasma $\beta \equiv p_{\rm gas}/ p_{\rm mag}$, the gas-magnetic pressure ratio, and two parameters $R_{\rm low}$ and $R_{\rm high}$, 
\begin{equation}\label{eq:M16}
	\frac{T_{\rm i}}{T_{\rm e}} = R_{\rm low}\frac{1}{1+\beta^2} + R_{\rm high}\frac{\beta^2}{1+\beta^2},
\end{equation}
which was introduced in \citet{2016A&A...586A..38M}. 
In this scheme, $T_{\rm e} \simeq T_{\rm i}/R_{\rm low}$ in the strongly magnetized region such as in the jets, while $T_{\rm e} \simeq T_{\rm i}/R_{\rm high}$ in the weakly magnetized, gas pressure dominant region such as in the midplane disk. 
Here, we adopt parameters of $R_{\rm low}=1$ and $R_{\rm high}=73$ for our fiducial model, corresponding with relatively high (or low) electron temperature in the jet (disk) region.\footnote{See Fig.~\ref{fig:GRMHDmaps} for maps of physical quantities including the electron temperature in our GRMHD model.}

While the electron temperature are thought to be lower in the disk than in the jet from comparison with spectral energy distributions and RMs \citep{2013A&A...559L...3M,2017MNRAS.468.2214M} and with two-temperature calculations \citep{2010MNRAS.409L.104H,2018ApJ...864..126R,2019PNAS..116..771K,2019MNRAS.486.2873C}, a wide range of the proton-electron temperature ratio both in the disk and in the jet is suggested. 
As far as the radiative cooling is not incorporated, recently, \citet{2021arXiv210609272M} demonstrated that this $R-\beta$ prescription and choice of parameters ($R_{\rm low}=1$, $R_{\rm high}=1-160$) are consistent with the turbulent- and magnetic reconnection- heating prescriptions in GRMHD simulations with electron thermodynamics, in comparison of images at 230~GHz obtained from GRRT calculations based on them. 
We discuss other choices for $R_{\rm high}$ in sub-subsection \ref{subsec:Rhigh}, focusing on the difference between the low-$T_{\rm e}$ and high-$T_{\rm e}$ disks.

\begin{figure*}[t!]
\plotone{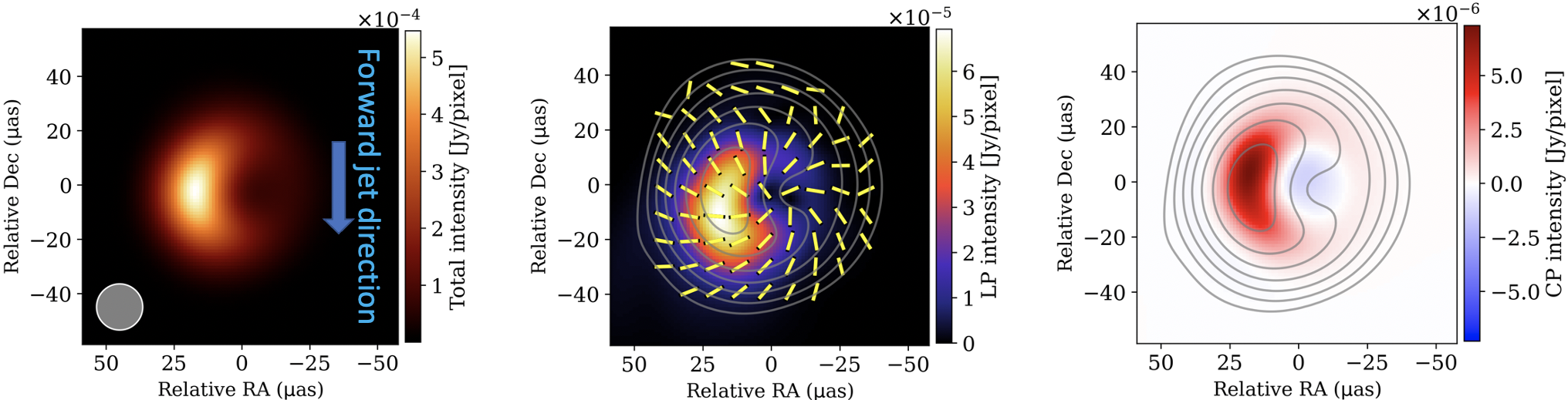}
\caption{Same as Fig.~\ref{fig:raw} but for the convolved images with Gaussian beam of $17~ {\rm \mu as}$. {The beam size is shown in the bottom left in the left image.} Line contour of the total intensity is overplotted on the LP and CP images. 
The total image shows an asymmetric ring feature with no visible jet feature. 
Although the LP fraction is not so large ($\approx 10-20\%$), compared with those of the raw image, the LP vectors show a much more ordered pattern. 
The CP image gives a ring feature, consisting of significant components ($\gtrsim 1\%$ in fraction) with positive signs. 
Note that the images consist of $100\times 100$ pixels, more coarsely than those in Fig. \ref{fig:raw}.
\label{fig:conv}}
\end{figure*}

\subsection{Polarimetric radiative transfer in Kerr metric}

We perform full polarimetric radiative transfer with the Stokes parameters $(\mathcal{I,Q,U,V})$ along light paths in Kerr metric determined by the general relativistic ray-tracing method, using our code developed and implemented in \citet{2020PASJ...72...32T,2020arXiv201205243T}. 
The polarized radiative coefficients for the ultrarelativisic thermal distribution of electrons, the synchrotron emissivities $(j_I,j_Q,j_U,j_V)$, synchrotron self absorption $(\alpha_I,\alpha_Q,\alpha_U,\alpha_V)$ and Faraday effects $(\rho_Q,\rho_U,\rho_V)$, are implemented into the code, based on previous works \citep{1996ApJ...465..327M,2008ApJ...688..695S,2016MNRAS.462..115D}. 
Further, the coefficient of Faraday rotation $\rho_V$ is modified for accurate descriptions in the low temperature and frequency ratio region, as discussed in \citet{2020MNRAS.494.4168D,2020MNRAS.498.5468R}. 

We adopt a black hole mass of $M_\bullet = 6.5\times 10^9 M_\odot$ and a distance of $16.7~{\rm Mpc}$ for M87* \citep{2007ApJ...655..144M,2011ApJ...729..119G,2019ApJ...875L...5E}, which give an angular diameter of $\simeq 3.8 ~{\rm \mu as}$ on the celestial sphere corresponding with the  gravitational radius $r_{\rm g}$. 
An inclination angle $i$ of the camera is set to $160^\circ$, nearly face-on to the midplane disk, while other inclinations are also discussed in subsection \ref{subsec:inclination}.
We set the camera at $r=10^4r_g$ and calculate radiative transfer within $r \le 100r_{\rm g}$,\footnote{We confirmed that significant radiative processes occur within $r \le 100r_{\rm g}$ for our models. (see, e.g., Fig.~\ref{fig:radcoeff} for estimation maps of the radiative coefficients.)} to present snapshot images with the ``fast-light'' approximation. 
We also scale a mass accretion rate onto the black hole $\dot{M}$ to reproduce the observed flux of $\approx 0.5~{\rm Jy}$ at 230~GHz \citep{2019ApJ...875L...4E}. 
$\dot{M} = 6 \times 10^{-4} M_\odot / {\rm yr}$ for our fiducial model, which is comparable with those in the ``passed'' MAD models in \citet{2021ApJ...910L..13E}. 

In the whole of this work, the sigma cut-off of $\sigma_{\rm cutoff} = 1$, removing the region with the plasma magnetization $\sigma \equiv B^2/4\pi \rho c^2 > \sigma_{\rm cutoff}$ in radiative transfer calculation, is adopted in order to avoid unphysical effects arising because of low density floors set in the MHD simulation. 
In subsection \ref{subsec:future}, we discuss the validity of our results with the sigma cutoff comparing to a case without the cutoff.

\section{LP-CP flux separation}\label{sec:results}

\begin{figure*}[t!]
\plotone{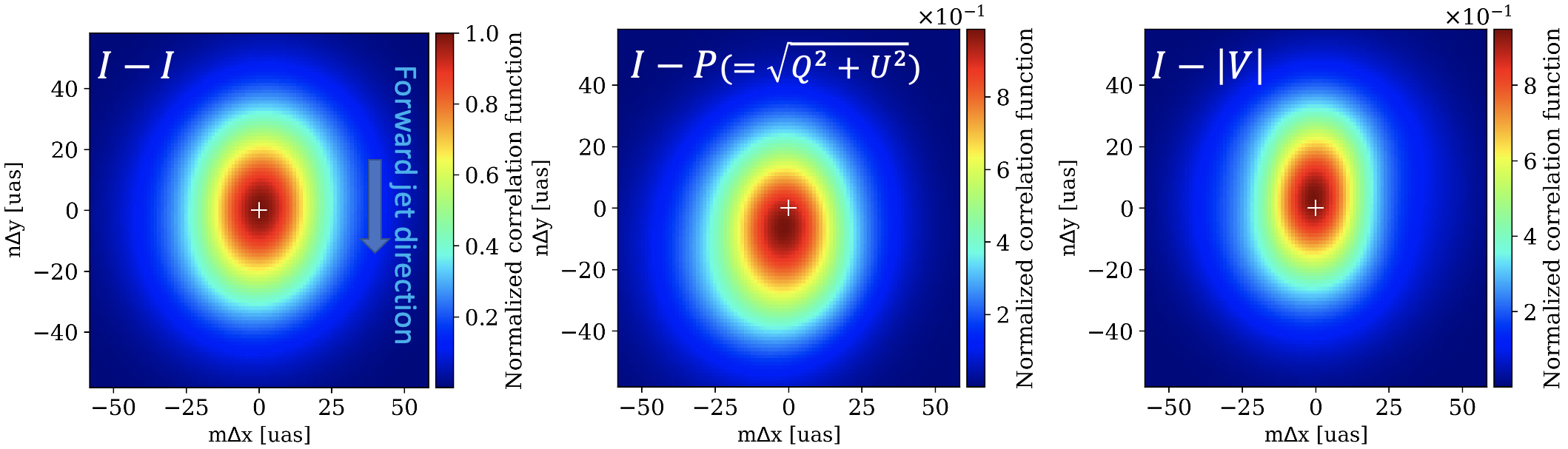}
\caption{Two-dimensional auto- and cross-correlation functions for the total, LP, and CP images in Fig.~\ref{fig:conv}. 
White ``+'' in the maps indicates the centroid position of the map, where we set $(m\Delta x, n\Delta y) = (0,0)$.
Left: auto-correlation of total intensity, Stokes $I$, with a peak at the origin by definition. 
Center: cross-correlation between $I$ and the LP intensity $P=\sqrt{Q^2+U^2}$. 
The position of the correlation maximum is shifted downwards with respect to the centroid position, reflecting the LP intensities distributed downwards relatively to the total intensities on the image.  
Right: cross-correlation between $I$ and the absolute CP intensity $|V|$. In contrast to $I-P$, it gives a peak in the upwards due to the CP intensities located upwards relatively to the total intensities. 
Three maps are normalized by the definition of the correlation coefficient, so that the auto-correlation of Stokes $I$ yields 1 at the origin.
\label{fig:correlations}}
\end{figure*}

\subsection{Polarization images\label{subsec:figures}}

The raw images of the total intensity ($I$), the LP intensity ($Q,U$), and the CP intensity ($V$) at 230~GHz obtained by the polarimetric radiative transfer calculation are shown in the left to right panels in Fig.~\ref{fig:raw}, respectively. 
The total intensity (Stokes $I$) image in the left panel gives the photon ring, which is a circle with a radius of $\approx 20~ {\rm \mu as}$ and is beamed in the left side due to the gravity of the spinning black hole and to the relativistic beaming effect by helical motion of plasma. In addition, a hint of tail-like jet extends downwards in the image. 

In the central panel, the LP intensity (Stokes $\sqrt{Q^2+U^2}$) distributes tracing the total intensity, with fractions of $\sqrt{Q^2+U^2}/I \sim 50 \%$ in individual pixels. 
The noteworthy features are that the LP vectors are not ordered but show chaotic features because of the Faraday rotation occurring within the disk. 
The CP (Stokes $V$) image in the right panel shows an asymmetric ring-like feature with positive sign, which traces the photon ring in the total intensity image. 
The CP components with a fraction up to $|V|/I \sim 10\%$ in individual pixels are significantly stronger than those of the synchrotron emission ($|V|/I \sim 1\%$), implying that these result from an amplification process through the Faraday conversion in hot region ($T_{\rm e} \gtrsim 10^{10}~{\rm K}$) near the black hole, as was firstly demonstrated in \cite{2020PASJ...72...32T}. 
These features of rotation of the LP vectors and amplified CP components in individual pixels  agree well with the results in our previous work \citep{2020PASJ...72...32T}, based on two-dimensional semi-MAD models.
Such monochromatic (uniform in +/- signs) CP ring features are also seen in the theoretical models in \citet{2020A&A...641A.126B,2021arXiv210300267M,2021MNRAS.tmp.1263R,2021arXiv210105327E}.\footnote{See also \citet{2021MNRAS.tmp.1263R} for a discussion about the sign-flipping sub-rings.}

At the same time of calculating of the images, we also calculate the intensity-weighted optical depths for each light ray to see the two Faraday effects and SSA; e.g., 
\begin{equation}
{\tau_{{\rm Frot},I} \equiv \int \rho_V I(s) {\rm d}s/I_{\rm fin}}
\end{equation}
for Faraday rotation depth, {where $I_{\rm fin}$ is a final value of Stokes $I$ in each pixel \citep{2021ApJ...910L..13E}}.  
We further average them over the image, weighting by the total intensity in each pixel, and  obtain the image-averaged, intensity-weighted optical depths, $\langle \tau_{{\rm Frot},I} \rangle \simeq 1.7 \times 10^2$, $\langle \tau_{{\rm Fcon},I} \rangle \simeq 1.1$, and $\langle \tau_{{\rm SSA},I} \rangle \simeq 0.1$. (Here $\tau_{{\rm Fcon},I} \equiv \int  \sqrt{\rho_Q^2+\rho_U^2} I(s) {\rm d}s/I_{\rm fin}$, $\tau_{{\rm SSA},I} \equiv \int \alpha_I I(s) {\rm d}s/I_{\rm fin}$.) 
From these we understand that plasma near the black hole is optically thick for the Faraday effects but thin for the SSA for the lights at 230~GHz, typically. 
As a result, we obtained clear photon-ring image but dim foreground jet image, the scrambled LP vectors, and the amplified CP components {(see Appendix \ref{app:RTplot} for GRRT process for a pixel on the image).}

Next, we show convolved (or blurred) images by Gaussian beam with size of $17~{\rm \mu as}$ in Fig.~\ref{fig:conv}. 
We have chosen this beam size, bearing the EHT observation at 230~GHz in mind. 
In the central panel, we can see an asymmetric ring feature without extended jet components, which is consistent with the EHT observation of M87* in 2017 \citep{2019ApJ...875L...1E,2019ApJ...875L...5E}.

While the Gaussian convolution on the whole tends to reduce the LP fraction in the central panel, with values of $\sim 10 - 20 \%$, it recovers a ``hidden'' ordered structure of the LP vectors in a hybrid pattern of azimuthal and radial ones, reflecting (i.e.~being perpendicular with) the magnetic field configurations at synchrotron emission. 
(Note that the magnetic field configuration is toroidally dominated in the disk region, while it has significant poloidal components in the jet region, roughly.)
Further, a distribution of the LP intensity is shifted downwards by $\lesssim 10~{\rm \mu as}$ in the image, compared with those of the total intensity. 
This is because the LP vectors which originate in the downstream region of the approaching jet are not affected by the Faraday rotation because of small Faraday rotation depth, and thus keep a well-ordered structure in emission. 
Those originating from the upstream or the photon ring are, by contrast, chaotically rotated in the disk region (see also \cite{2017MNRAS.468.2214M,2020MNRAS.498.5468R}) and drastically decrease their intensity by the convolution of observational beam (the beam depolarization). 
Such features as those seen in the convolved LP map agree with the observations of M87* in \citet{2021ApJ...910L..12E}, where the downward direction on the images in this work corresponds to the north-east on their observational images.

\begin{figure*}[t!]
\plotone{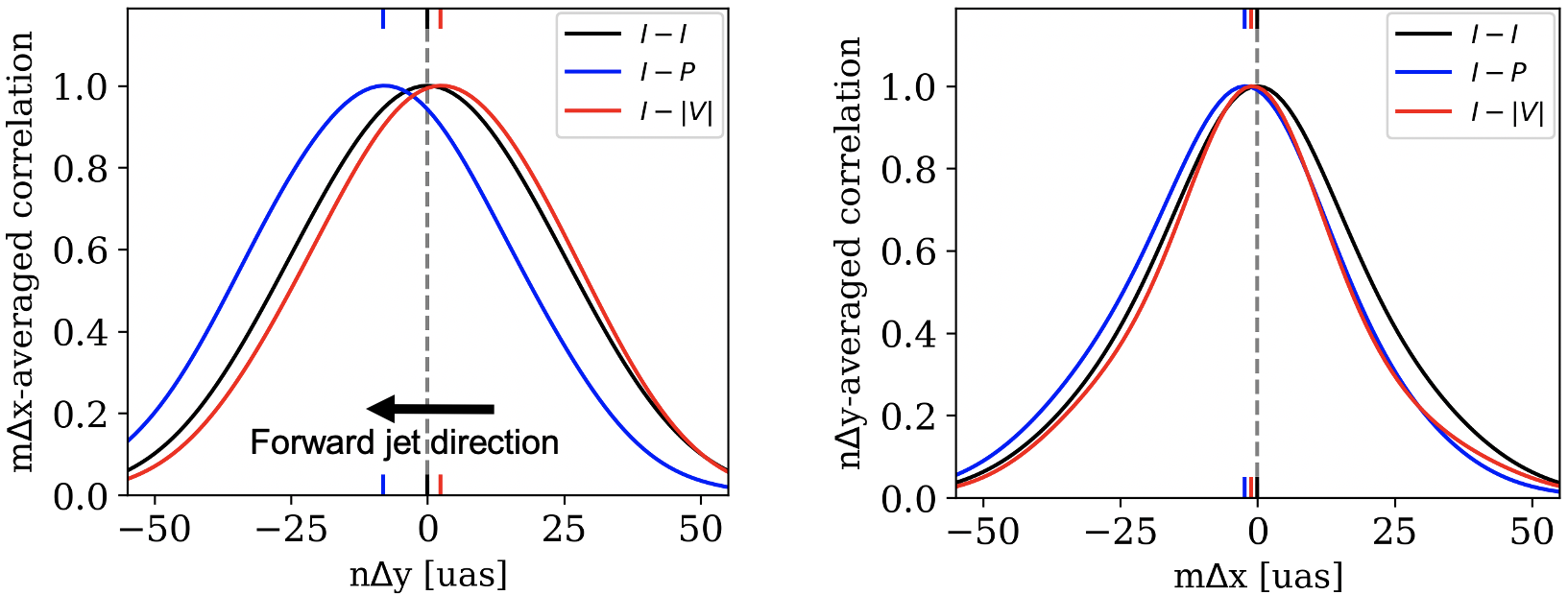}
\caption{One-dimensional auto- and cross- correlation functions calculated by integrating the two dimensional correlation functions in the $x$- (left) and $y$- (right) directions. 
The former corresponds to the vertical direction along the jet and BH spin axis, while the latter the horizontal direction perpendicular with the jet. 
Hatches in the upper and lower axes demarcate the places of the maximum correlation, {the relative offsets between two kinds of intensity distributions}.
Each profile is normalized by its maximum value. 
\label{fig:xyprofiles}}
\end{figure*}

Meanwhile, the CP intensity in the left panel is distributed around the photon ring in the total intensity image, with fractions of $|V|/I \sim $ a few percent. 
The centroid of the CP intensity is slightly shifted upwards by $\lesssim 5~{\rm \mu as}$, compared with that of the total intensity. 
This is because only those around the photon ring and from the receding jet can be amplified in energetic region near the black hole through the Faraday conversion from the LP components, and the emission from more background is more effectively converted with larger optical depth for the Faraday conversion. 

Further, we confirmed that these 230~GHz images give the net LP fraction of $2.6~\%$, the average LP fraction of $10.4~\%$ when convolved with $20~{\mu as}$ Gaussian beam, and the net CP fraction of $0.76~\%$. All of these fractions satisfy the observational constraints in the model scoring in \citet{2021ApJ...910L..13E}. 

Next, we describe the results for the total, LP and CP intensities to the nearly-face-on observer and their origin in subsection \ref{subsec:Faraday}.

\subsection{Correlation functions for the images}\label{subsec:correlations}

\subsubsection{Correlations in the Cartesian coordinates $(x,y)$}

\begin{figure*}[t!]
\plotone{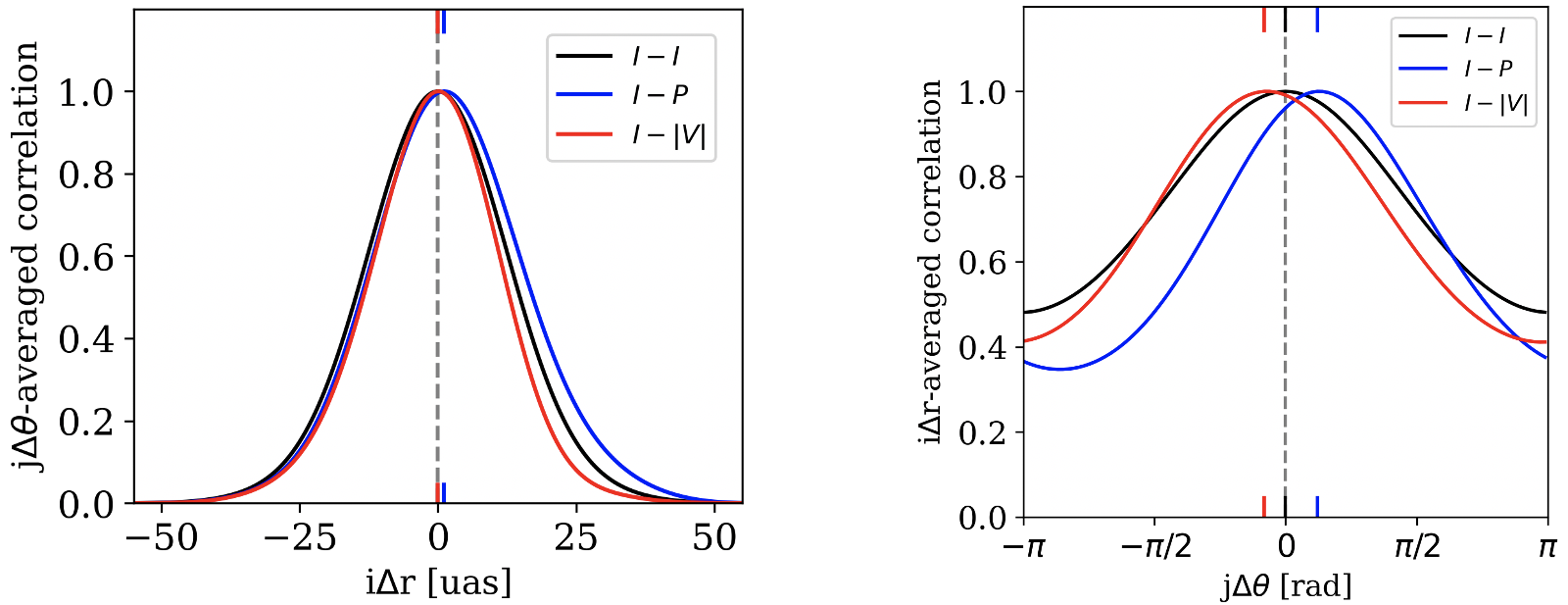}
\caption{{Same as the one-dimensional correlation profiles in Fig.~\ref{fig:xyprofiles} but in the polar coordinates.} 
The left panel shows the correlations in the radial direction, while the right one shows  those in the azimuthal angle. 
Hatches in the upper and lower axes demarcate the places of the maximum correlation, the relative offsets between two kinds of intensity distributions. 
Two-dimensional polar correlation maps are shown in Appendix \ref{app:polar}.
\label{fig:rthprofiles}}
\end{figure*}

In Fig.~\ref{fig:correlations}, we show correlation functions in the Cartesian coordinates $(x,y)$; that is, auto-correlation of total intensity (Stokes $I$), cross-correlations between $I$ and the LP intensity $P$ ($\equiv \sqrt{Q^2+U^2}$), and between $I$ and the absolute CP intensity ($|V|$), which are calculated from the convolved images in Fig.~\ref{fig:conv}. 
The correlation functions are calculated for a pair of $I$ and $S$ ($= I$, $P$, or $|V|$) at each pixel of $(x_i,y_j) = (i\Delta x, j\Delta y)$ by the following way:
\begin{equation}\label{eq:corr}
\begin{split}
	\{I-S\}(m\Delta x, n\Delta y) \equiv \frac{\{I-S\}_{\rm num}(m\Delta x, n\Delta y)}{\{I-S\}_{\rm den}}  \\
	(m,n=0,\pm1,\pm2,...), 
\end{split}
\end{equation}
where: 
\begin{equation}
	\{I-S\}_{\rm num}(m\Delta x, n\Delta y) \equiv \sum_{i=1}^N\sum_{j=1}^N I(x_i,y_j)S(x_{i+m},y_{j+n}), 
\end{equation}
and: 
\begin{equation}
	\{I-S\}_{\rm den} \equiv \sqrt{\left\{\sum_{i=1}^N\sum_{j=1}^N[I(x_i,y_j)]^2\right\}\left\{\sum_{i=1}^N\sum_{j=1}^N[S(x_i,y_j)]^2\right\}}, 
\end{equation}
so that $\{I-I\}(0,0) = 1$.
Here $\Delta x$ and $\Delta y$ are the size of pixels in the $x$- (horizontal) and $y$- (vertical) direction, respectively, $N=100$ is the number of pixels in each direction.\footnote{{Here, we take more coarse pixel composition in the convolved images than in the raw images in Fig. \ref{fig:raw} for faster calculation of the correlation functions. This does not change the results significantly because the size of convolutional beam is much larger than the pixel size.}}  

The left panel in Fig.~\ref{fig:correlations} represents the two-dimensional auto-correlation functions of Stokes $I$. We see that the correlation lengths (at half maximum $= 0.5$) of $20-30~ {\rm \mu as}$ in various directions on the $(m\Delta x, n\Delta y)$-plane, and that they show a vertically elongated shape. 
This reflects the vertically elongated emission profile in the left side of the ring (see the left panel of Fig.~\ref{fig:conv}). 
The cross-correlation between the total and LP intensities, $I-P$, in the central panel shows a shape similar to that of the auto-correlation $I-I$ in the left, but with a peak vertically shifted and at $\sim -8~ {\rm \mu as}$, meaning that the LP flux has a tendency to distribute downwards by $\sim 8~ {\rm \mu as}$ relatively to the total flux (see  Fig.~\ref{fig:conv}, see also the statements in subsection \ref{subsec:figures}). 
In the right panel, conversely, the cross-correlation map between the total and CP intensities $I-|V|$ yields a peak at $\sim +2~ {\rm \mu as}$ upwards in the $n\Delta y$-direction.
This is because the CP flux originates from the vicinity of the black hole and the counter-side (background) jet region, compared with the total flux (and LP flux).
In this way, we can quantitatively assess the distinction between the total, LP, and CP intensity distributions through the cross-correlation analyses.

In order to more clearly examine the auto- and cross- correlation functions, we display in Fig.~\ref{fig:xyprofiles} one-dimensional cross sections of the two-dimensional correlation functions displayed in Fig.~\ref{fig:correlations} in the vertical (left panel) and horizontal directions (right panel), respectively. 
The quantities are normalized by their maximum values.
In the left panel, the cross-correlations of $I-P$ and $I-|V|$ have their peaks at negative and positive $n\Delta y$ values (corresponding to the downward and upward direction), respectively, while 
those in the $m\Delta x$-direction in the right panel do not show significant deviation from the auto-correlation profile, except a small transition of $I-P$ to the left side corresponding to the tendency of the LP flux left-leaning relative to the total flux.
The autocorrelation profiles of $I-I$ in both of two panels have their peaks at the center of $(n\Delta y, m\Delta x) = (0,0)$ by definition.

\begin{figure*}[t!]
\epsscale{1.2}
\plotone{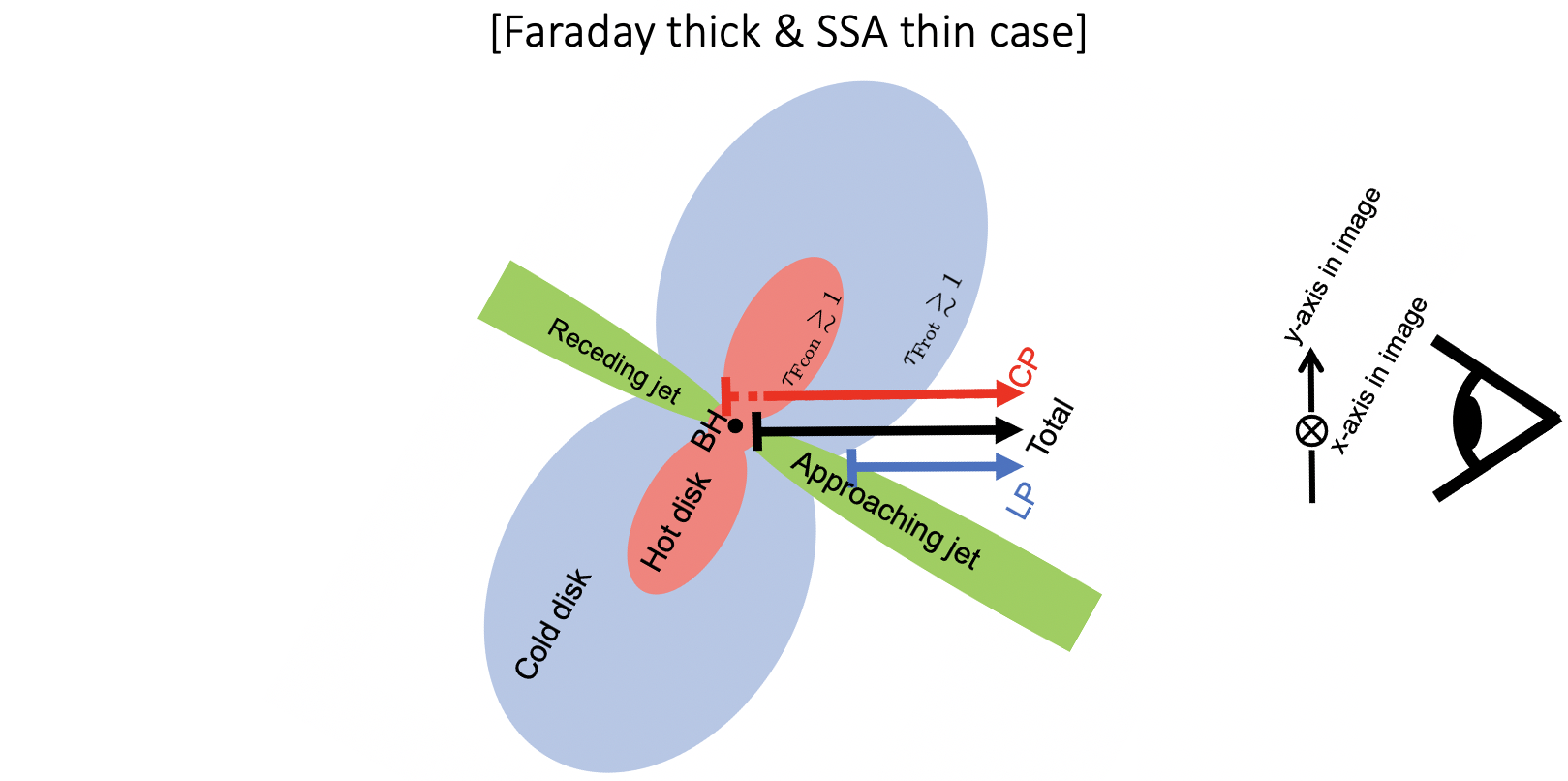}
\caption{
A schematic picture displaying the rough locations where the total, LP and CP fluxes are generated and their main propagation path to a distant observer located at the far right position. This illustrates the case where the system is Faraday thick but SSA thin (e.g., at 230~GHz for our fiducial model; see Appendix \ref{app:RTplot} for estimation maps of synchrotron emission and two Faraday effects). 
Here, the total flux is dominated by the emission from the jet base (green) and the inner hot disk (red) and the LP flux originates from the foreground jet, whereas the CP flux is dominated by the emission from the inner hot disk via the Faraday conversion (see subsection \ref{subsec:Faraday} for detail). 
Note that the original LP flux emitted from the inner hot disk and the background jet is strongly depolarized by the Faraday rotation when propagating through the inner hot, and the outer cold disk (blue). 
{The LP-CP separation becomes more enhanced at lower frequencies (say, 86~GHz; see also subsubsection \ref{subsubsec:86GHz} and Fig.~\ref{fig:opticaldepths}) or for higher mass accretion rates (see also subsection \ref{subsec:Mdot}).}
\label{fig:Faraday}}
\end{figure*}

\begin{table*}[]
\begin{center}
  \begin{tabular}{c||c|c||c|c||c|c||c}
    frequency & $I-P$ peak & $I-|V|$ peak & $P_{\rm tot}/I_{\rm tot}$ & $|V_{\rm tot}|/I_{\rm tot}$ & $\langle \tau_{{\rm Frot},I} \rangle$ & $\langle \tau_{{\rm Fcon},I} \rangle$ & figure number \\ \hline \hline
    230~GHz & $-8~{\rm \mu as}$ & $+2~{\rm \mu as}$ & $2.6\%$ & $0.76\%$ & $1.7 \times 10^2$ & $1.1$ & Figs.~\ref{fig:conv}, \ref{fig:correlations}, \ref{fig:xyprofiles} \\ \hline
     86~GHz & $-25~{\rm \mu as}$ & $+17~{\rm \mu as}$ & $4.2\%$ & $0.64\%$ & $1.3\times 10^3$ & $12$ & Figs.~\ref{fig:86GHz}, \ref{fig:86GHz_correlations}, \ref{fig:86GHz_xyprofiles} \\
  \end{tabular}
\end{center}
  \caption{Comparison between the various polarization quantities at 230~GHz and at 86~GHz; the vertical peak shifts of cross-correlation functions $I-P$ and $I-|V|$, the total LP and CP fractions, $P_{\rm tot}/I_{\rm tot}$ and $|V_{\rm tot}|/I_{\rm tot}$, and the image-averaged intensity-weighted optical depths for the Faraday rotation and conversion, $\langle \tau_{{\rm Frot},I} \rangle$ and $\langle \tau_{{\rm Fcon},I} \rangle$, from the left to the right. 
  \label{table:230and86}
  }
\end{table*}

\subsubsection{Correlations in the polar coordinates $(r,\theta)$}\label{subsubsec:polar}

In the previous subsection, we analyzed the correlations in the Cartesian coordinates $(x,y)$ on the images. 
This choice is reasonable for the M87 jet, because the direction, or position angle, of the approaching jet has been accurately constrained and established for a wide spatial  range (from $\sim {\rm \mu as}$- to $\sim {\rm kpc}$- scale) through multi-wavelength observations (e.g., \cite{2021arXiv210406855A}). 
{It can be, however, advantageous o use a polar coordinate system when performing this type of cross correlation analysis on a single epoch images of ring like features. 
With this in mind, here we introduce correlation analyses in the polar coordinates $(r,\theta)$ on the images with the origin at $(x,y) = (0,0)$. This is particularly useful for the M87* images at 230~GHz, because these total intensity images show ring-like features, symmetrical about the origin of the images, as seen in Fig.~\ref{fig:conv} or the actual observations by \cite{2019ApJ...875L...1E,2021ApJ...910L..12E}.}
The correlation functions in the polar coordinates are calculated from the Stokes parameters at each pixel of $(r_k,\theta_l) = (k\Delta r, l\Delta \theta)$ in the following way:
\begin{equation}\label{eq:rthdef}
\begin{split}
	\{I-S\}(i\Delta r, j\Delta \theta) \equiv \\
	\frac{\sum_{k=1}^N\sum_{l=1}^N r_k r_{k+i} I(r_k,y_l)S(r_{k+i},y_{l+j})}{\sqrt{\left\{\sum_{k,l=1}^Nr_k^2I(r_k,\theta_l)^2\right\}\left\{\sum_{k,l=1}^Nr_k^2S(r_k,\theta_l)^2\right\}}} \\
	(i, j = 0,\pm1,\pm2,...),
\end{split}
\end{equation}
so that $\{I-I\}(0,0)=1$.
Here $\Delta r$ and $\Delta \theta$ are the size of pixels in the $r$- and $\theta$- directions, respectively. The factors of $r_k$ and $r_{k+i}$ in the summation come from the area element in the two-dimensional polar coordinates, $r{\rm d}r{\rm d}\theta$.

We show the auto- and cross- correlations in the polar coordinates in Fig.~\ref{fig:rthprofiles}, respectively\footnote{See Appendix \ref{app:polar} for two-dimensional correlation maps in polar coordinates.}. 
Autocorrelation $I-I$ has peaks at $(i\Delta r, j\Delta \theta) = (0,0)$ by definition. 
In the left panel, two cross-correlations show similar profiles to that of the auto-correlation with correlation length (at half maximum) of $\sim 15~ {\rm \mu as}$, although the cross-correlation $I-P$ is slightly shifted in the larger $i\Delta r$-direction. 
We can see, in the right panel, narrow $I-P$ profile with its peak at $j\Delta \theta \sim +\pi/8$ and wide $I-|V|$ profile with its peak at $j\Delta \theta \sim - \pi/16$. 
Here positive (or negative) $j\Delta \theta$ corresponds to the clockwise (counter-clockwise) direction around the center of the images. 
Therefore, these results in polar coordinates quantitatively describe the fact that the total, LP and CP fluxes on the images in Fig.~\ref{fig:conv} are located roughly on the same circle in the order of the LP, total, and CP intensities, in the clockwise direction, if starting from 0 o'clock (i.e. $y$-axis on the images).

\subsection{Schematic of the Faraday rotation and conversion around the black hole}\label{subsec:Faraday}

In the previous subsections, we found a separation of the LP and CP intensities.
This is because the LP (or CP) components are mainly from the downward (upward) position with respect to the bright part of the total intensity distribution. 
Here, we interpret these polarimetric features by using a schematic picture of the Faraday rotation and conversion effects around the black hole.

Fig.~\ref{fig:Faraday} illustrates the case, in which the system is Faraday thick but SSA thin, as in the case that we encounter in subsection \ref{subsec:figures} (see upper row of Table \ref{table:230and86} for the Faraday optical depths at 230~GHz, {see also Appendix \ref{app:RTplot} for estimation maps of synchrotron emission and two Faraday effects}). 
Here, the total flux is dominated by the emission from the jet base (green) and the inner hot disk (red) around the black hole. 
As for polarization components, the contribution to the LP components from the downstream of the approaching (foreground) jet dominates over that from the receding (background) jet, since the latter is suppressed by strong Faraday rotation and depolarization when propagating through the inner hot and the outer cold parts of the disk. 
Meanwhile, the CP image is dominated by the components from the receding jet or the inner hot disk near the black hole, which are converted over larger Faraday conversion depths than those from the approaching jet.
Compared with the total intensity distribution, therefore, the LP flux is distributed in the downstream side of the jet, whereas the CP flux is distributed in the counter-side jet or around the photon ring. 

As introduced above, we find that the separation of the LP and CP components from the total intensity distribution can be understood in terms of the Faraday rotation and conversion in the jet-disk structure around the black hole. 
In the following section, we survey such a LP-CP flux separation for plasma and observational parameters.

\section{Conditions for the LP-CP flux separation}\label{sec:discussion}

\subsection{Dependence on the electron-temperature parameter $R_{\rm high}$}\label{subsec:Rhigh}

\begin{figure}[t!]
\epsscale{1.2}
\plotone{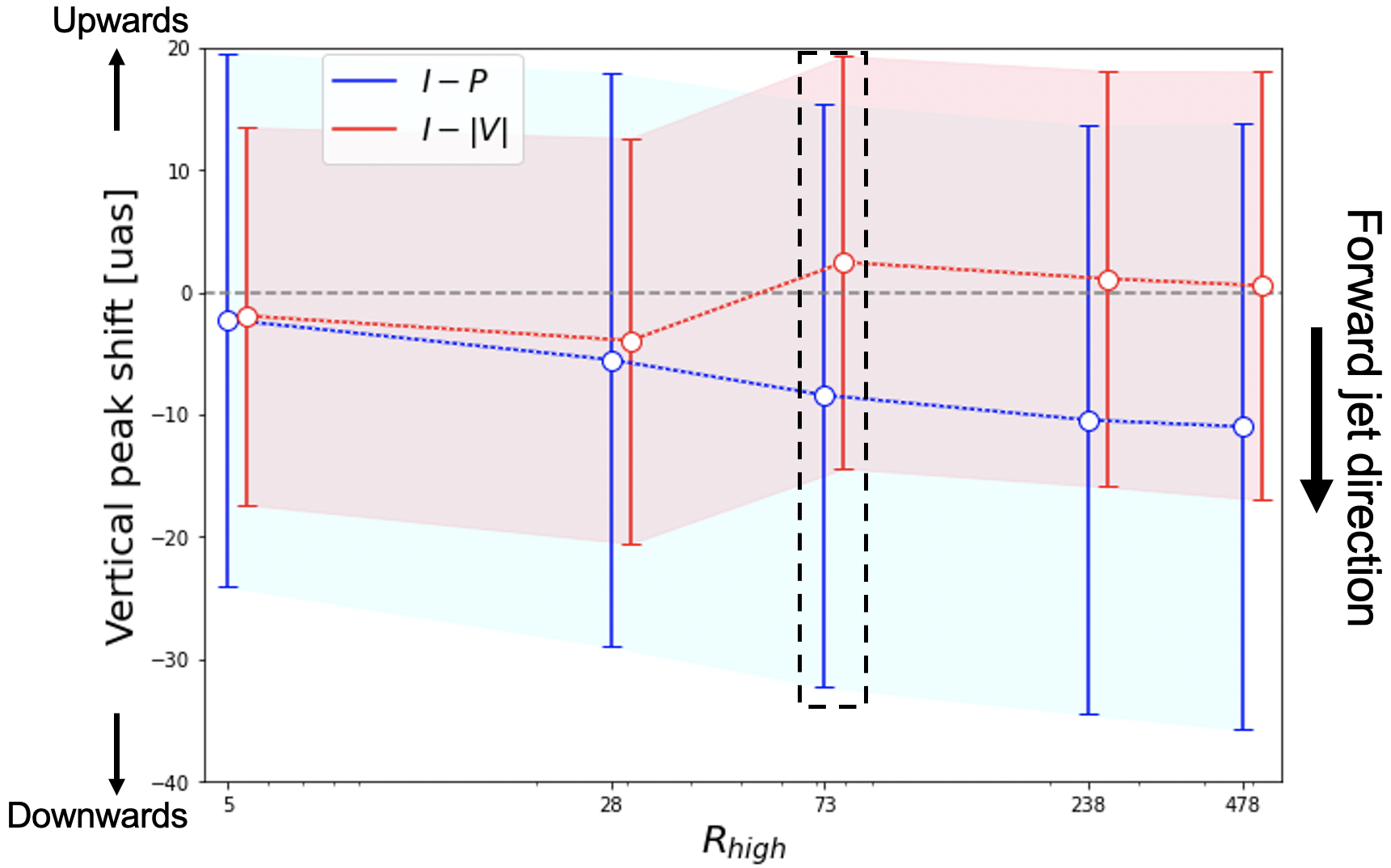}
\caption{Vertical peak shifts of cross-correlation functions $I-P$ and $I-|V|$ at 230~GHz for five electron-temperature parameters in the disk, $R_{\rm high}=5,28,73,238$, and $478$. Circles correspond to the peaks, and bars the $1\sigma$ ranges of fitted Gaussian functions. 
The one boxed with dotted line corresponds to the peak and width of the cross-correlation  profiles our fiducial model in Fig.~\ref{fig:xyprofiles}. 
All of the based profiles are shown in Fig.~\ref{fig:IPxyprofiles_Rhighs} in Appendix \ref{apdx:profiles}.
\label{fig:IPIV_Rh_ypeak}}
\end{figure}

In the model discussed so far, we fixed the parameters of $(R_{\rm low}, R_{\rm high}) = (1,73)$ as our fiducial ones to describe proton-electron coupling in the jet-disk structure (see Eq.~\ref{eq:M16}). 
While this choice seems reasonable in view of the recent GRMHD simulations including electron thermodynamics \citep{2021arXiv210609272M}, a large range of $R_{\rm high}$ values are also suggested.
To see how the results depend on this particular choice of the parameter, we calculate the images at 230~GHz for five models with $R_{\rm high} = 5, \ 28, \ 73 \, ({\rm fiducial}), \ 238, \ 478$, for which we scale the mass accretion rate of $\dot{M} = ( \, 1.2, \ 4, \ 6, \ 10, \ 12 \, ) \times 10^{-4} M_\odot / {\rm yr}$ to reproduce the total flux of $0.5 ~{\rm Jy}$ in M87*, respectively, and analyze their cross-correlation functions $I-P$ and $I-|V|$. 

In Fig.~\ref{fig:IPIV_Rh_ypeak}, we show vertical peak separation of the cross-correlations $I-P$ and $I-|V|$ as functions of $R_{\rm high}$. 
We also fit the vertical profiles with Gaussian function and plot their $1\sigma$ ranges as error bars on the figure (see Appendix \ref{apdx:profiles} for the raw profiles of the cross-correlations).
The peak positions of the $I-P$ correlation shift downwards (in the downstream of the jet) by up to $\sim 10~{\rm \mu as}$, as $R_{\rm high}$ increases. 
This is because the higher $R_{\rm high}$ is, the lower becomes electron temperature in the disk (with high plasma-$\beta$). 
The higher $R_{\rm high}$ value, hence, requires higher mass accretion rate to reproduce the observed flux. 
As a result, such lower temperature and higher mass accretion give rise to stronger Faraday rotation in the disk (since $\tau_{\rm Frot} \propto nBT_{\rm e}^{-2} \propto \dot{M}^{3/2} R_{\rm high}^2$ in the disk, as pointed out by \citet{2017MNRAS.468.2214M}). 
Therefore, the LP intensities originate in a more downstream region for higher $R_{\rm high}$, as shown in the picture of Fig.~\ref{fig:Faraday}.

\begin{figure*}[t!]
\plotone{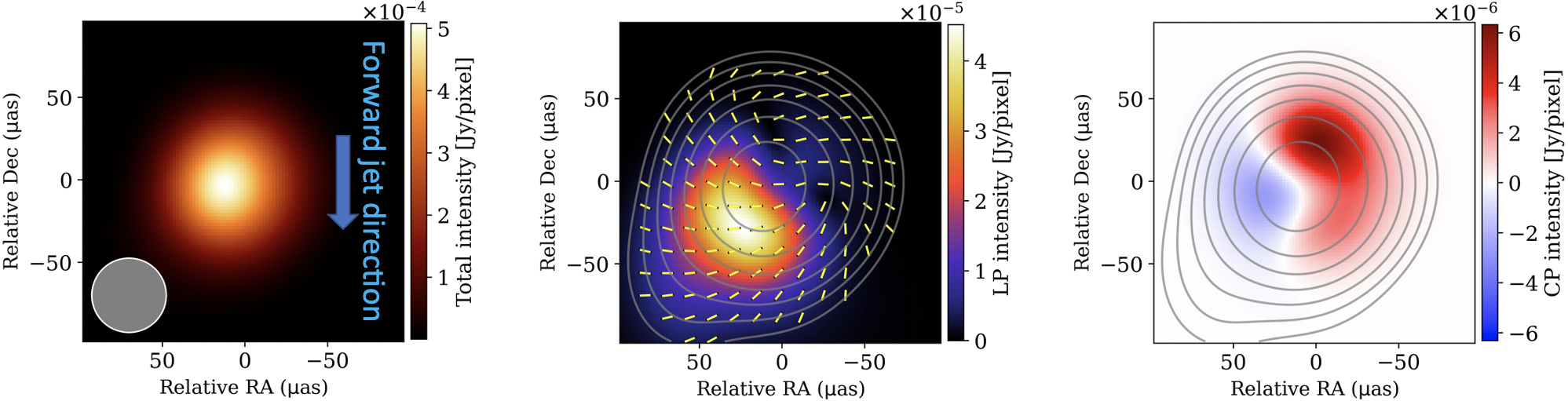}
\caption{Same as the convolved images in Fig.~\ref{fig:conv} but at 86~GHz with a larger field of view, being convolved with Gaussian beam of $45~ {\rm \mu as}$ {(shown in the bottom-left of the left image)}. The beam size is a bit smaller than those in the present global VLBI observations at 86~GHz such as GMVA, (e.g., $123 \times 51 {\rm \mu as}$; \citet{2018A&A...616A.188K}) with future observations in mind.   
\label{fig:86GHz}}
\end{figure*}

In contrast, the CP intensity distributions exhibit somewhat distinct trend; that is, the peak positions of the $I-|V|$ correlation behave irregularly for varying $R_{\rm high}$ values. 
This is because, unlike the Faraday rotation, the Faraday conversion is more enhanced for higher temperature and higher accretion rate, (since $\tau_{\rm Fcon} \propto nB^2T_{\rm e} \propto \dot{M}^2 R_{\rm high}^{-1}$ in the disk,) so that higher $R_{\rm high}$ and $\dot{M}$ values may not necessarily yield stronger Faraday conversion. 
In two models with the higher $R_{\rm high}$ values ($238$ and $478$), therefore, smaller  separations are seen in the red bars in Fig.~\ref{fig:IPIV_Rh_ypeak},  almost down to zero for $R_{\rm high} = 478$ (which contrast the $I-P$ correlations shown in blue bars).
In two models with low $R_{\rm high}$ values ($5$ and $28$), conversely, break the tendency of the upwards CP relative to the total intensity, with the $I-|V|$ peaks in $n \Delta y > 0$. 
We can attribute these results to the fact that in these high electron-temperature models, the polarized emission from the broad, high-temperature disk around the midplane, which has relatively turbulent magnetic structure, becomes comparable with that from the approaching and receding jets, and the description of the emission from the twin jets and inner disk is no longer applicable (see also Fig.~\ref{fig:IPIV_Rh_ypeak_ph180} in Appendix \ref{apdx:behind} for cases where we observe the object from behind about the $z$-axis, with $\phi_{\rm camera} = 180^\circ$).

These disk temperature survey in our model suggest that, if both the LP and CP separations would be observed in future observations, we could associate it with weaker proton-electron coupling (leading lower electron temperature) in the disk than in the jet. 
In the following subsections, we further survey the dependence of the LP-CP separation on observational frequencies $\nu$, observer's inclination angles $i$, and accretion rates onto the black hole $\dot{M}$, for our fiducial model $(R_{\rm low}, R_{\rm high}) = (1,73)$ with the low electron-temperature disk.

\subsection{The LP and CP separations at multi-frequencies}\label{subsec:mf}

So far we have seen clear tendencies in the cross-correlation functions between the total intensities and the LP or CP intensities for our fiducial model. 
These are caused by the Faraday effects which occur when radiation passes through the magnetized plasmas in the disk region, as described in subsection \ref{subsec:Faraday}.
Since the Faraday effects are known to be more enhanced for lower frequency (longer wavelength) observations ($\rho_V \propto \nu^{-2} \sim \lambda^2$ for the Faraday rotation and $\rho_{Q,U} \propto \nu^{-3} \sim \lambda^3$ for the Faraday conversion; \cite{2008ApJ...688..695S,2016MNRAS.462..115D}), 
we can expect that the LP-CP separation which we found at 230~GHz should be even clearer at lower frequencies, e.g., at 86~GHz. Therefore, we next survey the wavelength-dependence of the polarimetric correlations on the images, based on an angular resolution of global VLBI observations.

\subsubsection{Correlation maps of the images at 86~GHz}\label{subsubsec:86GHz}

\begin{figure*}[t!]
\plotone{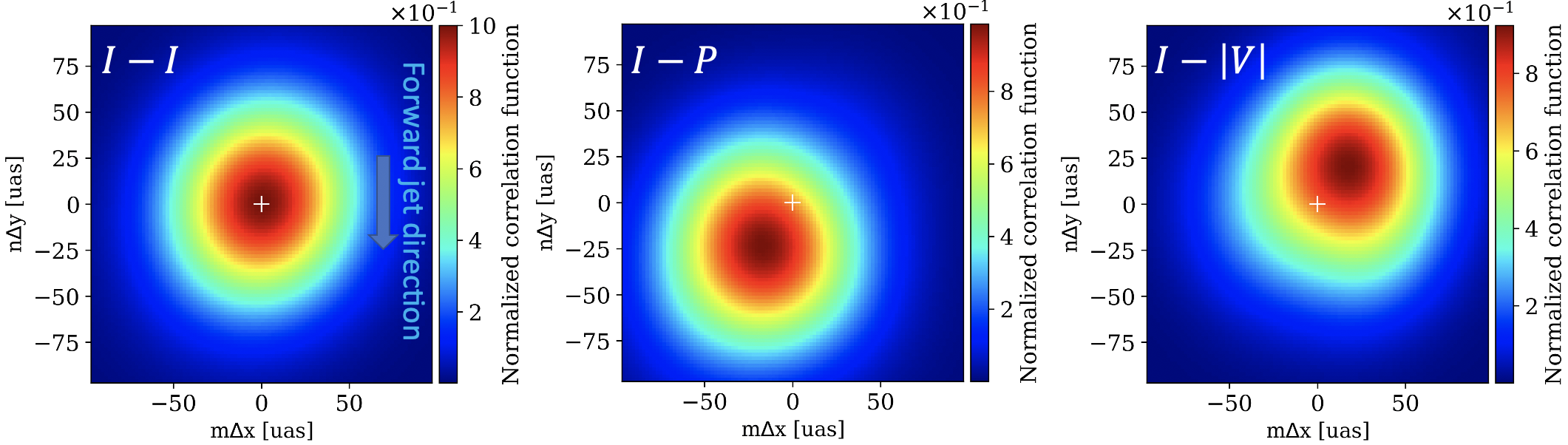}
\caption{Same as the correlation maps of Fig.~\ref{fig:correlations} but for the images at 86~GHz in Fig.~\ref{fig:86GHz}. 
\label{fig:86GHz_correlations}}
\end{figure*}

\begin{figure*}[t!]
\plotone{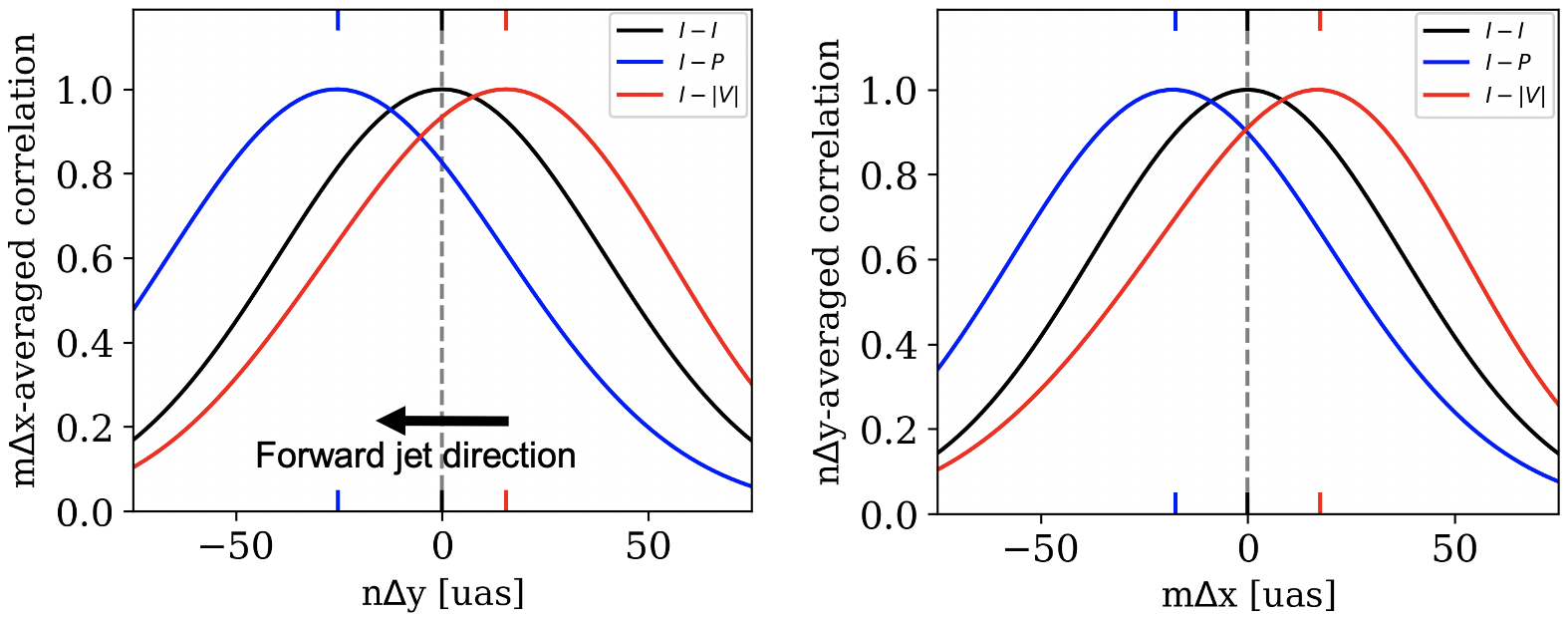}
\caption{Same as the correlation profile of Fig.~\ref{fig:xyprofiles} but for the images at 86~GHz in Fig.~\ref{fig:86GHz}. 
Hatches in the upper and lower axes demarcate the places of the maximum correlation, the relative offsets between two kinds of intensity distributions. 
\label{fig:86GHz_xyprofiles}}
\end{figure*}

We show the convolved polarimetric images at 86~GHz in Fig.~\ref{fig:86GHz}, as an example at a lower frequency. The size of Gaussian-beam, or angular resolution in VLBI observations, is assumed to be $45 \times 45 ~{\rm \mu as}$, which is extrapolated from the one by the EHT at 230~GHz with the scale rule of the diffraction limit, $\propto \lambda / D$, and a little optimistic compared with the existing VLBI observations at 86~GHz (e.g., $0.123\times0.051 ~{\rm mas} = 123\times51 ~{\rm \mu as}$; \cite{2018A&A...616A.188K}). 
The total intensity image in the left panel shows a round-shaped emission profile in the left side on a linear scale, due to round-shaped, larger-sized Gaussian beam profile and to the relativistic beaming effect. 
In the central panel, we can see the LP intensity distributed in the bottom-left area on the image, which is obviously located in the downward region, compared with that of the total intensity. Note that the typical LP fraction is $\sim 20 \%$. 
The CP components in the right panel shows a broad feature by Faraday conversion and the leftward ``separatrix'' due to the helical magnetic field configuration and the relativistic aberration effect, at which a sign reversal occurs from negative (in the left side) to positive sign (in the right side). The existence of such a separatrix was first noted by  \cite{2020arXiv201205243T}, although the separatrix is here overwritten and bent by the component from the approaching (foreground) jet around the origin. In absolute values, the positive CP components in the upper-right are brighter than other regions on the image.

Next, the two-dimensional correlation maps and their one-dimensional profiles in the Cartesian coordinates are shown in Figs.~\ref{fig:86GHz_correlations} and \ref{fig:86GHz_xyprofiles}, respectively. 
As mentioned above, we see that the peak shifts are larger at 86~GHz, compared with those at 230~GHz, because of the stronger Faraday effects. 
That is, the locations of the correlation peak in the left panel of Fig.~\ref{fig:86GHz_xyprofiles} are more separated from each other; peaks at $\sim -25~ {\rm \mu as}$ (more downwards) for $I-P$ and at $\sim +16~ {\rm \mu as}$ (more upwards) for $I-|V|$. 
In addition, the right panel of Fig.~\ref{fig:86GHz_xyprofiles} also shows a separated structure of $I-P$ with a peak at $\sim -17~ {\rm \mu as}$ (left-leaning) and $I-|V|$ with a peak at $\sim +17~ {\rm \mu as}$ (right-leaning). 
These results are direct consequences of the features seen in Fig.~\ref{fig:86GHz} that the LP (CP) intensity is located in the bottom-left (upper-right) area, relatively to the total intensity. 

We summarize the results of the correlation analyses at 230 and 86~GHz in Table~\ref{table:230and86} with the total LP and CP fractions and the image-averaged, intensity-weighted optical depths for the Faraday rotation and conversion. 
As was explained above, the stronger Faraday rotation (conversion) at 86~GHz results in the larger separation between the total and LP (CP) intensities, than at 230~GHz. 
Meanwhile, giving the higher total LP fraction and lower total CP fraction at 86~GHz in spite of the stronger Faraday rotation (depolarization) and conversion, the total polarization fractions do not predict the average Faraday depths, as also pointed out by \citet{2018MNRAS.478.1875J} for the LP maps at 230~GHz of Sgr A*.

\subsubsection{Dependence on frequencies}\label{subsubsec:wls}

\begin{figure}[t!]
\epsscale{1.2}
\plotone{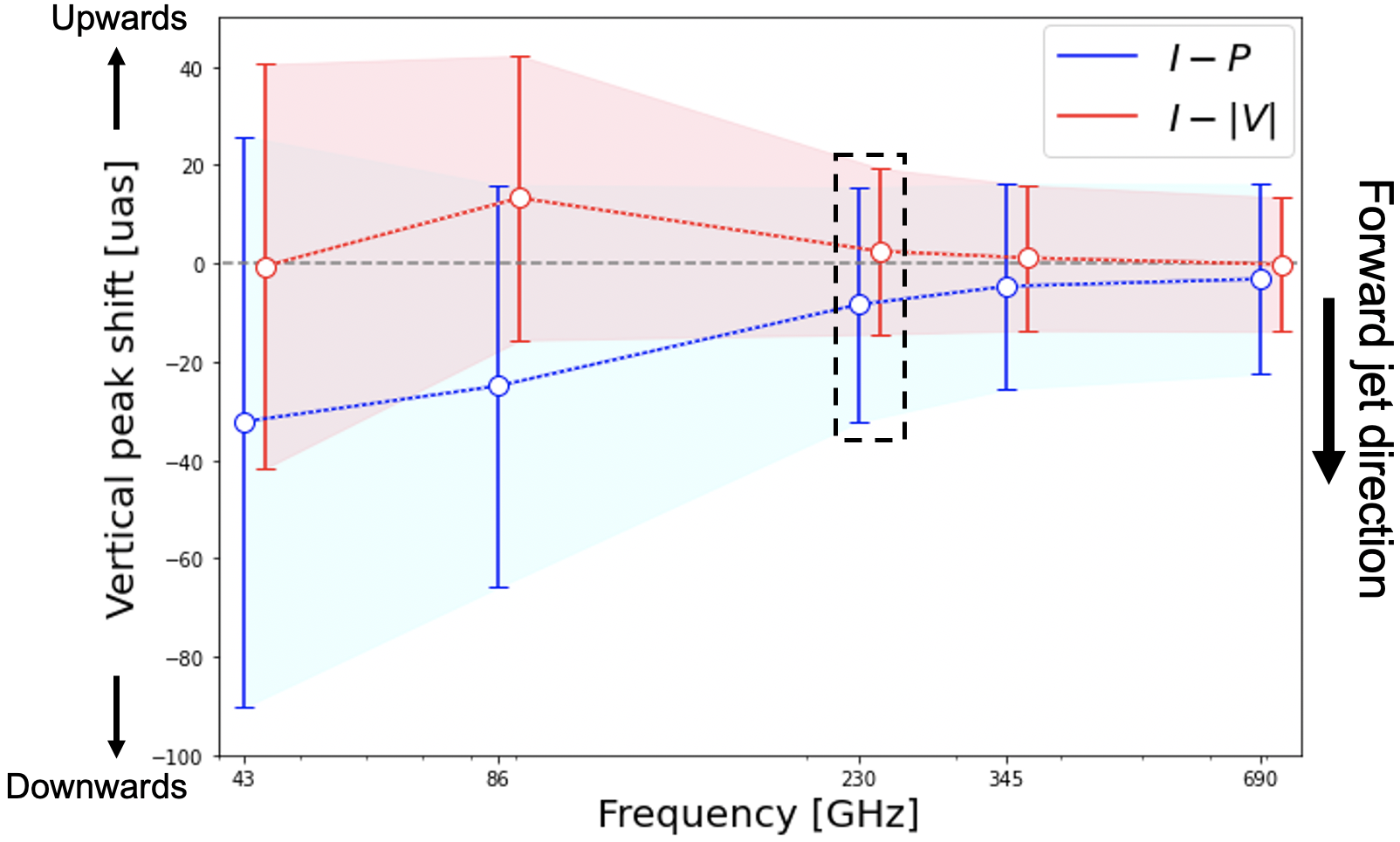}
\caption{Same as the vertical shifts of the cross-correlations in Fig.~\ref{fig:IPIV_Rh_ypeak}, but at five frequencies of 43, 86, 230, 345, and 690~GHz. 
The one boxed with dotted line corresponds to the image at 230~GHz in Fig.~\ref{fig:xyprofiles}. 
{While both of LP and CP show larger separations from the total intensity at lower frequency, the CP image at 43~GHz gives no separation because of strong SSA effect.}
See Fig.~\ref{fig:IPxyprofiles_wls} in Appendix \ref{apdx:profiles} for the correlation function profiles which are based on making this figure.
\label{fig:IPIV_nu_ypeak}}
\end{figure}

We show the frequency-dependence of the vertical peak shifts in Fig.~\ref{fig:IPIV_nu_ypeak}; at 43, 86, 230, 345, and 690~GHz, which are calculated by assuming Gaussian beam of 90, 45, 17, 10, and 5~${\rm \mu as}$, respectively.

As expected, the cross-correlation profiles at lower frequency show the larger tendency of the separation of the LP and CP components. 
That is, peaks of the cross-correlations $I-P$ (blue) leave off to the bottom-left up to $n\Delta y \sim 35~ {\rm \mu as}$ (and $m\Delta x \sim 25~ {\rm \mu as}$; see Appendix \ref{apdx:profiles}  for more detailed information and figures) as the frequency decreases, which demonstrate that the LP flux distribution is more shifted towards the bottom-left corner of the image along the beamed part of the approaching jet at lower frequency. 
Likewise, peaks of $I-|V|$ (red) tend to leave off to the top ($n\Delta y \sim 15~ {\rm \mu as}$), meaning that the CP flux at lower frequency down to 86~GHz is more separated from the total flux in the vertical direction. Exceptionally, the peak of the cross-correlation $I-|V|$ at 43~GHz behaves irregularly in the figure, showing coincidence of the total and CP intensity distributions (i.e., their cross-correlation show a peak at $n\Delta y \sim 0$). 
In the following subsection, we interpret these results in terms of the depths of the Faraday rotation and conversion, and of the SSA in radiative transfer process near the black hole.

\begin{figure*}[t!]
\epsscale{1.2}
\plotone{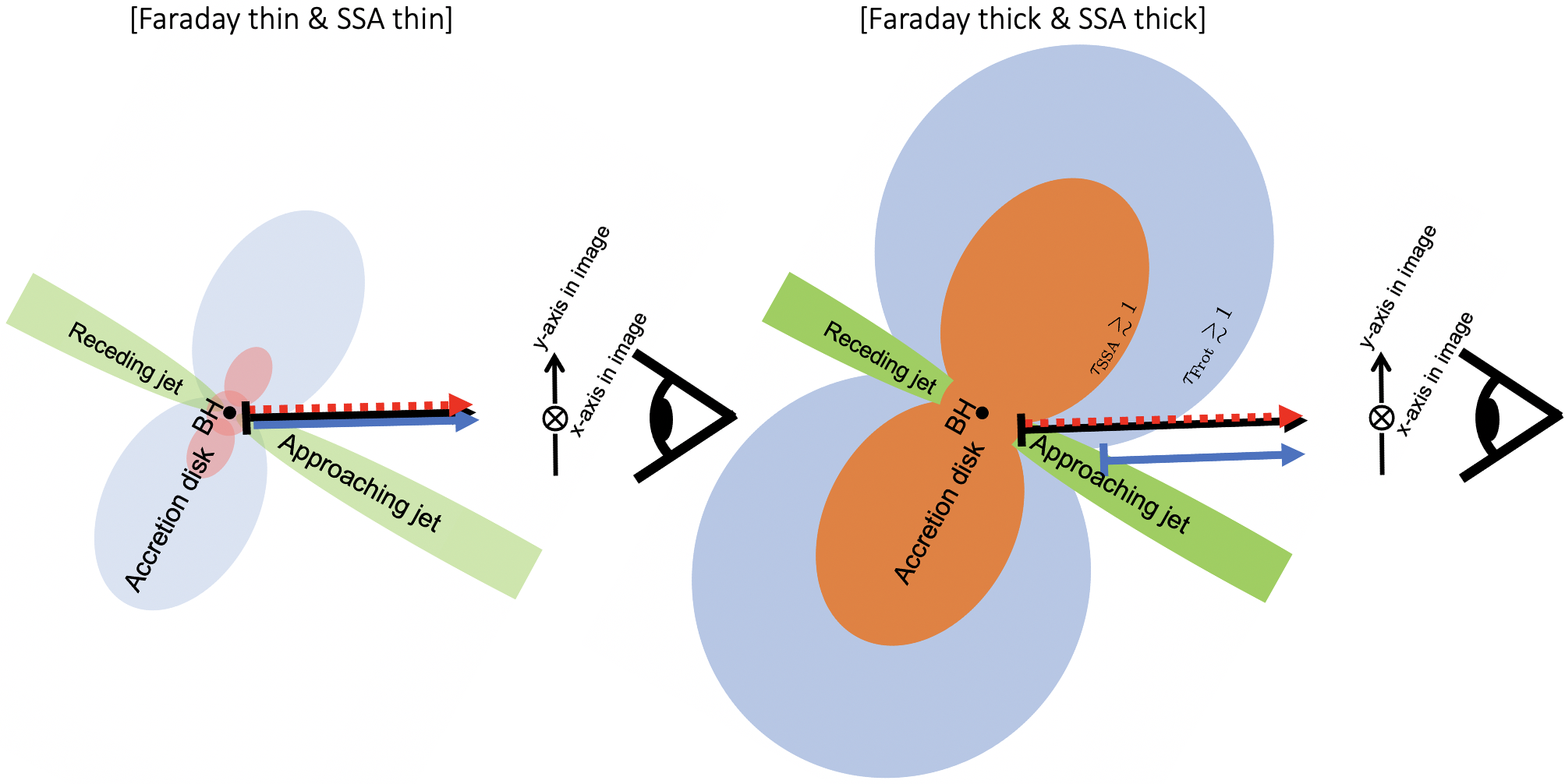}
\caption{
Same as the schematic picture of Fig.~\ref{fig:Faraday} but for the optically thin (left) and thick (right) cases for the Faraday rotation and conversion effects and the SSA. 
The left picture illustrates the case where the plasma near the black hole is optically thin both for the Faraday effects and for synchrotron self-absorption (SSA), at higher frequencies (say, 345 and 690~GHz; see also Fig.~\ref{fig:opticaldepths}) or for lower mass accretion rates. 
In this case, all of the total, strong LP and weak CP intensities at synchrotron emission directly come from near the black hole. 
The right picture illustrates the case where the system is Faraday thick and SSA thick at even lower frequencies (say, 43~GHz; see also Fig.~\ref{fig:opticaldepths}) or for even higher mass accretion rates. 
Here the total intensity and weak CP intensity originates from the surface of the photosphere (orange) of the disk-jet structure, while the LP flux is depolarized in the outer Faraday (rotation) thick plasma (blue) and is dominated by those from the downstream of the foreground jet. 
See subsection \ref{subsec:pic} for detail description.
\label{fig:picture}}
\end{figure*}

\begin{figure}[t!]
\epsscale{1.2}
\plotone{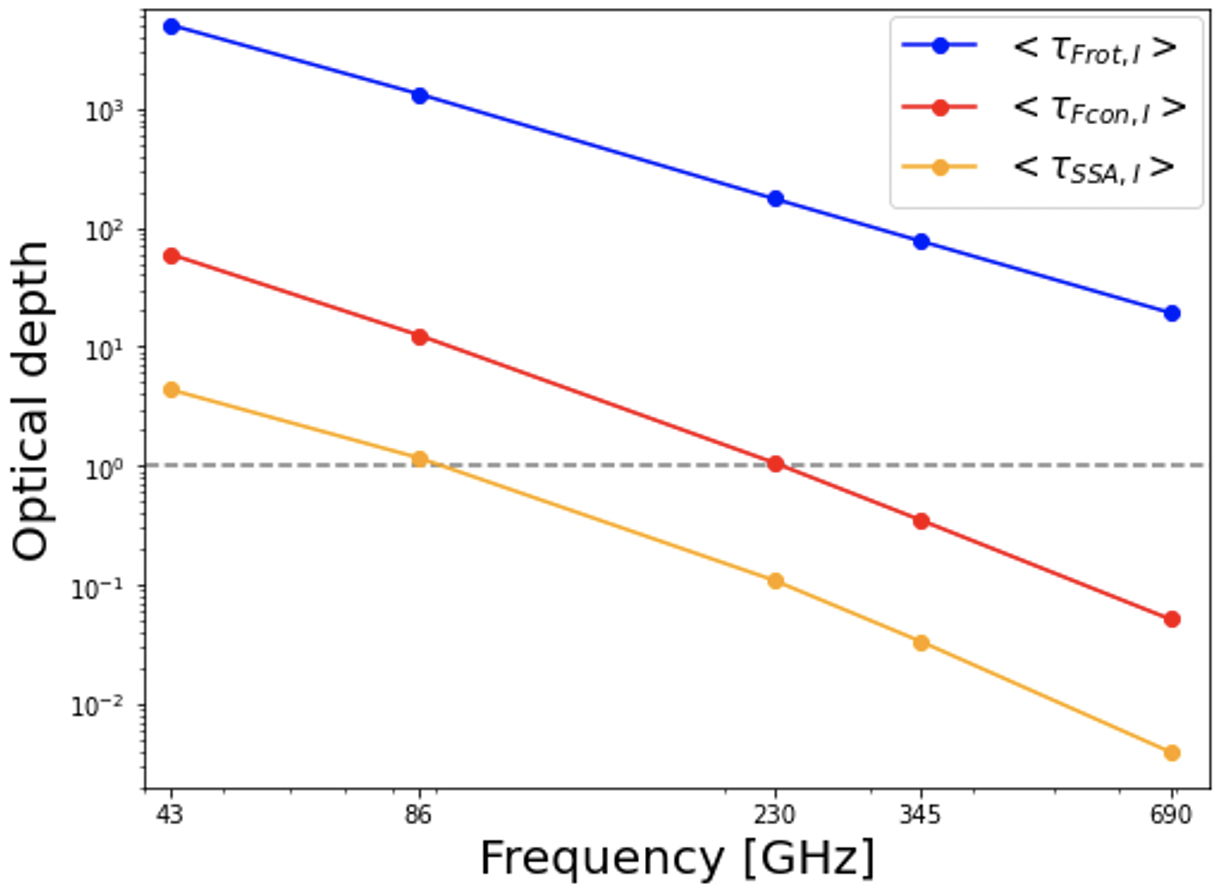}
\caption{Frequency-dependence of the image-averaged intensity-weighted optical depths for the Faraday rotation and conversion, and the synchrotron self-absorption, $\langle\tau_{\rm Frot,I}\rangle$, $\langle\tau_{\rm Fcon,I}\rangle$, and $\langle\tau_{\rm SSA,I}\rangle$. 
The grey dashed line corresponds to $\tau = 1$. 
$\langle\tau_{\rm Frot,I}\rangle$ and $\langle\tau_{\rm Fcon,I}\rangle$ roughly follow the rules of $\tau_{\rm Frot,I} \propto \nu^{-2}$ and $\tau_{\rm Fcon,I} \propto \nu^{-3}$, which reflect dependence of coefficients of the Faraday effects, $\rho_V\propto \nu^{-2}$ and $\rho_Q\propto \nu^{-3}$.
\label{fig:opticaldepths}}
\end{figure}

\subsection{Why CP separation disappears at 43~GHz?}\label{subsec:pic}

In the previous subsection, we saw the relationship between the polarized intensity distributions and that of the total intensity at multi-frequencies, finding peak separation increasing towards lower frequencies. 
We, however, noticed that such general tendency disappears for CPs at 43~GHz, why?. 
Here, we describe how we understand these results by using two schematic pictures of Fig.~\ref{fig:picture}, in comparison with Fig.~\ref{fig:Faraday}. 

The left picture in Fig.~\ref{fig:picture} illustrates the case, in which the disk-jet system is optically thin both for the Faraday effects (rotation and conversion) and for the SSA. 
This corresponds to the cases when the observed frequency is high (say, 345 or 690~GHz) and/or when the accretion rate is relatively low (e.g., $\langle \tau_{{\rm Frot},I} \rangle \simeq 19$ and $\langle \tau_{{\rm Fcon},I} \rangle \simeq 0.05$ at 690~GHz; see also the frequency-dependence of the three optical depths shown in Fig.~\ref{fig:opticaldepths}). 
Here, the inner hot disk (with $\gtrsim 10^{10} {\rm K}$), outer cold disk (with $ \lesssim 10^9 {\rm K}$), and the jet are indicated by the red, light blue, and green colors, respectively. 
We then see that all of the total, strong LP and weak CP intensities at synchrotron emission directly come from the region near the black hole without being affected by the Faraday effects nor the SSA, to reach the observer's camera. 
We thus understand that all of the total, LP and CP intensities originate in the same or close place, so that the peaks of cross-correlations $I-P$ and $I-|V|$ should be at zero shift; i.e., $(m\Delta x, n\Delta y) = (0,0)$ as the frequency becomes higher in Fig.~\ref{fig:IPIV_nu_ypeak}. 

Conversely, the right picture in Fig.~\ref{fig:picture} shows the case, in which the disk-jet system is Faraday thick and SSA thick. This corresponds to the cases when the observed frequency is low (say, 43~GHz) or when the accretion rate is relatively high, 
as was mentioned in subsection \ref{subsubsec:wls} and will be introduced in subsection \ref{subsec:Mdot}  (e.g., the image-averaged intensity-weighted SSA depth is $\langle \tau_{{\rm SSA},I} \rangle \simeq 4.4$ at 43~GHz, while $\simeq 0.11$ at 230~GHz; see also Fig.~\ref{fig:opticaldepths}). 
Here, we can understand the exceptional behavior at 43~GHz on Fig.~\ref{fig:IPIV_nu_ypeak} which arises because of a very large SSA depth near the black hole. 
In such a case, the polarized emissions come only from the surface of the photosphere (indicated by the orange color). 
Therefore, the emitted CP intensity is not amplified and is distributed in the similar way to the total intensity, while the LP intensity is depolarized by Faraday rotation in the outer cold disk (blue) and is dominated by those from the downstream of the approaching jet. 

{We also calculated the images and correlation functions at 22~GHz with circular convolution beam of $90~{\rm \mu as}$ same as at 43~GHz, because the beam size of $180~{\rm \mu as}$ extrapolated from the diffraction limit is too large compared to the field of view of $\approx 185 {\rm \mu as}$ for safe analyses. 
The resultant images at 22~GHz show downward LPs but no upward CPs as in those at 43~GHz, which are also consistent with the description in the Faraday- and SSA- thick case. 
Meanwhile, they give a little smaller separation between the total and LP intensities compared to those at 43~GHz. 
This can be because the SSA photosphere (orange) drastically expands and approaches to the sphere of the Faraday-rotation thick disk (blue) at 22~GHz. }

In summary, we classify the behavior of the total, LP and CP intensity distributions on the images as seen in Fig.~\ref{fig:IPIV_nu_ypeak} into three regimes based on the optical depths, as pictured in Figs.~\ref{fig:Faraday} and \ref{fig:picture}. 
At high frequencies at which the plasma is optically thin both for the Faraday effects and for the SSA, all of the total, dominant LP, and weak CP intensities are distributed in the similar way. 
At low frequencies at which the plasma is optically thick for the Faraday rotation and conversion, the LP distribution shifts upwards while the amplified CP components are distributed downwards compared with the total intensity distribution. 
At even lower frequencies at which the plasma is optically thick both for the Faraday effects and for the SSA, the CPs become distributed similarly to the total intensities while the LPs keep being distributed upwards relatively to the total intensity.

\subsection{Dependence on the inclination angle $i$}\label{subsec:inclination}

In the context described above, one may intuitively expect that the spatial gaps among the total, LP, and CP intensities should depend on the inclination angle (viewing angle) of the observer. 
That is, the larger (or smaller) is the inclination, or the closer is an observer to the edge-on (face-on) direction, the more (less) separated are among the total, LP, and CP intensity distribution, since the longer (shorter) becomes the distance projected on the observer's screen. 

To examine the inclination angle-dependence of the correlations, we show the vertical shifts of the peaks of the correlation profiles for inclinations of $i = 150^\circ$ and $170^\circ$ in Figs.~\ref{fig:IPIV_i_ypeak} (see Appendix \ref{apdx:profiles} for the profiles). 
Comparing with our fiducial model with $i=160^\circ$ displayed in Fig.~\ref{fig:xyprofiles}, we notice similar tendencies for other cases with different inclination angles; that is, downward (or upward) shift of the cross-correlations with LP (CP), but the larger (smaller) separations for the larger (smaller) inclination, demonstrating the above intuition. 
We summarize these results for the inclination angle-dependence in Table \ref{table:inclinations}. 

We can thus conclude that {\textit{the polarimetric correlation analyses are potentially important methods to give constraints on the inclination angle of the approaching jet in its base region}},  through the analyses of the separated polarization components on the images around the black hole, comparing the values constrained by observations of the larger-scaled jet at multi-frequencies (e.g., $i \approx 162^\circ - 163^\circ$ ($17^\circ - 18^\circ$) in \cite{2016A&A...595A..54M} and \cite{2018ApJ...855..128W}).

\begin{figure}[t!]
\epsscale{1.2}
\plotone{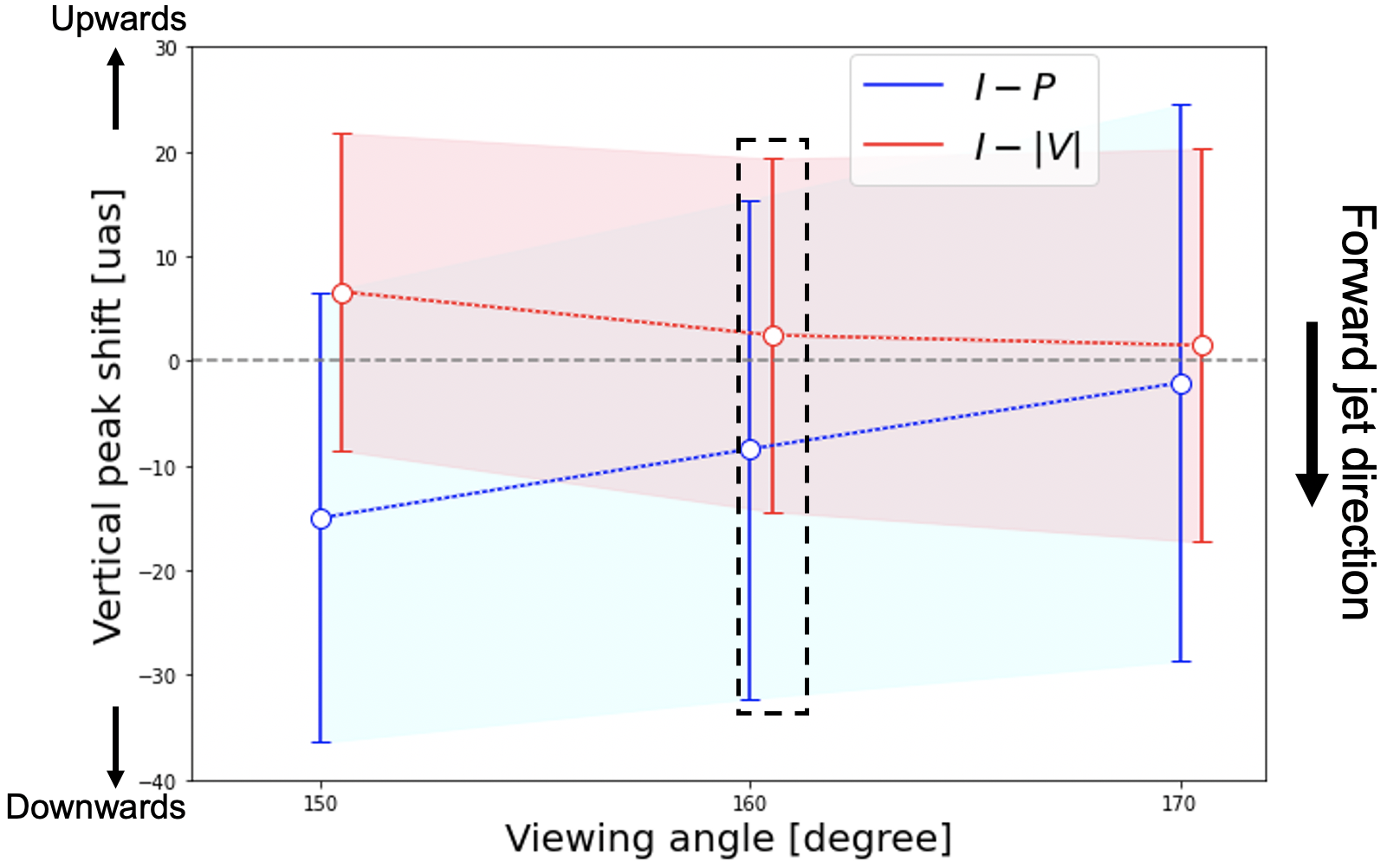}
\caption{Same as the vertical shifts of the cross-correlations in Fig.~\ref{fig:IPIV_Rh_ypeak}, but for three inclination (viewing) angles of observer, $i = 150, 160, 170^\circ$. 
The one boxed with dotted line corresponds to our fiducial model in Fig.~\ref{fig:xyprofiles}.
The based profiles are shown in Fig.~\ref{fig:highinclination_xyprofiles} and \ref{fig:lowinclination_xyprofiles} in Appendix \ref{apdx:profiles}. 
\label{fig:IPIV_i_ypeak}}
\end{figure}

\begin{table}[]
\begin{center}
  \begin{tabular}{c||c|c||c}
    Inclination angle & $I-P$ peak & $I-|V|$ peak & Fig. number \\ \hline \hline
    $150^\circ$ & $-15~{\rm \mu as}$ & $+7~{\rm \mu as}$ & Fig.~\ref{fig:highinclination_xyprofiles} \\ \hline
    $160^\circ$ & $-8~{\rm \mu as}$ & $+2~{\rm \mu as}$ & Figs.~\ref{fig:raw}, \ref{fig:conv}, \ref{fig:xyprofiles} \\ \hline
    $170^\circ$ & $-1~{\rm \mu as}$ & $+1~{\rm \mu as}$ & Fig.~\ref{fig:lowinclination_xyprofiles} \\ \hline
  \end{tabular}
\end{center}
  \caption{Comparison among the different inclination angles; the vertical peak shifts of cross-correlation functions $I-P$ and $I-|V|$, and corresponding figures, from the left to the right. 
  \label{table:inclinations}
  }
\end{table}

\begin{figure*}[t!]
\plotone{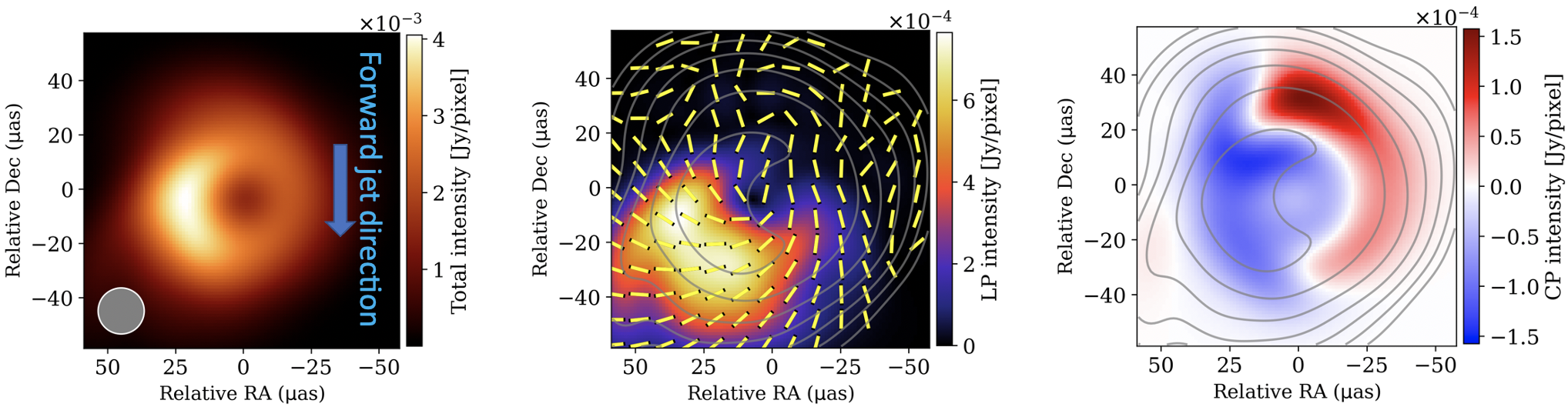}
\caption{Same as the convolved images in Fig.~\ref{fig:conv} but for a model with ten times  higher accretion rate of $\dot{M} = 6\times 10^{-3} M_\odot / {\rm yr}$. 
\label{fig:highmdot}}
\end{figure*}

\subsection{Dependence on accretion rates onto the black hole $\dot{M}$}\label{subsec:Mdot}

In subsection \ref{subsec:Rhigh}, we changed the parameter $R_{\rm high}$ and accordingly scaled the mass accretion rate onto the black hole, $\dot{M}$, to reproduce the observed flux of M87*. 
Here, we only change the accretion rate $\dot{M}$ for a fixed $R_{\rm high}$ ($= 73$), bearing application to a variety of LLAGN jets in mind. 

\begin{figure*}[t!]
\plotone{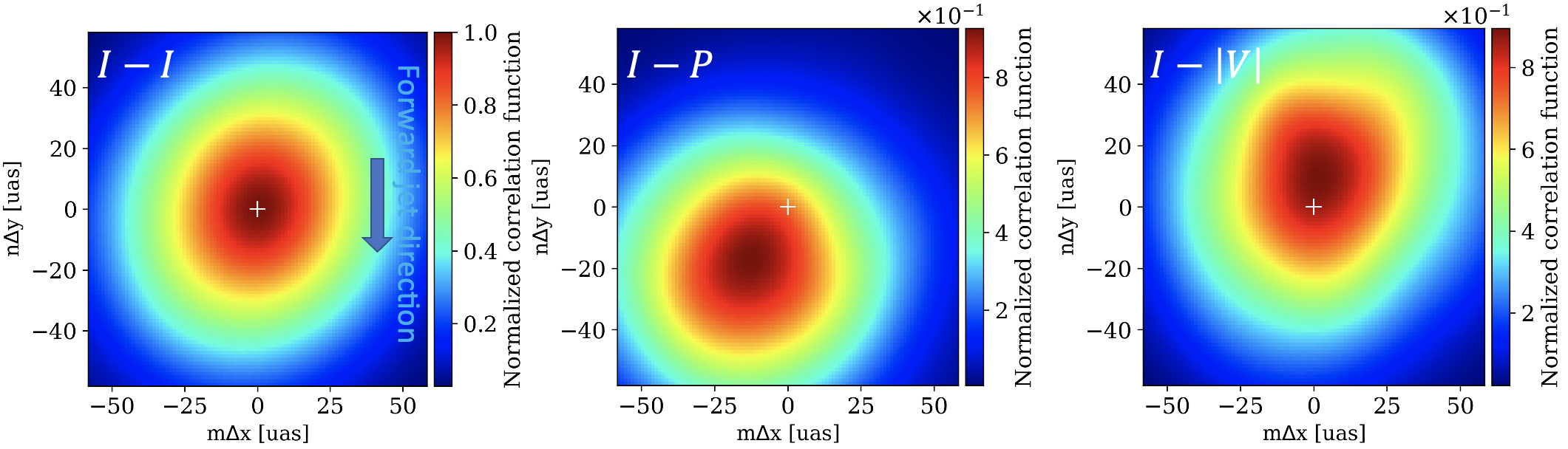}
\caption{Same as the correlation maps of Fig.~\ref{fig:correlations} but for the images for high accretion model in Fig.~\ref{fig:highmdot}. 
\label{fig:highmdot_correlations}}
\end{figure*}

We calculate the images for $\dot{M} = 6 \times 10^{-3} M_\odot / {\rm yr}$, ten times higher accretion rate than our fiducial model, and show the convolved images with $17~{\rm \mu as}$ Gaussian beam in Fig.~\ref{fig:highmdot}. 
Compared with Fig.~\ref{fig:conv}, they show a broader emission profile consisting of the photon ring and the foreground jet, dominance of the LP intensity in the jet, and stronger CP components in the photon ring and from the background jet with the sign-flipping separatrix (see in Fig.~\ref{fig:86GHz} for the images at 86~GHz, see also \cite{2020arXiv201205243T}).
We also show three maps of the auto- and cross- correlation functions in Fig.~\ref{fig:highmdot_correlations}. 
They reflect the polarimetric features described above and give larger separation between the total and LP and between the total and CP intensity distributions, than our fiducial model displayed in Fig.~\ref{fig:correlations}.

\begin{figure}[t!]
\epsscale{1.2}
\plotone{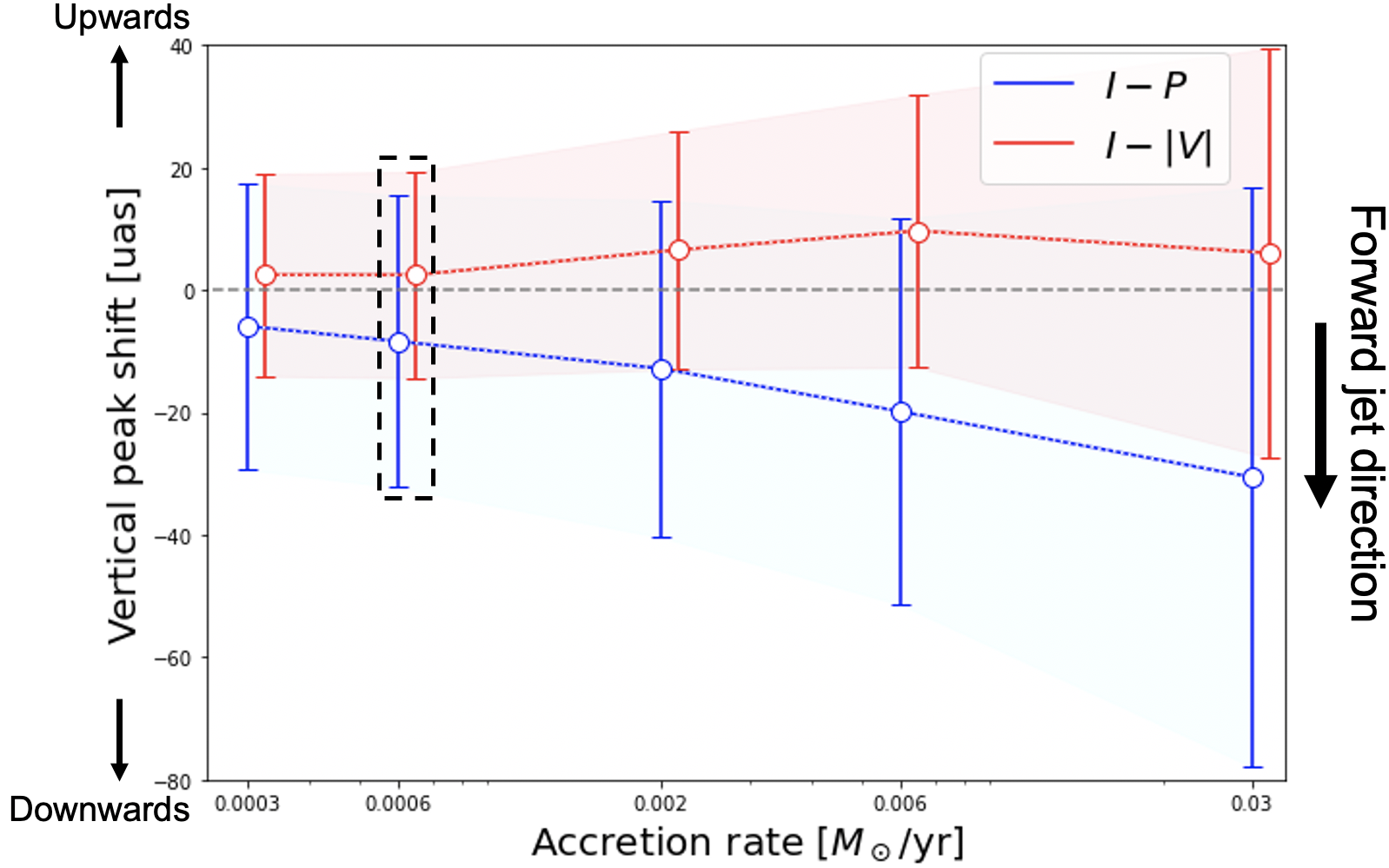}
\caption{Same as the vertical peak shifts of cross-correlations in Fig.~\ref{fig:IPIV_Rh_ypeak} but for four mass accretion rates onto the black holes $\dot{M} = (3,6,20,60,300)\times 10^{-4} M_\odot{\rm /yr}$. 
The one boxed with dotted line corresponds to our fiducial model in Fig.~\ref{fig:xyprofiles}. 
{While both of LP and CP show larger separations from the total intensity for larger mass  accretion rate, the CP image for $\dot{M} = 3\times 10^{-2} M_\odot{\rm /yr}$ gives a small  separation because of strong SSA effect.}
The based profiles are shown in Fig.~\ref{fig:IPxyprofiles_Mdots} in Appendix \ref{apdx:profiles}. 
\label{fig:IPIV_Mdot_ypeak}}
\end{figure}

In Fig.~\ref{fig:IPIV_Mdot_ypeak}, we show the vertical shifts of the peaks of the cross-correlation functions $I-P$ and $I-|V|$ for four mass accretion rates, $\dot{M} = ( \, 3, \ 6, \ 20, \ 60, \ 300 \, ) \times 10^{-4}M_\odot / {\rm yr}$.
Both of $I-P$ and $I-|V|$ give monotonic increases in their peaks as the accretion rate increase, demonstrating that the LP (or CP) intensity on the image for higher accretion rate originates from in more downward (upward) regions, relative to the total intensity  emitting region (see Fig.~\ref{fig:Faraday} and the left picture in Fig.~\ref{fig:picture}, see also their explanation in subsections \ref{subsec:Faraday} and \ref{subsec:pic}). 
This is because higher mass accretion rate $\dot{M}$ leads to higher particle density and stronger magnetic fields in non-radiative GRMHD simulations with a fixed black hole mass $M_\bullet$, giving rise to stronger Faraday effects, as was shown in subsection \ref{subsec:Rhigh} ($\tau_{\rm Frot} \propto \dot{M}^{3/2}$ and $\tau_{\rm Fcon} \propto \dot{M}^{2}$).
The highest accretion-rate model with $\dot{M} = 3 \times 10^{-2} M_\odot / {\rm yr}$ shows a somewhat different behavior, that is, it gives a small peak shift in $I-|V|$. 
This is because the highly accreted plasma becomes optically thick not only for the Faraday effects but also for the SSA, with $\langle \tau_{{\rm SSA},I} \rangle \simeq 21$, and the polarized images are dominated by emission from the foreground photosphere (see the case at 43~GHz in Fig.~\ref{fig:IPIV_nu_ypeak}, see also the right picture in Fig.~\ref{fig:picture} and its explanation in subsection \ref{subsec:pic}).

{The above results show that higher-mass accretion rates give larger LP-CP separations, but even higher mass accretion suppresses the separation of the CPs due to SSA effect, if the other parameters are fixed to those of M87*. This can be analogous with the LLAGNs with large-scale jets, such as 3C~279 or Cen~A, because we here assume that the electrons are hotter in the jet than in the disk and emission in the jet dominates over that in the disk. 
Meanwhile, we should be careful to apply these discussion to the LLAGNs without large jet, like Sgr~A*. Such LLAGNs can be modeled with the hotter disk, so that the disk emission becomes dominant. 
Our M87 models with higher disk temperature, as shown in subsection 4.1, do not necessarily present the separation of CPs. 
In future works, we should statistically check the hot disk cases with various BH masses and inclination angles, bearing a variety of LLAGNs in mind. }

\subsection{Comparison with observations}\label{subsec:observation}

\begin{figure}[t!]
\epsscale{1.0}
\plotone{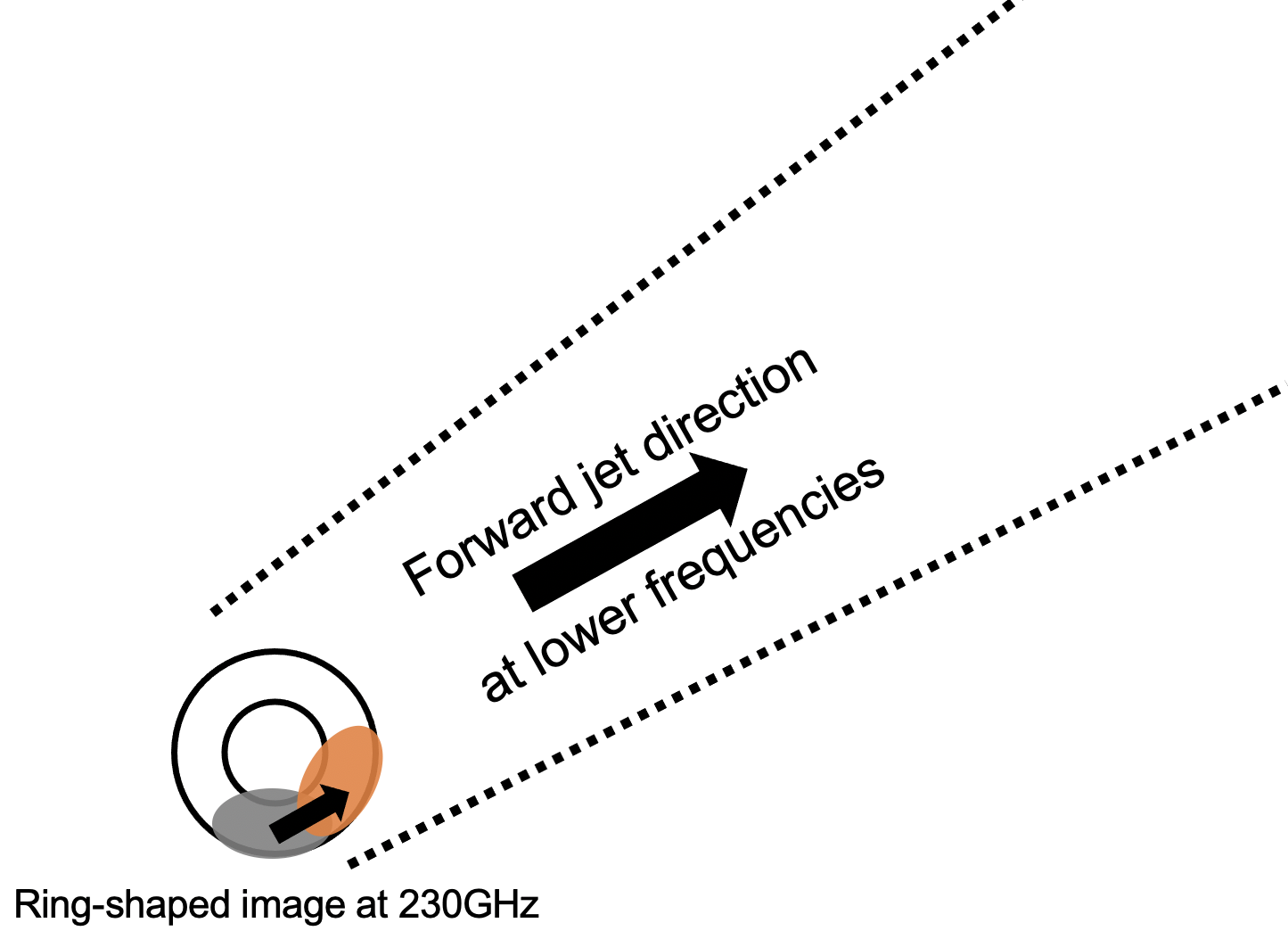}
\caption{{A schematic picture showing the relationship between the total intensity and LP images at 230 GHz and the total intensity image at lower frequencies (e.g. 86 GHz) with our interpretation.
The ring in the lower-left corner corresponds to the EHT image at 230 GHz and the brightest region in the total intensity image and that in the LP map are indicated by the grey color (in the south part of the ring) and by the orange color (in the south-west part), respectively. 
The jet, which is observed at lower frequencies, is indicated by the two dotted lines 
extending to the north-west direction (the downward direction in our images; see, e.g., Fig.~\ref{fig:conv} and \ref{fig:86GHz}).
Thus, we can interpret that the LP flux is mainly distributed in the downstream side of the jet, compared with the total flux distribution.} 
\label{fig:ring_jet}}
\end{figure}

Here, we compare our results at multi-wavelengths with existing observations including linear-polarimetry. 
As mentioned in subsection \ref{subsec:figures} and also pointed out in \citet{2021ApJ...910L..12E}, the linear-polarimetric images at 230~GHz obtained by the EHT persistently show strong LP components in the south-west region on the ring feature. 
{This region corresponds to the downstream side of the large-scale jet, extending from the bright region (south part of the asymmetric ring) in total intensity image of our study, as pictured in Fig.~\ref{fig:ring_jet} (see the middle panel of Fig.~\ref{fig:conv}; note that the jet direction is downward in this plot). 
In this sense, our simulated images at 230 GHz are consistent with the observational
features as was already discussed (see, e.g., subsection \ref{subsec:figures}). 
(Note, however, that it is observationally unclear if this region really corresponds to a jet.)}

We furthermore infer that this region may extend to the north-west jet, which was observed at lower frequencies (e.g. at 86~GHz). 
\cite{2016ApJ...817..131H} observed M87 jet at 86~GHz by the Very Long Baseline Array (VLBA) and the Green Bank Telescope, and presented the first 86~GHz polarimetric image in their Figure 10. 
They detected a polarized feature at $\approx 0.1~{\rm mas}$ ($= 100~{\rm \mu as}$) downstream from the M87 core with LP fraction of $3-4~\%$. 
\cite{2018ApJ...855..128W} presented the LP maps of M87 jet at 43~GHz by VLBA in their Figure 15, showing the peak of LP intensity at $\approx 0.15 {\rm mas}$ ($= 150~ {\rm \mu as}$) southwest of the core with fractional LP of $1-4~\%$. 
\cite{2020A&A...637L...6K} also gave the LP maps at 43 (and 24) GHz by VLBA in their Figure 1, with the LP emission peaks at $\sim 0.1-0.2 {\rm mas}$ ($= 100-200~ {\rm \mu as}$) downstream with LP fraction of $2-3~\%$ over a long period (2007 - 2018).

Our results at 86~GHz (and at 43~GHz) in subsection \ref{subsec:mf} suggests that the LP intensities are distributed left-downward by $25-30~{\rm \mu as}$ ($30-40~{\rm \mu as}$) relatively to the total intensity, with LP fraction of $\approx 20~\%$. 
These are qualitatively consistent with the observations in that the LP maps at lower frequencies give larger separations from the total intensity images, suggesting that the LP components at multi-wavelengths from near the black hole and the base region of the extended jet can be unifiedly explained by a persistent description, as pictured in Fig.~\ref{fig:picture}.
Meanwhile, the values of distances and LP fractions differ by factors from the observations. 
These deviations can be resolved by future observations with higher resolution, since we here assumed smaller beam size than existing observations (e.g., $45\times 45~{\rm \mu as}$ at 86~GHz)\footnote{We can point out that the total (image-integrated) LP fractions in our model of $4.2\%$ at 86~GHz and $4.6\%$ at 43~GHz are comparable with the observed values in the peak LP regions.}. 
Furthermore, combination between the linear- and circular-polarimetry in future observations  will improve the situation.

In \citet{2014ApJ...794..150J} and \citet{2015Sci...350.1242J}, they showed that the offset between the centroids of the total and linearly polarized flux can be estimated from the visibility on a short baseline. 
Thus we can expect to extract the information about the separation of polarized fluxes from even a single or a few interferometric baselines in present and future observations, to give a constraint on the plasma properties by the description introduced above.

\subsection{Future prospects}\label{subsec:future}

Whereas we adapt the $R-\beta$ prescription by Eq.~\ref{eq:M16} in determination of the electron temperature distribution, \citet{2021ApJ...910L..13E} pointed out that the temperature ratio is not necessarily well described by this prescription in comparison with their fully radiative simulations.

Actually, the polarization components from near the black hole should be affected by the temperature prescription in the jet-disk region through the Faraday effects.
In future works, we should verify the validity of the present results through comparison with those based on the fluid calculation incorporating the radiative cooling effect, which should significantly affect both of the ion and electron temperature distribution.

\bigskip

Related to the above discussion, \citet{2021ApJ...910L..13E} also suggested $R_{\rm low}>1$, characterized as low electron-temperature in the jet region, for M87* from radiative simulations. 
To survey this parameter domain, we calculate a same model as the above but with $R_{\rm low}=10$. 
We confirm the LP-CP separation feature at 230~GHz with an increased mass accretion rate of $\dot{M} = 1.5 \times 10^{-3} M_\odot {\rm /yr}$. 

\bigskip

To examine the uncertainty in the sigma cutoff $\sigma_{\rm cutoff} < 1$, we also calculated a test model without the sigma cutoff. 
The resultant images at 230~GHz give only the downward LPs but not the upward CPs, because a lower mass accretion rate of $\dot{M} = 2.5 \times 10^{-4} M_\odot {\rm /yr}$ leads to small Faraday conversion depths, $\langle\tau_{\rm Fcon,I}\rangle \sim 0.1$. 
Meanwhile, the images at 86~GHz show both of the LP and CP separations due to large Faraday rotation and conversion depths. 

\bigskip

In regards to the fluid model, we showed the MRI Q-values of $(Q_r, Q_\theta, Q_\phi) = (3.23, 3.97, 11.0)$ in subsection \ref{subsec:GRMHD}. 
The $Q_\phi$ seems sufficient compared with the fiducial value $Q\sim6$ in \citet{2004ApJ...605..321S}, although $Q_r$ and $Q_\theta$ seem a bit insufficient. 
Meanwhile, these three values are insufficient compared to $Q_z \sim 10$ and $Q_\phi \sim 20$ in \citet{2011ApJ...738...84H}. 
Based on that depolarization by turbulent magnetic fields in small scale can make quantitative difference, we will perform highly resolved GRMHD simulations and polarized GRRT, and quantitatively analyze the results in future work.

\bigskip

{In this work, we suggested the LP-CP separation features for the images based on semi-MAD models. 
It should be checked in future works whether and to what extent the LP-CP separation would be obtained for SANE or MAD models. 
The tendency of the LP-CP separation might be complicated by two conflicting factors: 
(1) We could assume that SANE models might give larger separations due to the larger Faraday depth with a higher mass accretion rate to reproduce the flux of M87*, while MADs might show smaller ones because of a lower mass accretion. 
(2) In contrast, another possibility is that the stronger magnetic field and higher jet velocity in MADs could result in the stronger LP flux in the approaching jet, i.e., larger LP-CP separation, which could be expected from fig.4 in EHTC (2021b; paper VIII). 
In addition to those mentioned above, the separations can also be affected by the jet-disk structure and its time-variability, in particular to the MADs. Thus, it should be statistically tested both for the SANE-MAD regime and for various model parameters such as the BH spin, the electron-temperature prescription, observer's inclination angle.}

\bigskip

Finally, all of the results and discussions above are based on one snapshot of the GRMHD model with different parameters at multi-frequencies. 
To check the validity of the results for the choice of GRMHD snapshot\footnote{We here distinguish the term of ``choice of snapshot'' from ``time-variability'', in that we adapt different scaling factor from simulation- to cgs- units for each snapshot to reproduce the M87* flux of $0.5~{\rm Jy}$ in 2017.}, we newly pick up three snapshots in the quasi-steady state, in addition to the above one. 
Here, we calculate these four models for four different azimuthal angles of the observer's camera, $\phi_{\rm camera} = 0, 90^\circ, 180^\circ$, and $270^\circ$, thus sixteen images at 230~GHz in total. 

As a result, we confirm the LP-CP separation with the shifts of up to $\sim 15 {\mu as}$ in 13 out of 16 images, while the remaining three images show only $I-P$ peak shift but give the both of LP and CP separation in the images at 86~GHz. 
{(See also Fig.~\ref{fig:scat_histo} in appendix for a scatter diagram with histogram of $I-P$ and $I-|V|$ for these images.)}
Thus, we conclude that the results are robust for the choice of the GRMHD snapshot, although more statistical analyses including the time-variability should be performed in future works.

\section{Conclusion}\label{sec:conclusion}

While the LP and CP emissions from near the black hole and the base region of the jet can be a good tool to survey the magnetic field configuration possibly driving the LLAGN jets such as M87, both of observational and theoretical studies have suggested that they can be affected by the Faraday effects in magnetized plasma. 
In particular to M87 jet with a nearly face-on viewing angle ($i \approx 160^\circ$) assumed, the LP vectors, especially from the background (receding) jet, can be scrambled by the Faraday rotation in the midplane disk, as pointed out by \citet{2017MNRAS.468.2214M,2020MNRAS.498.5468R}. 
In addition, the CP components can be amplified by the Faraday conversion in energetic region near the black hole through the medium of the Faraday rotation and twist of the fields, imprinting the direction and configuration of the magnetic fields  \citep{2020PASJ...72...32T,2020arXiv201205243T,2021arXiv210300267M,2021MNRAS.tmp.1263R}. 

To examine and quantify the relationship between these polarization components and the plasma properties near the black hole, we calculated theoretical polarization images based on a moderately-magnetized (semi-MAD) GRMHD model (with a magnetic flux in the intermediate range of $5 \lesssim \phi \lesssim 50$ which was not explicitly examined in \citet{2021ApJ...910L..13E}),  and analyzed the correlation relations among the total intensity, LP, and CP components on the images. 
By surveying the peak shifts of correlation functions at multi-wavelengths and for different model parameters, we established a unified description by three schematic pictures as in Figs.~\ref{fig:Faraday} and \ref{fig:picture}: 
\begin{itemize}

	\item {\bf Faraday thin and SSA thin case}: at higher frequencies (say, 345 and 690~GHz for our fiducial model) and for lower mass accretion onto the black hole, the polarized synchrotron emission reaches to us without suffering the Faraday effects because both of the Faraday rotation and conversion are weaker. As a result, we observe the intrinsic polarization components consisting of dominant LP and weak CP with a distribution similar to the total intensity image.
	
	\item {\bf Faraday thick and SSA thin case}: the LP vectors from the background jet and the inner disk are strongly scrambled by the Faraday rotation in the disk and are depolarized after convolved by observational beam, while the CP components are amplified by the Faraday conversion near the black hole. As a result, the LP components from the downstream of the foreground (approaching) jet dominate over those from the upstream, the counter-side jet or the photon ring, whereas the CP components are distributed around the photon ring and the counter-side jet. Thus, the downwards LPs and upwards CPs, relatively to the total intensity distribution, are observed on the images (e.g., at 230 and 86~GHz for our fiducial model). These tendencies become more enhanced at lower frequency or for higher mass accretion rate, as long as the SSA is not significant.
	
	\item {\bf Faraday thick and SSA thick case}: at even lower frequencies (say 43~GHz for our fiducial model) or for even higher mass accretion rate, the SSA becomes significant in addition to the Faraday effects. In this case, the polarized emission comes from the surface of the photosphere. Therefore, the intrinsic CPs are observed in similar distribution to the total intensities, while the LPs are depolarized in the outer disk and are dominated by those from the downstream.  

\end{itemize}

We found that high electron-temperature disk (low $R_{\rm high}$) models also show a downwards LP distribution, but do not necessarily give an upwards CP distribution. 
This is because the emission from the midplane disk, where the plasma structure is relatively turbulent, is dominant in these models, and thus the CP image is affected by the disk structure in small scale rather than the up- and down- stream structure of the jet. 
Thus we can propose the LP-CP separation feature as a possible test of the proton-electron coupling in the jet-disk structure. 
We also confirmed that larger viewing angle (i.e.~more edge-on observer) gives larger separation among the total, LP and CP intensities because of larger projected distance on the screen.

Comparing these results with existing observations of M87 by the EHT and other VLBIs, we can see a persistent tendency at multi-frequencies of the LP components distributed in the downstream of the jet. 
We can further expect that future observations including both of the linear and circular polarimetries with high angular resolution at a large range of frequencies will give a strong constraint on the plasma properties such as the optical thickness for the Faraday effects and the SSA, the density/temperature distribution and magnetic field structure near the black hole and the jet base region.

In future works, we will examine the description obtained in this work in the context of the time-variable fluid model. 
As a precursor, we calculated the images of different snapshots with different azimuthal angles of the camera $\phi_{\rm camera} = 0 - 360^\circ$ (rotating the camera about the z-axis), and obtained the features variable but qualitatively consistent with the description for our fiducial one (e.g., Figs.~\ref{fig:IPIV_Rh_ypeak_ph180} or \ref{fig:scat_histo} in appendix, see also a discussion in subsection \ref{subsec:future}). 
The contribution of non-thermal electrons to the synchrotron emission should be also discussed in future works, which is thought to be important especially for the images at lower frequencies. 
In addition, we should also verify the validity of the determination of the electron temperature, here by the $R-\beta$ prescription, through comparison with fluid calculations incorporating the radiative cooling.

\acknowledgments

{The authors wish to acknowledge Andrew Chael, Nicholas MacDonald and the members of the Event Horizon Telescope Collaboration Publication Committee for their constructive comments and suggestions.} 
This work was supported in part by JSPS KAKENHI Grant Number JP20J22986 (YT) and JP18K13594 (TK), JSPS Grant-in-Aid for Scientific Research (A) JP21H04488 (KO), same but for Scientific Research (C) JP20K04026 (SM), JP18K03710 (KO), and JP20K11851, JP20H01941 (HRT).
This work was also supported by MEXT as ``Program for Promoting Researches on the Supercomputer Fugaku'' (Toward a unified view of the universe: from large scale structures to planets, JPMXP1020200109) (KO, TK, and HRT), and by Joint Institute for Computational Fundamental Science (JICFuS, KO). 
KA is financially supported in part by grants from the National Science Foundation  (AST-1440254, AST-1614868, AST-2034306). 
Numerical computations were in part carried out on Cray XC50 at Center for Computational Astrophysics, National Astronomical Observatory of Japan.

\appendix

\renewcommand{\thefigure}{A\arabic{figure}}
\setcounter{figure}{0}

\section{Maps of plasma quantities in the GRMHD model}

\begin{figure*}[t!]
\plotone{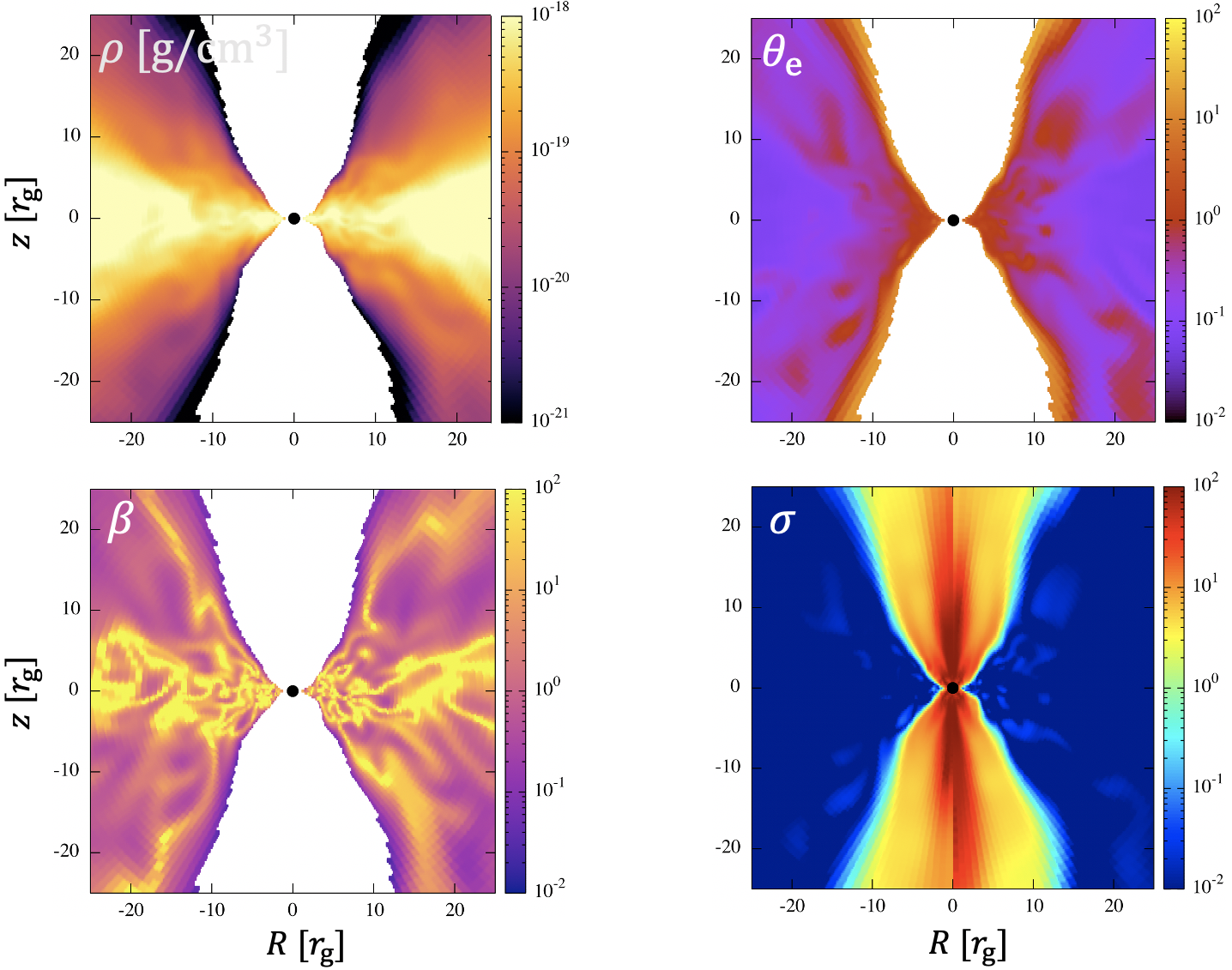}
\caption{Maps of four plasma quantities in our GRMHD model with $R_{\rm low}=1, R_{\rm high}=73$. 
Upper-left: the plasma density $\rho$ in ${\rm g/cm^3}$. Upper-right: the dimensionless electron temperature $\theta_{\rm e} \equiv k_{\rm B}T_{\rm e}/m_{\rm e}c^2$. Bottom-left: the plasma-$\beta$ parameter. Bottom-right: the plasma magnetization $\sigma$. 
Each map consists of a snapshot at $t=9000t_{\rm g}$ for $\phi=\pi$ in the left half  and for $\phi=0$ in the right half.
In the former three maps, only the region with $\sigma < \sigma_{\rm cutoff} = 1$ is plotted. 
\label{fig:GRMHDmaps}}
\end{figure*}

In Fig.~\ref{fig:GRMHDmaps}, we show the poloidal maps of four plasma quantities for our fiducial model, the plasma density $\rho$ in ${\rm g/cm^3}$, the dimensionless electron temperature $\theta_{\rm e} \equiv k_{\rm B}T_{\rm e}/m_{\rm e}c^2$, the plasma-$\beta$ parameter, and the plasma magnetization $\sigma$. 
The particle density is scaled with the black hole mass $M_\bullet = 6.5\times 10^9 M_\odot$ and the accretion rate $\dot{M} = 6 \times 10^{-4} M_\odot / {\rm yr}$.
We take $(R_{\rm low}, R_{\rm high}) = (1,73)$ in the determination of the electron temperature by Eq.~\ref{eq:M16}. 
The other two quantities are independent of the model parameters.

\renewcommand{\thefigure}{B\arabic{figure}}
\setcounter{figure}{0}

{\section{Radiative coefficient maps and transfer plots along a light path}\label{app:RTplot}}

\begin{figure}[t!]
\epsscale{1.2}
\plotone{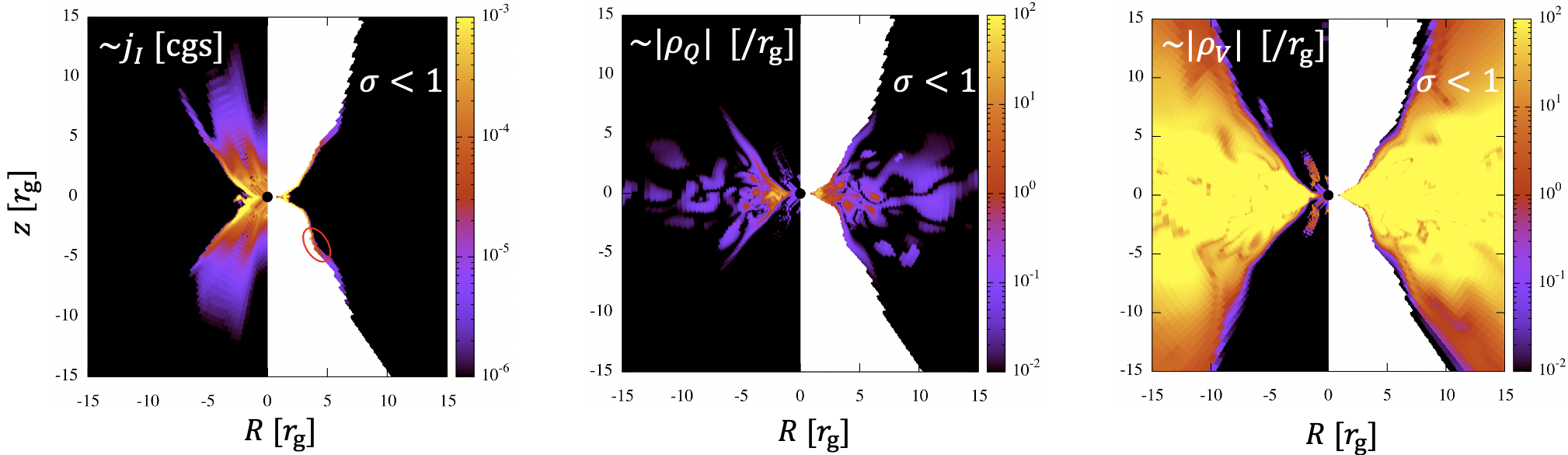}
\caption{{Three maps of the synchrotron emissivity $j_I$, Faraday conversion coefficient $\rho_V$, and Faraday rotation coefficient $\rho_Q$ at 230~GHz, left to right.
The values are estimated from the plasma density, electron temperature, and magnetic strength at $t=9000t_{\rm g}$, ignoring the relativistic effects and the angle effect between the light path and magnetic field. Each map consists of no sigma cutoff case in the left half ($\phi=\pi$) and sigma cutoff case in the right half ($\phi=0$). 
The jet emission is dominant over the disk emission, except the region in the vicinity of the BH $r \lesssim 3 r_{\rm g}$.
The Faraday effects are stronger in the disk than in the jet. 
A red circle in the left panel corresponds to the ``hump''-like feature introduced in step (4) in Fig.~\ref{fig:RTplot} and Appendix \ref{app:RTplot}.}
\label{fig:radcoeff}}
\end{figure}

{We show three maps of the synchrotron emissivity $j_I$, and coefficients of Faraday conversion and rotation, $\rho_Q$ and $\rho_V$ at 230~GHz in Fig.~\ref{fig:radcoeff}, which are estimated from the plasma density, electron temperature, and magnetic field strength. They consist of no sigma cutoff case in the left half ($\phi=\pi$) and sigma cutoff case in the right half ($\phi=0$).
These estimation maps demonstrate that emissions in the edge of jet within a range of  $-5r_{\rm g} < z < 5r_{\rm g}$ dominate over those in the disk, except the region in the vicinity of the black hole $r \lesssim 3 r_{\rm g}$, even with the sigma cutoff, while Faraday conversion and rotation are strong in the inner and outer disk, respectively, as pictured in Fig.~\ref{fig:Faraday}. }

\begin{figure}[t!]
\epsscale{1.2}
\plotone{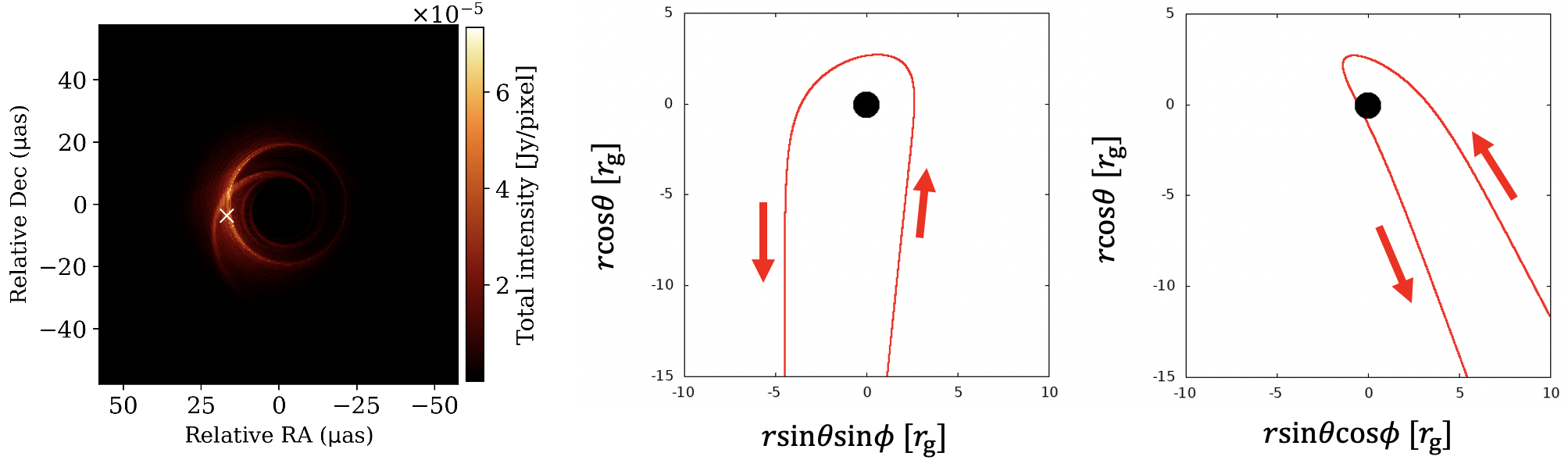}
\caption{{Left: The total intensity image at 230~GHz of our fiducial model (same with the left panel of Fig.~\ref{fig:raw}). We pick up a pixel around the ``cross-section'' between the photon ring and the tail-like jet feature, shown by a white ``x''. 
Center and right: the light path corresponding to the pixel, projected to the y-z and x-z plane in the simulation coordinates, respectively.}
\label{fig:path}}
\end{figure}

\begin{figure}[h!]
\epsscale{0.6}
\plotone{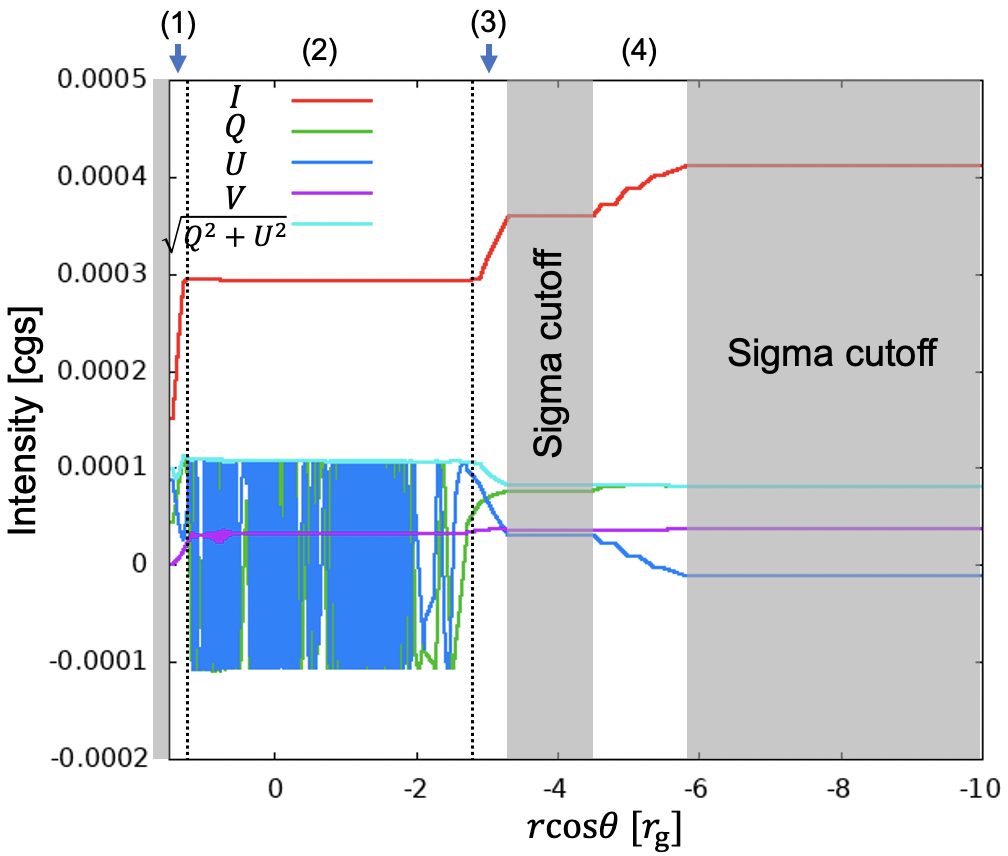}
\caption{{The radiative transfer plots of Stokes parameters $(I,Q,U,V)$ and $\sqrt{Q^2+U^2}$ along the z-coordinate of the light path in Fig.~\ref{fig:path}. 
The areas skipped by the sigma cutoff are marked with grey. 
The radiative process can be followed up by four steps (1) - (4), as described in Appendix \ref{app:RTplot}. }
\label{fig:RTplot}}
\end{figure}

{Further, we pick up a pixel pointed by a white ``x'' in the left image of Fig.~\ref{fig:path}, and show the radiative transfer plots along the light path (shown in the central and right panel of of Fig.~\ref{fig:path}) in Fig.~\ref{fig:RTplot}. 
The pixel is located in the brightest region in the total intensity image, and around a  ``cross-section'' between the photon ring and the tail-like jet feature. }

{We can follow up the radiative transfer plot lines of Stokes parameters in Fig.~\ref{fig:RTplot} by four steps, referring the radiative coefficients in Fig.~\ref{fig:radcoeff}, as follows: 
\begin{itemize}
\item[(1)] the synchrotron emission occurs in the jet-edge in the north ($z>0$) side simultaneously  with the Faraday rotation and conversion processes. 
Combination of these effects leads to increase of both LP ($\sqrt{Q^2+U^2}$) and CP ($V$), in addition to the total intensity ($I$), as we also introduced in \cite{2020PASJ...72...32T}. 
\item[(2)] Entering the disk region around the equatorial plane, Faraday rotation becomes dominant. 
Thus, the LP vector is drastically rotated, giving rise to rapid oscillations of $Q$ and $U$. 
\item[(3)] In the jet-edge in the south ($z<0$) side, the emission arises again. While the total intensity increases, the rotated LP vector are partly canceled out with the new emission component. The CP does not change significantly due to weak Faraday conversion, because the light is now passing through the outer or downstream region relatively to the prior northern jet-edge. After leaving this region, the light enters the sigma cutoff region in the southern funnel region. 
\item[(4)] There is a low-$\sigma$ region in the funnel distributed in a spiral shape in the three-dimensional fluid model. 
This feature can be seen, for example, as a ``hump''-like feature along the jet-edge around $(5r_{\rm g}, -5r_{\rm g})$ in the left panel of Fig.~\ref{fig:radcoeff}.
Here, the total intensity increases and the LP is overwritten in similar way to (3) in the south jet-edge, since the synchrotron emission occurs again. 
\end{itemize}}

{As a result, we obtain the total intensity increased in the inner jet-edges and the downstream spiral low-$\sigma$ component, which can be seen as the photon ring and the tail-like jet on the image. 
Further, the obtained LP vector consists of the rotated components from the north (counter-side) jet-edge and the overwriting emission from the south (approaching) jet-edge. 
Finally, the obtained CP is originated from those increased in the north (counter-side) jet-edge. 
Therefore, the LP map is dominated by the contributions from the approaching jet while the CP image by those from the counter-side jet, after the observational beam convolution. 
In this way, we demonstrate that the scenario pictured in Fig.~\ref{fig:Faraday} actually occurs in the radiative transfer calculation.}

\renewcommand{\thefigure}{C\arabic{figure}}
\setcounter{figure}{0}

\section{Correlation maps in polar coordinates}\label{app:polar}

\begin{figure*}[t!]
\plotone{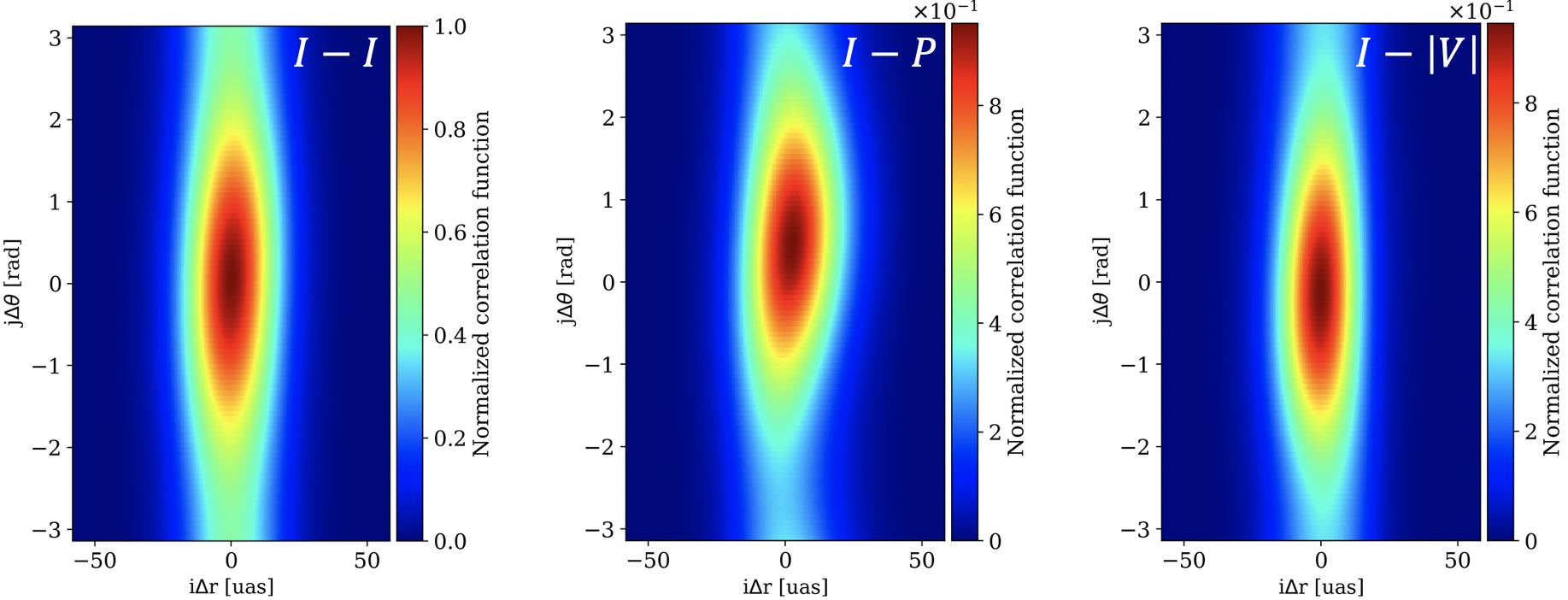}
\caption{Distribution maps of correlation functions in polar coordinates $(r,\theta)$, for the polarization images in Fig. \ref{fig:conv}. 
Right: auto-correlation of Stokes $I$. Center: cross-correlation between $I$ and $P=\sqrt{Q^2+U^2}$. Right: cross-correlation between $I$ and $|V|$. Three maps are normalized so that auto-correlation of Stokes $I$ yields 1 in the origin. We average the central and right maps in vertical (horizontal) direction and show them as $i\Delta r$- ($j\Delta \theta$-) profile in Fig.~\ref{fig:rthprofiles}.
\label{fig:correlations_rth}}
\end{figure*}

In Fig.~\ref{fig:correlations_rth}, we show three maps of auto- and cross- correlation functions $I-I$, $I-P$, and $I-|V|$ for polar coordinates $(r,\theta)$ on the images at 230GHz, defined by Eq.~\ref{eq:rthdef}. 
The positive (negative) $j\Delta \theta$ corresponds to counterclockwise (clockwise) direction on the images.
The maps have a period of $2\pi$ in the $j \Delta \theta$ direction, so that they have same values in the top ($j \Delta \theta = + \pi$) and bottom ($j \Delta \theta = -\pi$).

The auto correlation for the total intensity, $I-I$ has a peak at $(r,\theta)=(0,0)$ by definition. 
In the radial, $i\Delta r$- direction, two cross-correlations $I-P$ and $I-|V|$ have little deviations from the center, reflecting the fact that most of the total, LP and CP intensities are distributed on the common ring. 
Meanwhile, $I-P$ ($I-|V|$) gives a peak at positive (negative) region in the azimuthal, $j\Delta \theta$- direction. 
Now the total intensity image is brighter in the left side of the asymmetric ring feature, so this results quantify the tendency that the LP (CP) intensities are distributed in the lower-left (upper-left) of the common ring.

\renewcommand{\thefigure}{D\arabic{figure}}
\setcounter{figure}{0}

\section{Vertical peak shifts of $I-P$ and $I-|V|$ for the cases seeing from behind}\label{apdx:behind}

\begin{figure}[t!]
\plotone{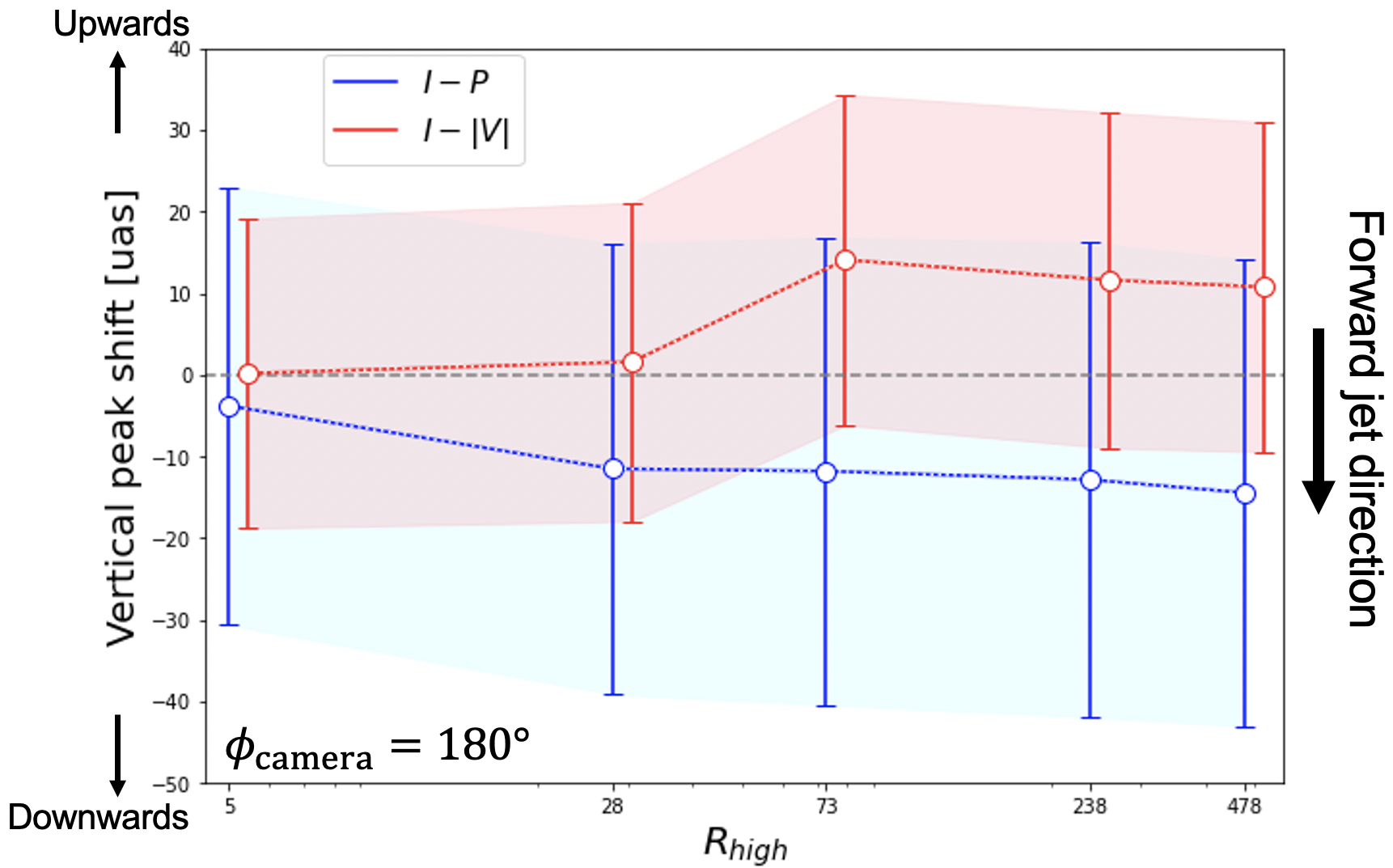}
\caption{Same as Fig.~\ref{fig:IPIV_Rh_ypeak} but for the cases with $\phi_{\rm camera} = 180^\circ$.
\label{fig:IPIV_Rh_ypeak_ph180}}
\end{figure}

In Fig.~\ref{fig:IPIV_Rh_ypeak_ph180}, We show the vertical peak shifts of the cross-correlation functions for different $R_{\rm high}$ parameters, as in Fig.~\ref{fig:IPIV_Rh_ypeak}, but for the azimuthal angle position of the camera $\phi_{\rm camera}=180^\circ$, which corresponds to the observer in the opposite side with respect to the jet (z-) axis.

\renewcommand{\thefigure}{E\arabic{figure}}
\setcounter{figure}{0}

\section{Vertical and horizontal profiles of cross-correlation functions $I-P$ and $I-|V|$ for different model parameters}\label{apdx:profiles}

\begin{figure*}[h!]
\epsscale{0.9}
\plotone{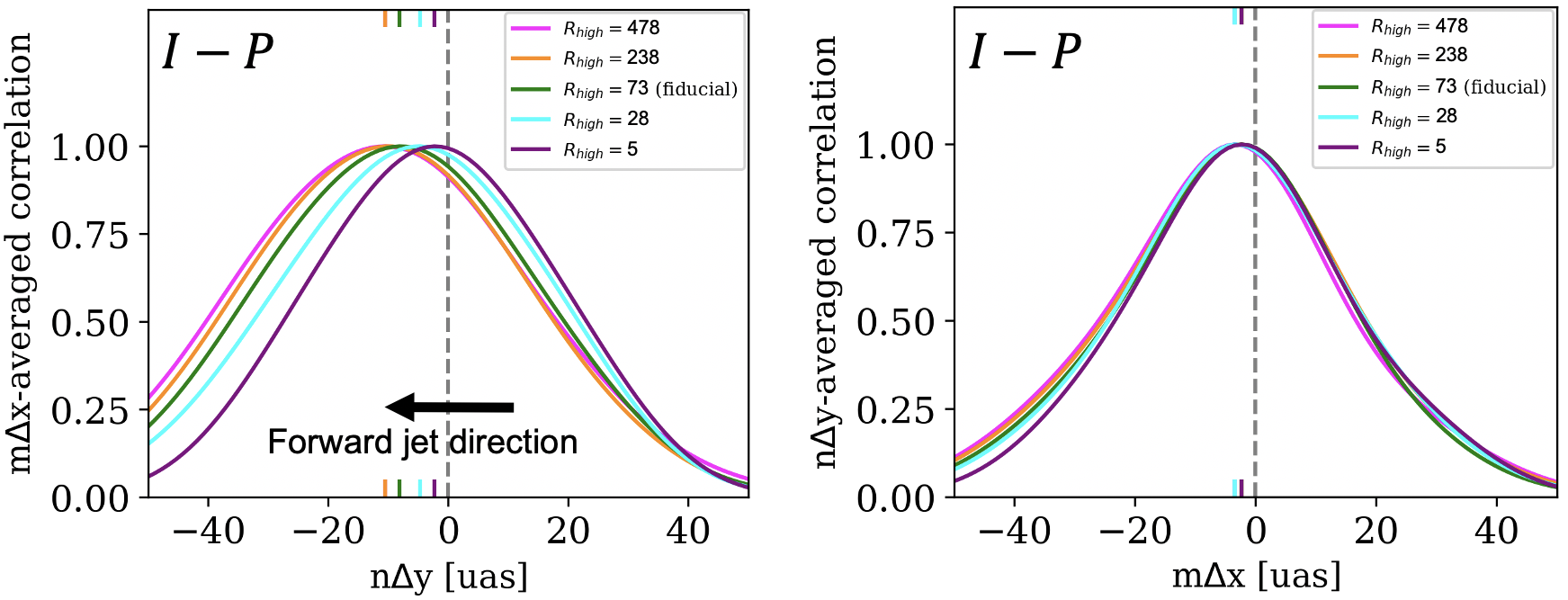}
\plotone{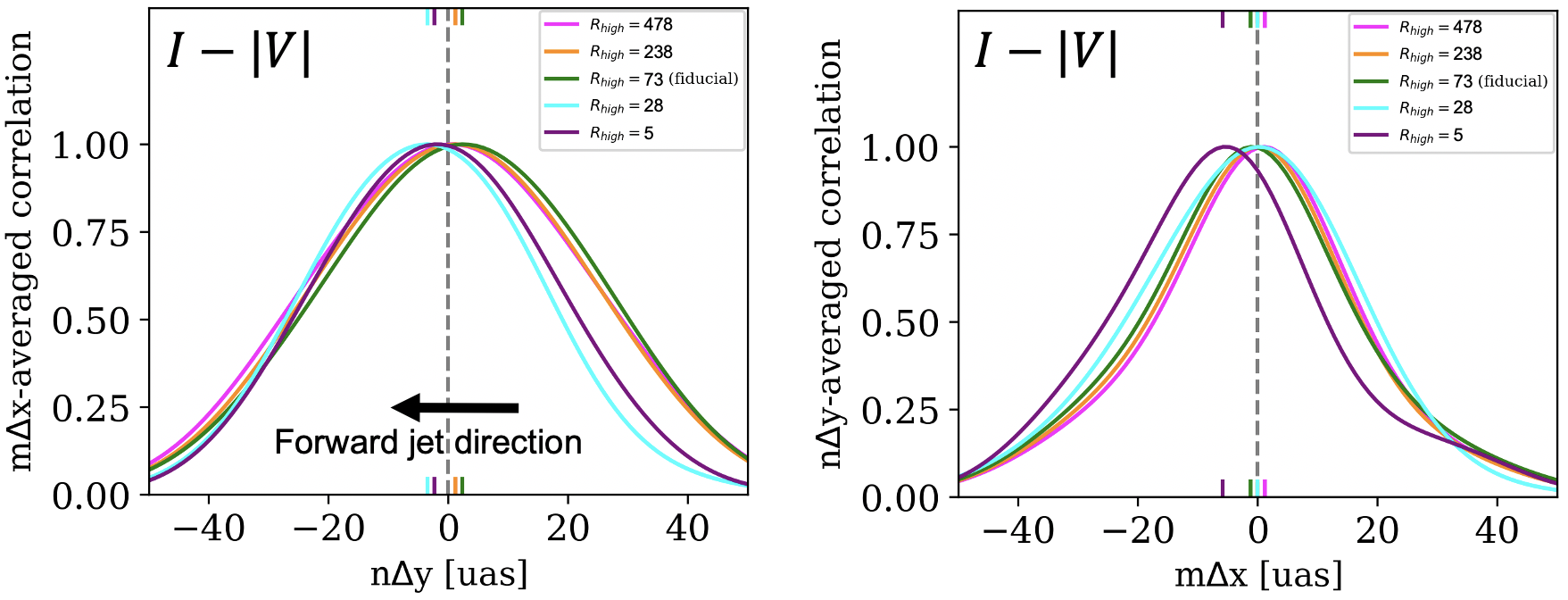}
\caption{$n\Delta y$- (left) and $m\Delta x$- (right) profiles of cross-correlations $I-P$ (top) and $I-|V|$ (bottom) for five parameters $R_{\rm high} = 2,10,25,80,$ and $160$.
\label{fig:IPxyprofiles_Rhighs}}
\end{figure*}

\begin{figure*}[h!]
\epsscale{0.9}
\plotone{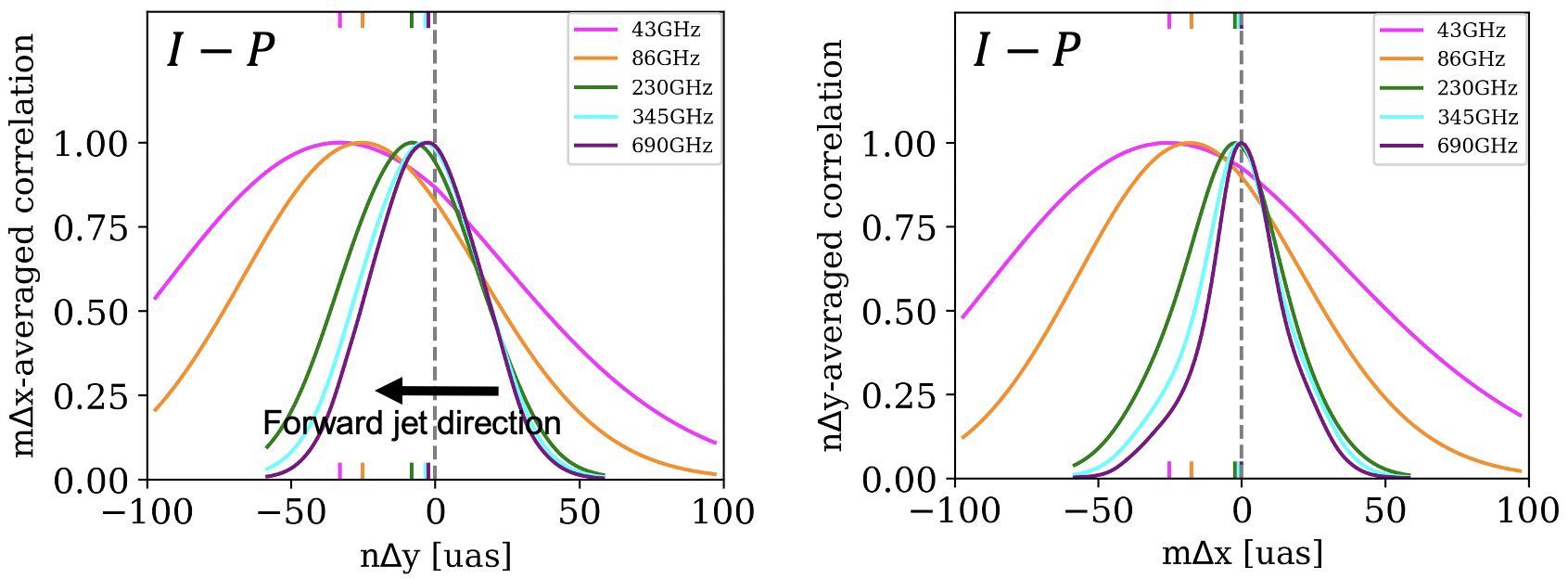}
\plotone{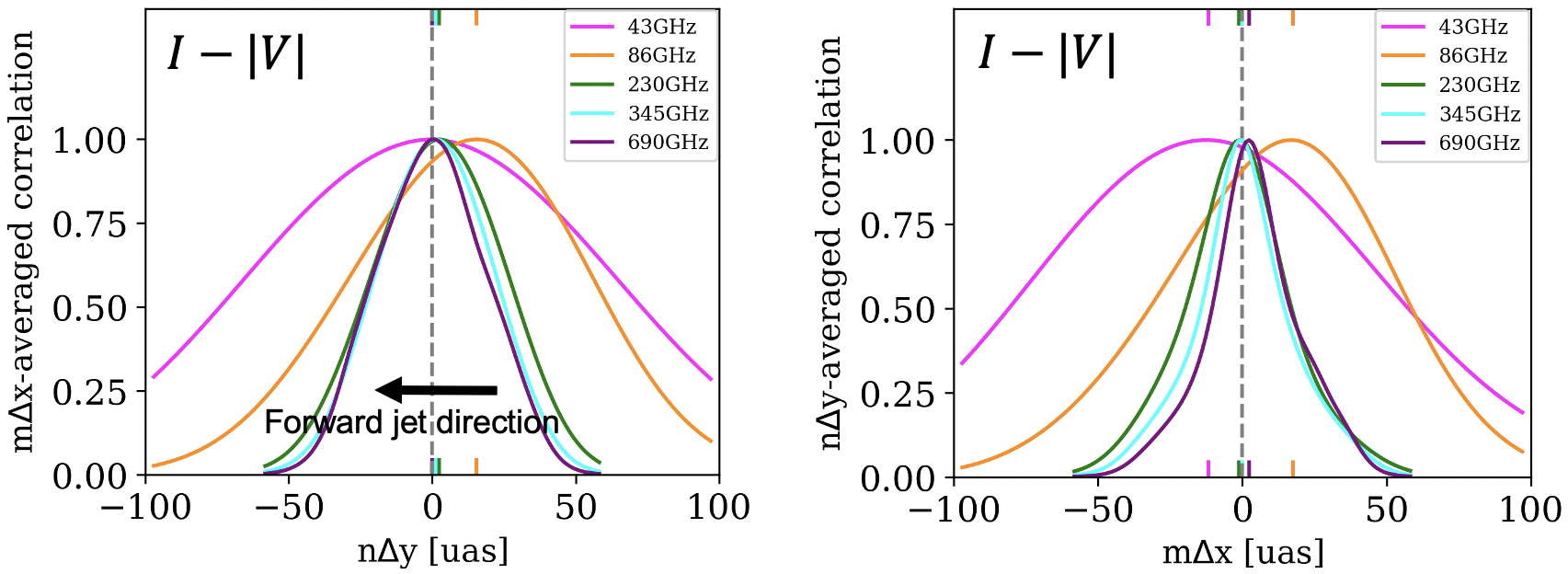}
\caption{Same as Fig.~\ref{fig:IPxyprofiles_Rhighs} but at five wavelengths of 43, 86, 230, 345, and 690~GHz.
\label{fig:IPxyprofiles_wls}}
\end{figure*}

\begin{figure*}[h!]
\plotone{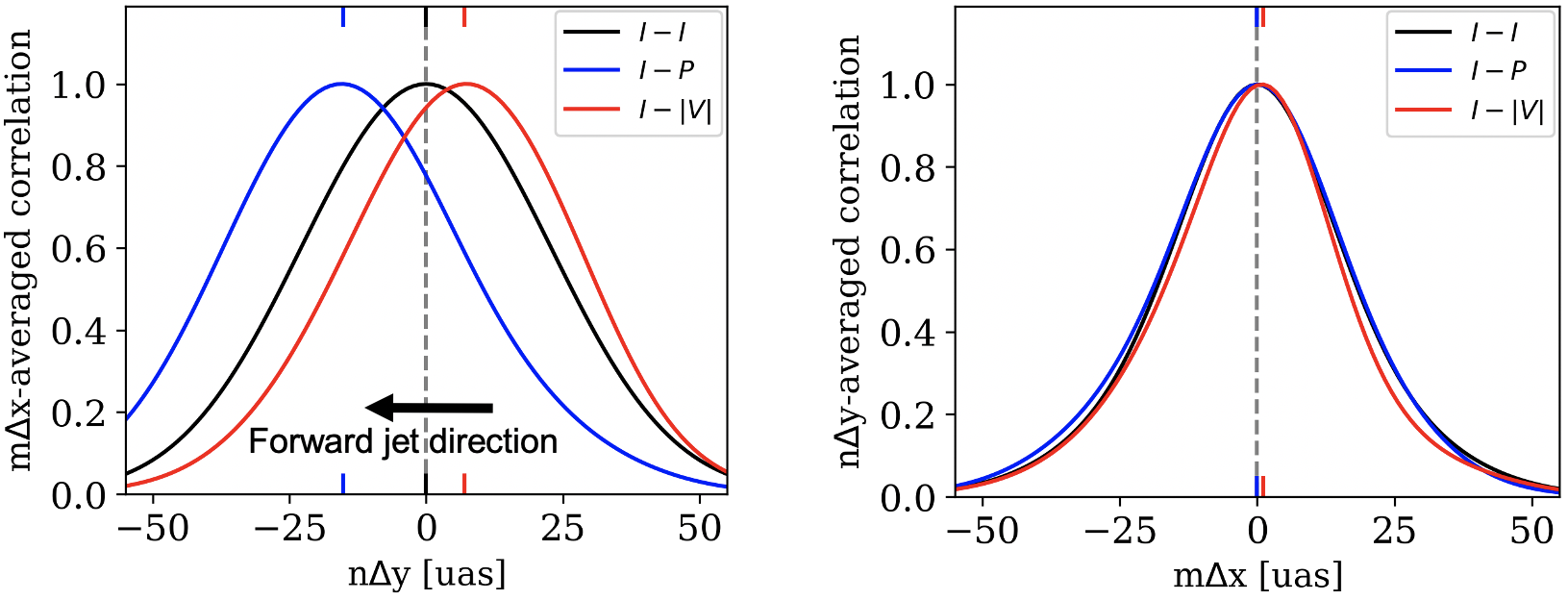}
\caption{Same as Fig.~\ref{fig:xyprofiles} but for a high inclination angle of $i = 150^\circ$. 
\label{fig:highinclination_xyprofiles}}
\end{figure*}

\begin{figure*}[h!]
\plotone{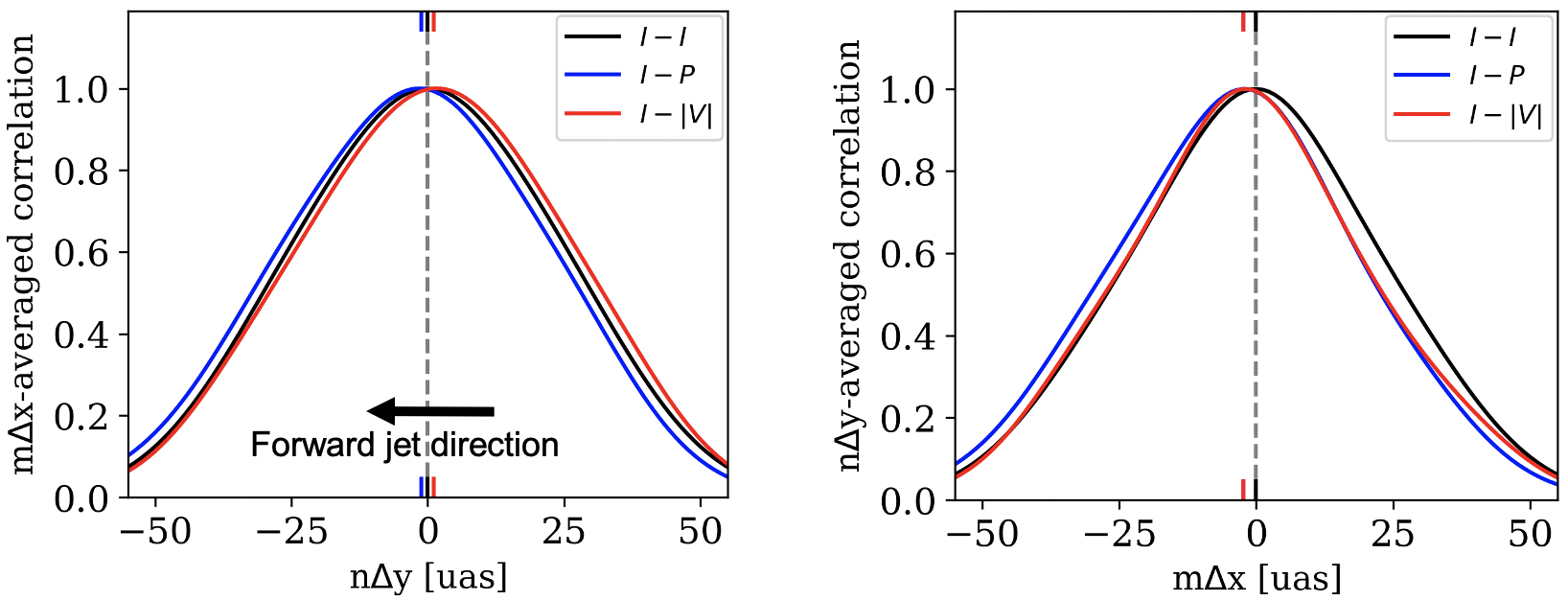}
\caption{Same as Fig.~\ref{fig:xyprofiles} but for a low inclination angle of $i = 170^\circ$. 
\label{fig:lowinclination_xyprofiles}}
\end{figure*}

\begin{figure*}[h!]
\plotone{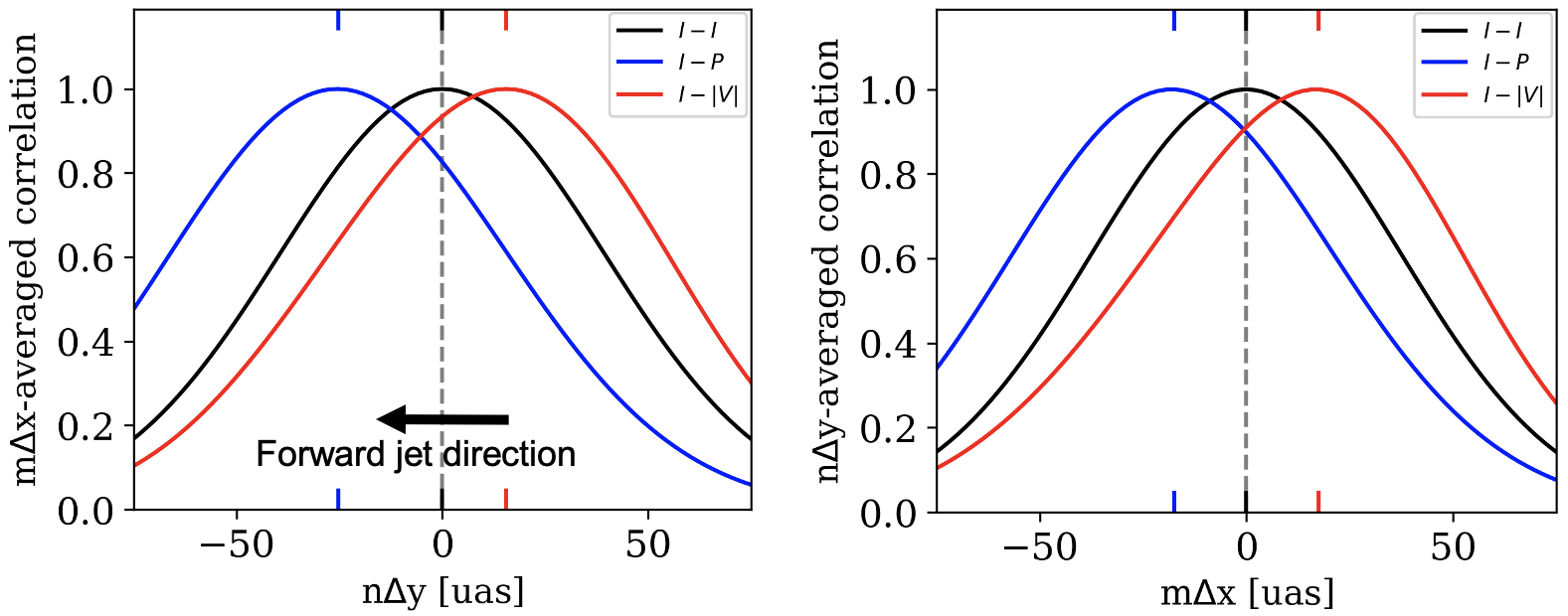}
\caption{Same as Fig.~\ref{fig:xyprofiles} but for the images for high accretion model in Fig.~\ref{fig:highmdot}. 
\label{fig:highmdot_xyprofiles}}
\end{figure*}

\begin{figure*}[h!]
\plotone{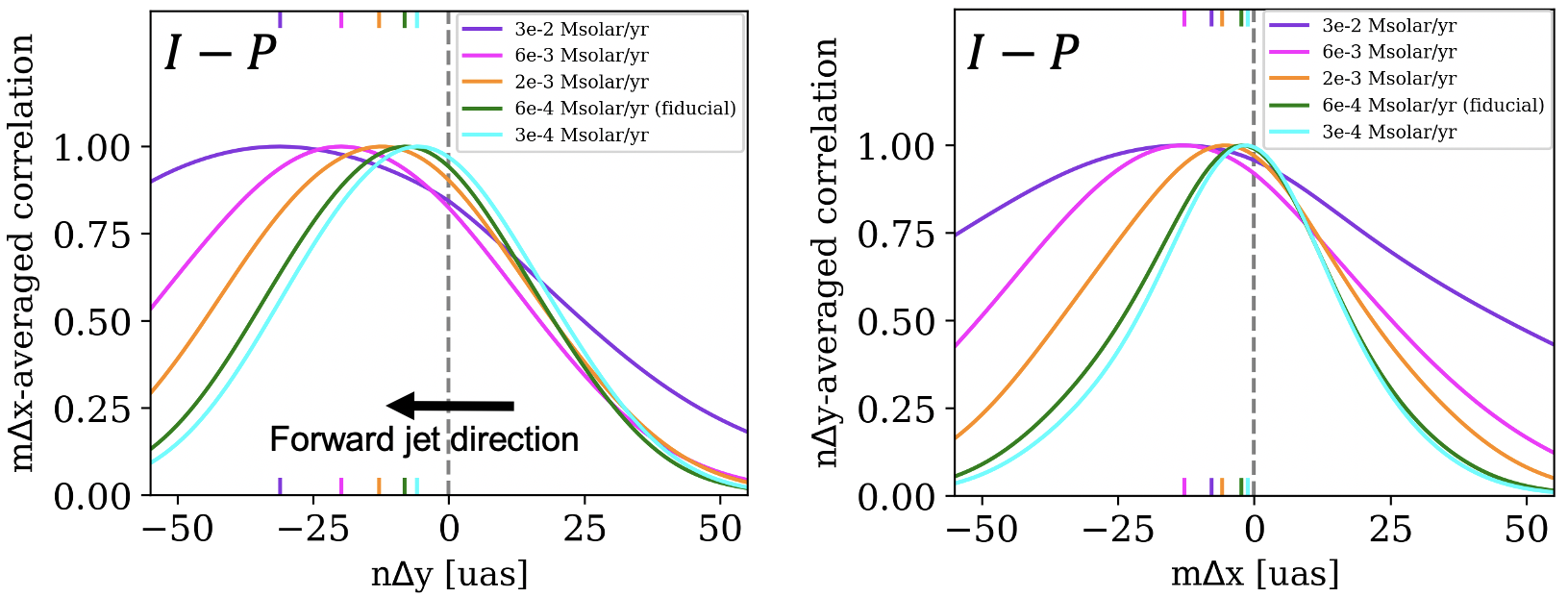}
\plotone{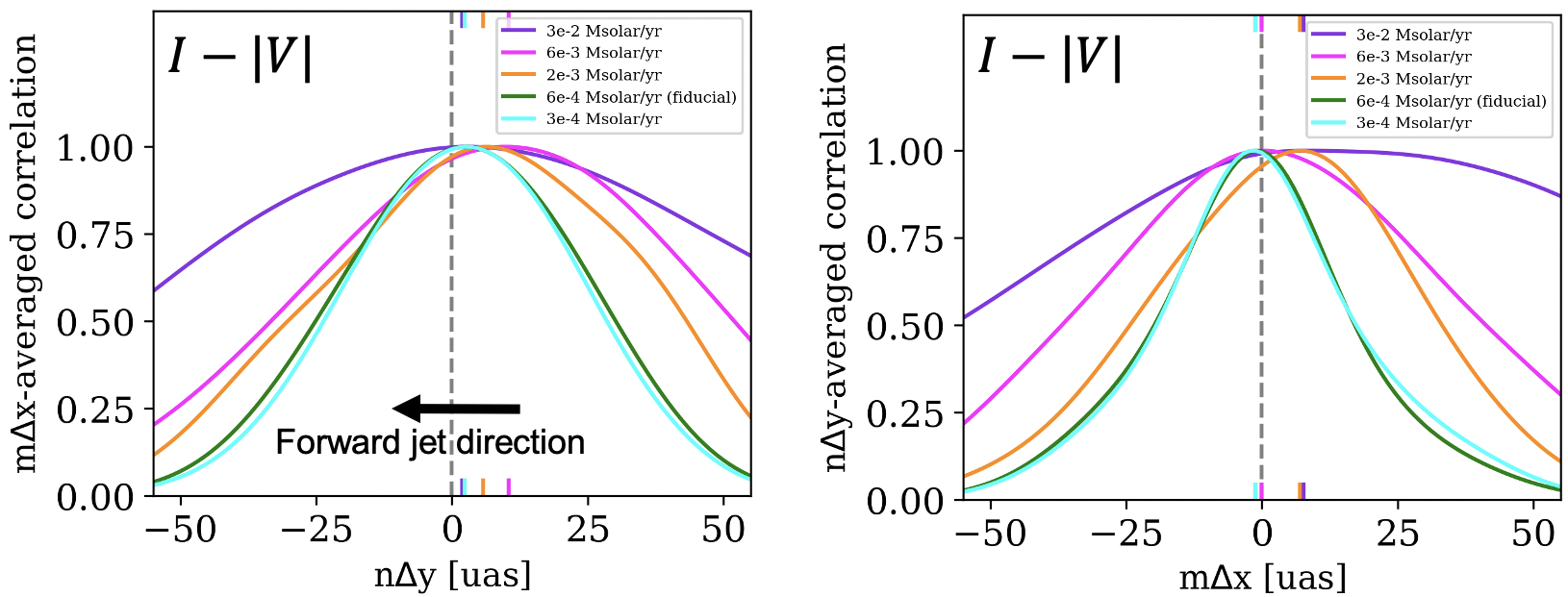}
\caption{$n\Delta y$- (left) and $m\Delta x$- (right) profiles of cross-correlations $I-P$ (top) and $I-|V|$ (bottom) for five mass accretion rates onto the black holes of $\dot{M} = (3,6,20,60,300)\times 10^{-4} M_\odot{\rm /yr}$.
\label{fig:IPxyprofiles_Mdots}}
\end{figure*}

In the text, we showed only the vertical peak shifts of the correlation functions at the higher and lower frequencies except 230 and 86~GHz, and for various model parameters except a high accretion model with $\dot{M} = 6 \times 10^{-3} M_\odot / {\rm yr}$. 
Here, we show the vertical ($y$-) and horizontal ($x$-) profiles of the correlation functions, in Figs.\ref{fig:IPxyprofiles_Rhighs} to \ref{fig:IPxyprofiles_Mdots}.

\renewcommand{\thefigure}{F\arabic{figure}}
\setcounter{figure}{0}

{\section{A scatter diagram with histogram of $I-P$ and $I-|V|$ vertical peaks for sixteen images}\label{apdx:scat_histo}}

\begin{figure*}[h!]
\plotone{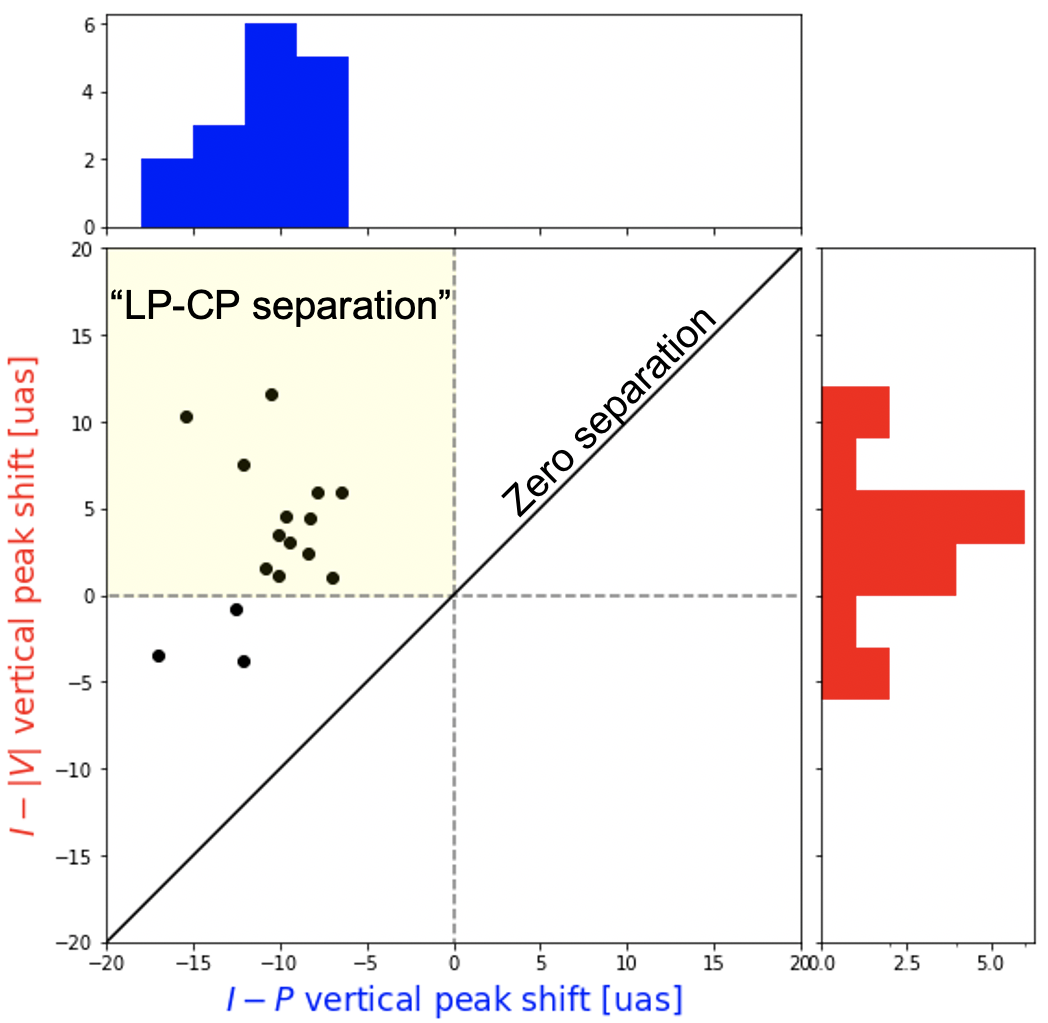}
\caption{{A scatter diagram with histogram of vertical peak shifts of cross-correlations $I-P$ and $I-|V|$ on 16 images, for four snapshots (at $t=9000t_{\rm g}, 9500t_{\rm g}, 10000t_{\rm g}$, and $11000t_{\rm g}$) and for four observer's azimuthal angles ($\phi_{\rm camera} = 0 ^\circ, 90^\circ, 180^\circ$, and $270^\circ$). 
Thirteen out of sixteen images show the LP-CP separation (i.e., positive $I-P$ peak and negative $I-|V|$ peak; yellow-marked region in the diagram), while the remaining three images do not present negative $I-|V|$ peak shifts. 
Furthermore, nine images give $I-P$ peak shifts larger than $10~{\rm \mu as}$, while five images yield $I-|V|$ peak shift larger than $5~{\rm \mu as}$.}
\label{fig:scat_histo}}
\end{figure*}

{In Fig.~\ref{fig:scat_histo}, we show a scatter diagram with histogram of the peak shifts of $I-P$ and $I-|V|$ on the sixteen images introduced in subsection \ref{subsec:future}, where 13 out of 16 images show the LP-CP separation (in the yellow-marked region in the diagram). 
It also shows that 9 of 16 images give $I-P$ peak shifts larger than $10~{\rm \mu as}$, while 5 images yield $I-|V|$ peak shift larger than $5~{\rm \mu as}$. }

\clearpage

\bibliography{correlation_paper}{}

\begin{thebibliography}{}
\expandafter\ifx\csname natexlab\endcsname\relax\def\natexlab#1{#1}\fi
\providecommand{\url}[1]{\href{#1}{#1}}
\providecommand{\dodoi}[1]{doi:~\href{http://doi.org/#1}{\nolinkurl{#1}}}
\providecommand{\doeprint}[1]{\href{http://ascl.net/#1}{\nolinkurl{http://ascl.net/#1}}}
\providecommand{\doarXiv}[1]{\href{https://arxiv.org/abs/#1}{\nolinkurl{https://arxiv.org/abs/#1}}}

\bibitem[{{Abramowski} {et~al.}(2012){Abramowski}, {Acero}, {Aharonian},
  {Akhperjanian}, {Anton}, {Balzer}, {Barnacka}, {Barres de Almeida},
  {Becherini}, {Becker}, {Behera}, {Bernl{\"o}hr}, {Birsin}, {Biteau},
  {Bochow}, {Boisson}, {Bolmont}, {Bordas}, {Brucker}, {Brun}, {Brun}, {Bulik},
  {B{\"u}sching}, {Carrigan}, {Casanova}, {Cerruti}, {Chadwick}, {Charbonnier},
  {Chaves}, {Cheesebrough}, {Clapson}, {Coignet}, {Cologna}, {Conrad},
  {Dalton}, {Daniel}, {Davids}, {Degrange}, {Deil}, {Dickinson},
  {Djannati-Ata{\"\i}}, {Domainko}, {Drury}, {Dubus}, {Dutson}, {Dyks},
  {Dyrda}, {Egberts}, {Eger}, {Espigat}, {Fallon}, {Farnier}, {Fegan},
  {Feinstein}, {Fernandes}, {Fiasson}, {Fontaine}, {F{\"o}rster},
  {F{\"u}{\ss}ling}, {Gallant}, {Gast}, {G{\'e}rard}, {Gerbig}, {Giebels},
  {Glicenstein}, {Gl{\"u}ck}, {Goret}, {G{\"o}ring}, {H{\"a}ffner}, {Hague},
  {Hampf}, {Hauser}, {Heinz}, {Heinzelmann}, {Henri}, {Hermann}, {Hinton},
  {Hoffmann}, {Hofmann}, {Hofverberg}, {Holler}, {Horns}, {Jacholkowska}, {de
  Jager}, {Jahn}, {Jamrozy}, {Jung}, {Kastendieck}, {Katarzy{\'n}ski}, {Katz},
  {Kaufmann}, {Keogh}, {Khangulyan}, {Kh{\'e}lifi}, {Klochkov}, {Klu{\'z}niak},
  {Kneiske}, {Komin}, {Kosack}, {Kossakowski}, {Laffon}, {Lamanna}, {Lennarz},
  {Lohse}, {Lopatin}, {Lu}, {Marandon}, {Marcowith}, {Masbou}, {Maurin},
  {Maxted}, {Mayer}, {McComb}, {Medina}, {M{\'e}hault}, {Moderski}, {Moulin},
  {Naumann}, {Naumann-Godo}, {de Naurois}, {Nedbal}, {Nekrassov}, {Nguyen},
  {Nicholas}, {Niemiec}, {Nolan}, {Ohm}, {de O{\~n}a Wilhelmi}, {Opitz},
  {Ostrowski}, {Oya}, {Panter}, {Paz Arribas}, {Pedaletti}, {Pelletier},
  {Petrucci}, {Pita}, {P{\"u}hlhofer}, {Punch}, {Quirrenbach}, {Raue},
  {Rayner}, {Reimer}, {Reimer}, {Renaud}, {de los Reyes}, {Rieger}, {Ripken},
  {Rob}, {Rosier-Lees}, {Rowell}, {Rudak}, {Rulten}, {Ruppel}, {Sahakian},
  {Sanchez}, {Santangelo}, {Schlickeiser}, {Sch{\"o}ck}, {Schulz}, {Schwanke},
  {Schwarzburg}, {Schwemmer}, {Sheidaei}, {Skilton}, {Sol}, {Spengler},
  {Stawarz}, {Steenkamp}, {Stegmann}, {Stinzing}, {Stycz}, {Sushch}, {Szostek},
  {Tavernet}, {Terrier}, {Tluczykont}, {Valerius}, {van Eldik}, {Vasileiadis},
  {Venter}, {Vialle}, {Viana}, {Vincent}, {V{\"o}lk}, {Volpe}, {Vorobiov},
  {Vorster}, {Wagner}, {Ward}, {White}, {Wierzcholska}, {Zacharias}, {Zajczyk},
  {Zdziarski}, {Zech}, {Zechlin}, {H.~E.~S.~S. Collaboration}, {Aleksi{\'c}},
  {Antonelli}, {Antoranz}, {Backes}, {Barrio}, {Bastieri}, {Becerra
  Gonz{\'a}lez}, {Bednarek}, {Berdyugin}, {Berger}, {Bernardini}, {Biland},
  {Blanch}, {Bock}, {Boller}, {Bonnoli}, {Borla Tridon}, {Braun}, {Bretz},
  {Ca{\~n}ellas}, {Carmona}, {Carosi}, {Colin}, {Colombo}, {Contreras},
  {Cortina}, {Cossio}, {Covino}, {Dazzi}, {De Angelis}, {De Cea del Pozo}, {De
  Lotto}, {Delgado Mendez}, {Diago Ortega}, {Doert}, {Dom{\'\i}nguez}, {Dominis
  Prester}, {Dorner}, {Doro}, {Elsaesser}, {Ferenc}, {Fonseca}, {Font},
  {Fruck}, {Garc{\'\i}a L{\'o}pez}, {Garczarczyk}, {Garrido}, {Giavitto},
  {Godinovi{\'c}}, {Hadasch}, {H{\"a}fner}, {Herrero}, {Hildebrand},
  {H{\"o}hne-M{\"o}nch}, {Hose}, {Hrupec}, {Huber}, {Jogler}, {Klepser},
  {Kr{\"a}henb{\"u}hl}, {Krause}, {La Barbera}, {Lelas}, {Leonardo},
  {Lindfors}, {Lombardi}, {L{\'o}pez}, {Lorenz}, {Makariev}, {Maneva},
  {Mankuzhiyil}, {Mannheim}, {Maraschi}, {Mariotti}, {Mart{\'\i}nez}, {Mazin},
  {Meucci}, {Miranda}, {Mirzoyan}, {Miyamoto}, {Mold{\'o}n}, {Moralejo},
  {Munar}, {Nieto}, {Nilsson}, {Orito}, {Oya}, {Paneque}, {Paoletti}, {Pardo},
  {Paredes}, {Partini}, {Pasanen}, {Pauss}, {Perez-Torres}, {Persic},
  {Peruzzo}, {Pilia}, {Pochon}, {Prada}, {Prada Moroni}, {Prandini}, {Puljak},
  {Reichardt}, {Reinthal}, {Rhode}, {Rib{\'o}}, {Rico}, {R{\"u}gamer},
  {Saggion}, {Saito}, {Saito}, {Salvati}, {Satalecka}, {Scalzotto}, {Scapin},
  {Schultz}, {Schweizer}, {Shayduk}, {Shore}, {Sillanp{\"a}{\"a}}, {Sitarek},
  {Sobczynska}, {Spanier}, {Spiro}, {Stamerra}, {Steinke}, {Storz}, {Strah},
  {Suri{\'c}}, {Takalo}, {Takami}, {Tavecchio}, {Temnikov}, {Terzi{\'c}},
  {Tescaro}, {Teshima}, {Thom}, {Tibolla}, {Torres}, {Treves}, {Vankov},
  {Vogler}, {Wagner}, {Weitzel}, {Zabalza}, {Zandanel}, {Zanin}, {MAGIC
  Collaboration}, {Arlen}, {Aune}, {Beilicke}, {Benbow}, {Bouvier}, {Bradbury},
  {Buckley}, {Bugaev}, {Byrum}, {Cannon}, {Cesarini}, {Ciupik}, {Connolly},
  {Cui}, {Dickherber}, {Duke}, {Errando}, {Falcone}, {Finley}, {Finnegan},
  {Fortson}, {Furniss}, {Galante}, {Gall}, {Godambe}, {Griffin}, {Grube},
  {Gyuk}, {Hanna}, {Holder}, {Huan}, {Hui}, {Kaaret}, {Karlsson}, {Kertzman},
  {Khassen}, {Kieda}, {Krawczynski}, {Krennrich}, {Lang}, {LeBohec}, {Maier},
  {McArthur}, {McCann}, {Moriarty}, {Mukherjee}, {Nu{\~n}ez}, {Ong}, {Orr},
  {Otte}, {Park}, {Perkins}, {Pichel}, {Pohl}, {Prokoph}, {Ragan}, {Reyes},
  {Reynolds}, {Roache}, {Rose}, {Ruppel}, {Schroedter}, {Sembroski},
  {{\c{S}}ent{\"u}rk}, {Telezhinsky}, {Te{\v{s}}i{\'c}}, {Theiling},
  {Thibadeau}, {Varlotta}, {Vassiliev}, {Vivier}, {Wakely}, {Weekes},
  {Williams}, {Zitzer}, {VERITAS Collaboration}, {Barres de Almeida}, {Cara},
  {Casadio}, {Cheung}, {McConville}, {Davies}, {Doi}, {Giovannini},
  {Giroletti}, {Hada}, {Hardee}, {Harris}, {Junor}, {Kino}, {Lee}, {Ly},
  {Madrid}, {Massaro}, {Mundell}, {Nagai}, {Perlman}, {Steele}, {Walker}, \&
  {Wood}}]{2012ApJ...746..151A}
{Abramowski}, A., {Acero}, F., {Aharonian}, F., {et~al.} 2012, \apj, 746, 151,
  \dodoi{10.1088/0004-637X/746/2/151}

\bibitem[{{Aharonian} {et~al.}(2006){Aharonian}, {Akhperjanian}, {Bazer-Bachi},
  {Beilicke}, {Benbow}, {Berge}, {Bernl{\"o}hr}, {Boisson}, {Bolz}, {Borrel},
  {Braun}, {Brown}, {B{\"u}hler}, {B{\"u}sching}, {Carrigan}, {Chadwick},
  {Chounet}, {Coignet}, {Cornils}, {Costamante}, {Degrange}, {Dickinson},
  {Djannati-Ata{\"\i}}, {Drury}, {Dubus}, {Egberts}, {Emmanoulopoulos},
  {Espigat}, {Feinstein}, {Ferrero}, {Fiasson}, {Fontaine}, {Funk}, {Funk},
  {F{\"u}{\ss}ling}, {Gallant}, {Giebels}, {Glicenstein}, {Goret},
  {Hadjichristidis}, {Hauser}, {Hauser}, {Heinzelmann}, {Henri}, {Hermann},
  {Hinton}, {Hoffmann}, {Hofmann}, {Holleran}, {Hoppe}, {Horns},
  {Jacholkowska}, {de Jager}, {Kendziorra}, {Kerschhaggl}, {Kh{\'e}lifi},
  {Komin}, {Konopelko}, {Kosack}, {Lamanna}, {Latham}, {Le Gallou},
  {Lemi{\`e}re}, {Lemoine-Goumard}, {Lenain}, {Lohse}, {Martin},
  {Martineau-Huynh}, {Marcowith}, {Masterson}, {Maurin}, {McComb}, {Moulin},
  {de Naurois}, {Nedbal}, {Nolan}, {Noutsos}, {Orford}, {Osborne}, {Ouchrif},
  {Panter}, {Pelletier}, {Pita}, {P{\"u}hlhofer}, {Punch}, {Ranchon},
  {Raubenheimer}, {Raue}, {Rayner}, {Reimer}, {Ripken}, {Rob}, {Rolland},
  {Rosier-Lees}, {Rowell}, {Sahakian}, {Santangelo}, {Saug{\'e}}, {Schlenker},
  {Schlickeiser}, {Schr{\"o}der}, {Schwanke}, {Schwarzburg}, {Schwemmer},
  {Shalchi}, {Sol}, {Spangler}, {Spanier}, {Steenkamp}, {Stegmann}, {Superina},
  {Tam}, {Tavernet}, {Terrier}, {Tluczykont}, {van Eldik}, {Vasileiadis},
  {Venter}, {Vialle}, {Vincent}, {V{\"o}lk}, {Wagner}, \&
  {Ward}}]{2006Sci...314.1424A}
{Aharonian}, F., {Akhperjanian}, A.~G., {Bazer-Bachi}, A.~R., {et~al.} 2006,
  Science, 314, 1424, \dodoi{10.1126/science.1134408}

\bibitem[{{Akiyama} {et~al.}(2015){Akiyama}, {Lu}, {Fish}, {Doeleman},
  {Broderick}, {Dexter}, {Hada}, {Kino}, {Nagai}, {Honma}, {Johnson}, {Algaba},
  {Asada}, {Brinkerink}, {Blundell}, {Bower}, {Cappallo}, {Crew}, {Dexter},
  {Dzib}, {Freund}, {Friberg}, {Gurwell}, {Ho}, {Inoue}, {Krichbaum},
  {Loinard}, {MacMahon}, {Marrone}, {Moran}, {Nakamura}, {Nagar}, {Ortiz-Leon},
  {Plambeck}, {Pradel}, {Primiani}, {Rogers}, {Roy}, {SooHoo}, {Tavares},
  {Tilanus}, {Titus}, {Wagner}, {Weintroub}, {Yamaguchi}, {Young}, {Zensus}, \&
  {Ziurys}}]{2015ApJ...807..150A}
{Akiyama}, K., {Lu}, R.-S., {Fish}, V.~L., {et~al.} 2015, \apj, 807, 150,
  \dodoi{10.1088/0004-637X/807/2/150}

\bibitem[{{Akiyama} {et~al.}(2017{\natexlab{a}}){Akiyama}, {Kuramochi},
  {Ikeda}, {Fish}, {Tazaki}, {Honma}, {Doeleman}, {Broderick}, {Dexter},
  {Mo{\'s}cibrodzka}, {Bouman}, {Chael}, \& {Zaizen}}]{2017ApJ...838....1A}
{Akiyama}, K., {Kuramochi}, K., {Ikeda}, S., {et~al.} 2017{\natexlab{a}}, \apj,
  838, 1, \dodoi{10.3847/1538-4357/aa6305}

\bibitem[{{Akiyama} {et~al.}(2017{\natexlab{b}}){Akiyama}, {Ikeda}, {Pleau},
  {Fish}, {Tazaki}, {Kuramochi}, {Broderick}, {Dexter}, {Mo{\'s}cibrodzka},
  {Gowanlock}, {Honma}, \& {Doeleman}}]{2017AJ....153..159A}
{Akiyama}, K., {Ikeda}, S., {Pleau}, M., {et~al.} 2017{\natexlab{b}}, \aj, 153,
  159, \dodoi{10.3847/1538-3881/aa6302}

\bibitem[{{Algaba} {et~al.}(2016){Algaba}, {Asada}, \&
  {Nakamura}}]{2016ApJ...823...86A}
{Algaba}, J.~C., {Asada}, K., \& {Nakamura}, M. 2016, \apj, 823, 86,
  \dodoi{10.3847/0004-637X/823/2/86}

\bibitem[{{Algaba} {et~al.}(2021){Algaba}, {Anczarski}, {Asada}, {Balokovic},
  {Chandra}, {Cui}, {Falcone}, {Giroletti}, {Goddi}, {Hada}, {Haggard},
  {Jorstad}, {Kaur}, {Kawashima}, {Keating}, {Kim}, {Kino}, {Komossa},
  {Kravchenko}, {Krichbaum}, {Lee}, {Lu}, {Lucchini}, {Markoff}, {Neilsen},
  {Nowak}, {Park}, {Principe}, {Ramakrishnan}, {Reynolds}, {Sasada},
  {Savchenko}, {Williamson}, {The Event Horizon Telescope Collaboration}, {The
  Fermi Large Area Telescope Collaboration}, {H.~E.~S.~S. Collaboration},
  {MAGIC Collaboration}, {VERITAS Collaboration}, \& {EAVN
  Collaboration}}]{2021arXiv210406855A}
{Algaba}, J.~C., {Anczarski}, J., {Asada}, K., {et~al.} 2021, arXiv e-prints,
  arXiv:2104.06855.
\newblock \doarXiv{2104.06855}

\bibitem[{{Anantua} {et~al.}(2020){Anantua}, {Emami}, {Loeb}, \&
  {Chael}}]{2020ApJ...896...30A}
{Anantua}, R., {Emami}, R., {Loeb}, A., \& {Chael}, A. 2020, \apj, 896, 30,
  \dodoi{10.3847/1538-4357/ab9103}

\bibitem[{{Asada} \& {Nakamura}(2012)}]{2012ApJ...745L..28A}
{Asada}, K., \& {Nakamura}, M. 2012, \apjl, 745, L28,
  \dodoi{10.1088/2041-8205/745/2/L28}

\bibitem[{{Asada} {et~al.}(2014){Asada}, {Nakamura}, {Doi}, {Nagai}, \&
  {Inoue}}]{2014ApJ...781L...2A}
{Asada}, K., {Nakamura}, M., {Doi}, A., {Nagai}, H., \& {Inoue}, M. 2014,
  \apjl, 781, L2, \dodoi{10.1088/2041-8205/781/1/L2}

\bibitem[{{Balbus} \& {Hawley}(1991)}]{Balbus_Hawley_1991}
{Balbus}, S.~A., \& {Hawley}, J.~F. 1991, \apj, 376, 214,
  \dodoi{10.1086/170270}

\bibitem[{{Biretta} {et~al.}(1995){Biretta}, {Zhou}, \&
  {Owen}}]{1995ApJ...447..582B}
{Biretta}, J.~A., {Zhou}, F., \& {Owen}, F.~N. 1995, \apj, 447, 582,
  \dodoi{10.1086/175901}

\bibitem[{{Blandford} \& {Begelman}(1999)}]{1999MNRAS.303L...1B}
{Blandford}, R.~D., \& {Begelman}, M.~C. 1999, \mnras, 303, L1,
  \dodoi{10.1046/j.1365-8711.1999.02358.x}

\bibitem[{{Blandford} \& {K{\"o}nigl}(1979)}]{1979ApJ...232...34B}
{Blandford}, R.~D., \& {K{\"o}nigl}, A. 1979, \apj, 232, 34,
  \dodoi{10.1086/157262}

\bibitem[{{Blandford} \& {Payne}(1982)}]{1982MNRAS.199..883B}
{Blandford}, R.~D., \& {Payne}, D.~G. 1982, \mnras, 199, 883,
  \dodoi{10.1093/mnras/199.4.883}

\bibitem[{{Blandford} \& {Znajek}(1977)}]{1977MNRAS.179..433B}
{Blandford}, R.~D., \& {Znajek}, R.~L. 1977, \mnras, 179, 433,
  \dodoi{10.1093/mnras/179.3.433}

\bibitem[{{Bridle} \& {Perley}(1984)}]{1984ARA&A..22..319B}
{Bridle}, A.~H., \& {Perley}, R.~A. 1984, \araa, 22, 319,
  \dodoi{10.1146/annurev.aa.22.090184.001535}

\bibitem[{{Broderick} \& {Loeb}(2009)}]{2009ApJ...697.1164B}
{Broderick}, A.~E., \& {Loeb}, A. 2009, \apj, 697, 1164,
  \dodoi{10.1088/0004-637X/697/2/1164}

\bibitem[{{Bronzwaer} {et~al.}(2020){Bronzwaer}, {Younsi}, {Davelaar}, \&
  {Falcke}}]{2020A&A...641A.126B}
{Bronzwaer}, T., {Younsi}, Z., {Davelaar}, J., \& {Falcke}, H. 2020, \aap, 641,
  A126, \dodoi{10.1051/0004-6361/202038573}

\bibitem[{{Chael} {et~al.}(2019){Chael}, {Narayan}, \&
  {Johnson}}]{2019MNRAS.486.2873C}
{Chael}, A., {Narayan}, R., \& {Johnson}, M.~D. 2019, \mnras, 486, 2873,
  \dodoi{10.1093/mnras/stz988}

\bibitem[{{Chael} {et~al.}(2016){Chael}, {Johnson}, {Narayan}, {Doeleman},
  {Wardle}, \& {Bouman}}]{2016ApJ...829...11C}
{Chael}, A.~A., {Johnson}, M.~D., {Narayan}, R., {et~al.} 2016, \apj, 829, 11,
  \dodoi{10.3847/0004-637X/829/1/11}

\bibitem[{{Davelaar} {et~al.}(2019){Davelaar}, {Olivares}, {Porth},
  {Bronzwaer}, {Janssen}, {Roelofs}, {Mizuno}, {Fromm}, {Falcke}, \&
  {Rezzolla}}]{2019A&A...632A...2D}
{Davelaar}, J., {Olivares}, H., {Porth}, O., {et~al.} 2019, \aap, 632, A2,
  \dodoi{10.1051/0004-6361/201936150}

\bibitem[{{Dexter}(2016)}]{2016MNRAS.462..115D}
{Dexter}, J. 2016, \mnras, 462, 115, \dodoi{10.1093/mnras/stw1526}

\bibitem[{{Dexter} {et~al.}(2012){Dexter}, {McKinney}, \&
  {Agol}}]{2012MNRAS.421.1517D}
{Dexter}, J., {McKinney}, J.~C., \& {Agol}, E. 2012, \mnras, 421, 1517,
  \dodoi{10.1111/j.1365-2966.2012.20409.x}

\bibitem[{{Dexter} {et~al.}(2020){Dexter}, {Jim{\'e}nez-Rosales}, {Ressler},
  {Tchekhovskoy}, {Baub{\"o}ck}, {de Zeeuw}, {Eisenhauer}, {von Fellenberg},
  {Gao}, {Genzel}, {Gillessen}, {Habibi}, {Ott}, {Stadler}, {Straub}, \&
  {Widmann}}]{2020MNRAS.494.4168D}
{Dexter}, J., {Jim{\'e}nez-Rosales}, A., {Ressler}, S.~M., {et~al.} 2020,
  \mnras, 494, 4168, \dodoi{10.1093/mnras/staa922}

\bibitem[{{Di Matteo} {et~al.}(2003){Di Matteo}, {Allen}, {Fabian}, {Wilson},
  \& {Young}}]{2003ApJ...582..133D}
{Di Matteo}, T., {Allen}, S.~W., {Fabian}, A.~C., {Wilson}, A.~S., \& {Young},
  A.~J. 2003, \apj, 582, 133, \dodoi{10.1086/344504}

\bibitem[{{Doeleman} {et~al.}(2012){Doeleman}, {Fish}, {Schenck}, {Beaudoin},
  {Blundell}, {Bower}, {Broderick}, {Chamberlin}, {Freund}, {Friberg},
  {Gurwell}, {Ho}, {Honma}, {Inoue}, {Krichbaum}, {Lamb}, {Loeb}, {Lonsdale},
  {Marrone}, {Moran}, {Oyama}, {Plambeck}, {Primiani}, {Rogers}, {Smythe},
  {SooHoo}, {Strittmatter}, {Tilanus}, {Titus}, {Weintroub}, {Wright}, {Young},
  \& {Ziurys}}]{2012Sci...338..355D}
{Doeleman}, S.~S., {Fish}, V.~L., {Schenck}, D.~E., {et~al.} 2012, Science,
  338, 355, \dodoi{10.1126/science.1224768}

\bibitem[{{Emami} {et~al.}(2021){Emami}, {Anantua}, {Chael}, \&
  {Loeb}}]{2021arXiv210105327E}
{Emami}, R., {Anantua}, R., {Chael}, A.~A., \& {Loeb}, A. 2021, arXiv e-prints,
  arXiv:2101.05327.
\newblock \doarXiv{2101.05327}

\bibitem[{{En{\ss}lin}(2003)}]{2003A&A...401..499E}
{En{\ss}lin}, T.~A. 2003, \aap, 401, 499, \dodoi{10.1051/0004-6361:20030162}

\bibitem[{{Event Horizon Telescope Collaboration}
  {et~al.}(2019{\natexlab{a}}){Event Horizon Telescope Collaboration},
  {Akiyama}, {Alberdi}, {Alef}, {Asada}, {Azulay}, {Baczko}, {Ball},
  {Balokovi{\'c}}, {Barrett}, {Bintley}, {Blackburn}, {Boland}, {Bouman},
  {Bower}, {Bremer}, {Brinkerink}, {Brissenden}, {Britzen}, {Broderick},
  {Broguiere}, {Bronzwaer}, {Byun}, {Carlstrom}, {Chael}, {Chan}, {Chatterjee},
  {Chatterjee}, {Chen}, {Chen}, {Cho}, {Christian}, {Conway}, {Cordes}, {Crew},
  {Cui}, {Davelaar}, {De Laurentis}, {Deane}, {Dempsey}, {Desvignes}, {Dexter},
  {Doeleman}, {Eatough}, {Falcke}, {Fish}, {Fomalont}, {Fraga-Encinas},
  {Freeman}, {Friberg}, {Fromm}, {G{\'o}mez}, {Galison}, {Gammie},
  {Garc{\'\i}a}, {Gentaz}, {Georgiev}, {Goddi}, {Gold}, {Gu}, {Gurwell},
  {Hada}, {Hecht}, {Hesper}, {Ho}, {Ho}, {Honma}, {Huang}, {Huang}, {Hughes},
  {Ikeda}, {Inoue}, {Issaoun}, {James}, {Jannuzi}, {Janssen}, {Jeter}, {Jiang},
  {Johnson}, {Jorstad}, {Jung}, {Karami}, {Karuppusamy}, {Kawashima},
  {Keating}, {Kettenis}, {Kim}, {Kim}, {Kim}, {Kino}, {Koay}, {Koch}, {Koyama},
  {Kramer}, {Kramer}, {Krichbaum}, {Kuo}, {Lauer}, {Lee}, {Li}, {Li},
  {Lindqvist}, {Liu}, {Liuzzo}, {Lo}, {Lobanov}, {Loinard}, {Lonsdale}, {Lu},
  {MacDonald}, {Mao}, {Markoff}, {Marrone}, {Marscher}, {Mart{\'\i}-Vidal},
  {Matsushita}, {Matthews}, {Medeiros}, {Menten}, {Mizuno}, {Mizuno}, {Moran},
  {Moriyama}, {Moscibrodzka}, {M{\"u}ller}, {Nagai}, {Nagar}, {Nakamura},
  {Narayan}, {Narayanan}, {Natarajan}, {Neri}, {Ni}, {Noutsos}, {Okino},
  {Olivares}, {Ortiz-Le{\'o}n}, {Oyama}, {{\"O}zel}, {Palumbo}, {Patel}, {Pen},
  {Pesce}, {Pi{\'e}tu}, {Plambeck}, {PopStefanija}, {Porth}, {Prather},
  {Preciado-L{\'o}pez}, {Psaltis}, {Pu}, {Ramakrishnan}, {Rao}, {Rawlings},
  {Raymond}, {Rezzolla}, {Ripperda}, {Roelofs}, {Rogers}, {Ros}, {Rose},
  {Roshanineshat}, {Rottmann}, {Roy}, {Ruszczyk}, {Ryan}, {Rygl},
  {S{\'a}nchez}, {S{\'a}nchez-Arguelles}, {Sasada}, {Savolainen}, {Schloerb},
  {Schuster}, {Shao}, {Shen}, {Small}, {Sohn}, {SooHoo}, {Tazaki}, {Tiede},
  {Tilanus}, {Titus}, {Toma}, {Torne}, {Trent}, {Trippe}, {Tsuda}, {van
  Bemmel}, {van Langevelde}, {van Rossum}, {Wagner}, {Wardle}, {Weintroub},
  {Wex}, {Wharton}, {Wielgus}, {Wong}, {Wu}, {Young}, {Young}, {Younsi},
  {Yuan}, {Yuan}, {Zensus}, {Zhao}, {Zhao}, {Zhu}, {Algaba}, {Allardi},
  {Amestica}, {Anczarski}, {Bach}, {Baganoff}, {Beaudoin}, {Benson},
  {Berthold}, {Blanchard}, {Blundell}, {Bustamente}, {Cappallo},
  {Castillo-Dom{\'\i}nguez}, {Chang}, {Chang}, {Chang}, {Chen}, {Chilson},
  {Chuter}, {C{\'o}rdova Rosado}, {Coulson}, {Crawford}, {Crowley}, {David},
  {Derome}, {Dexter}, {Dornbusch}, {Dudevoir}, {Dzib}, {Eckart}, {Eckert},
  {Erickson}, {Everett}, {Faber}, {Farah}, {Fath}, {Folkers}, {Forbes},
  {Freund}, {G{\'o}mez-Ruiz}, {Gale}, {Gao}, {Geertsema}, {Graham}, {Greer},
  {Grosslein}, {Gueth}, {Haggard}, {Halverson}, {Han}, {Han}, {Hao},
  {Hasegawa}, {Henning}, {Hern{\'a}ndez-G{\'o}mez}, {Herrero-Illana},
  {Heyminck}, {Hirota}, {Hoge}, {Huang}, {Impellizzeri}, {Jiang}, {Kamble},
  {Keisler}, {Kimura}, {Kono}, {Kubo}, {Kuroda}, {Lacasse}, {Laing}, {Leitch},
  {Li}, {Lin}, {Liu}, {Liu}, {Lu}, {Marson}, {Martin-Cocher}, {Massingill},
  {Matulonis}, {McColl}, {McWhirter}, {Messias}, {Meyer-Zhao}, {Michalik},
  {Monta{\~n}a}, {Montgomerie}, {Mora-Klein}, {Muders}, {Nadolski}, {Navarro},
  {Neilsen}, {Nguyen}, {Nishioka}, {Norton}, {Nowak}, {Nystrom}, {Ogawa},
  {Oshiro}, {Oyama}, {Parsons}, {Paine}, {Pe{\~n}alver}, {Phillips}, {Poirier},
  {Pradel}, {Primiani}, {Raffin}, {Rahlin}, {Reiland}, {Risacher}, {Ruiz},
  {S{\'a}ez-Mada{\'\i}n}, {Sassella}, {Schellart}, {Shaw}, {Silva}, {Shiokawa},
  {Smith}, {Snow}, {Souccar}, {Sousa}, {Sridharan}, {Srinivasan}, {Stahm},
  {Stark}, {Story}, {Timmer}, {Vertatschitsch}, {Walther}, {Wei}, {Whitehorn},
  {Whitney}, {Woody}, {Wouterloot}, {Wright}, {Yamaguchi}, {Yu}, {Zeballos},
  {Zhang}, \& {Ziurys}}]{2019ApJ...875L...1E}
{Event Horizon Telescope Collaboration}, {Akiyama}, K., {Alberdi}, A., {et~al.}
  2019{\natexlab{a}}, \apjl, 875, L1, \dodoi{10.3847/2041-8213/ab0ec7}

\bibitem[{{Event Horizon Telescope Collaboration}
  {et~al.}(2019{\natexlab{b}}){Event Horizon Telescope Collaboration},
  {Akiyama}, {Alberdi}, {Alef}, {Asada}, {Azulay}, {Baczko}, {Ball},
  {Balokovi{\'c}}, {Barrett}, {Bintley}, {Blackburn}, {Boland}, {Bouman},
  {Bower}, {Bremer}, {Brinkerink}, {Brissenden}, {Britzen}, {Broderick},
  {Broguiere}, {Bronzwaer}, {Byun}, {Carlstrom}, {Chael}, {Chan}, {Chatterjee},
  {Chatterjee}, {Chen}, {Chen}, {Cho}, {Christian}, {Conway}, {Cordes}, {Crew},
  {Cui}, {Davelaar}, {De Laurentis}, {Deane}, {Dempsey}, {Desvignes}, {Dexter},
  {Doeleman}, {Eatough}, {Falcke}, {Fish}, {Fomalont}, {Fraga-Encinas},
  {Friberg}, {Fromm}, {G{\'o}mez}, {Galison}, {Gammie}, {Garc{\'\i}a},
  {Gentaz}, {Georgiev}, {Goddi}, {Gold}, {Gu}, {Gurwell}, {Hada}, {Hecht},
  {Hesper}, {Ho}, {Ho}, {Honma}, {Huang}, {Huang}, {Hughes}, {Ikeda}, {Inoue},
  {Issaoun}, {James}, {Jannuzi}, {Janssen}, {Jeter}, {Jiang}, {Johnson},
  {Jorstad}, {Jung}, {Karami}, {Karuppusamy}, {Kawashima}, {Keating},
  {Kettenis}, {Kim}, {Kim}, {Kim}, {Kino}, {Koay}, {Koch}, {Koyama}, {Kramer},
  {Kramer}, {Krichbaum}, {Kuo}, {Lauer}, {Lee}, {Li}, {Li}, {Lindqvist}, {Liu},
  {Liuzzo}, {Lo}, {Lobanov}, {Loinard}, {Lonsdale}, {Lu}, {MacDonald}, {Mao},
  {Markoff}, {Marrone}, {Marscher}, {Mart{\'\i}-Vidal}, {Matsushita},
  {Matthews}, {Medeiros}, {Menten}, {Mizuno}, {Mizuno}, {Moran}, {Moriyama},
  {Moscibrodzka}, {Mul{\ensuremath{\ddot{}}}ler}, {Nagai}, {Nagar}, {Nakamura},
  {Narayan}, {Narayanan}, {Natarajan}, {Neri}, {Ni}, {Noutsos}, {Okino},
  {Olivares}, {Oyama}, {{\"O}zel}, {Palumbo}, {Patel}, {Pen}, {Pesce},
  {Pi{\'e}tu}, {Plambeck}, {PopStefanija}, {Porth}, {Prather},
  {Preciado-L{\'o}pez}, {Psaltis}, {Pu}, {Ramakrishnan}, {Rao}, {Rawlings},
  {Raymond}, {Rezzolla}, {Ripperda}, {Roelofs}, {Rogers}, {Ros}, {Rose},
  {Roshanineshat}, {Rottmann}, {Roy}, {Ruszczyk}, {Ryan}, {Rygl},
  {S{\'a}nchez}, {S{\'a}nchez-Arguelles}, {Sasada}, {Savolainen}, {Schloerb},
  {Schuster}, {Shao}, {Shen}, {Small}, {Sohn}, {SooHoo}, {Tazaki}, {Tiede},
  {Tilanus}, {Titus}, {Toma}, {Torne}, {Trent}, {Trippe}, {Tsuda}, {van
  Bemmel}, {van Langevelde}, {van Rossum}, {Wagner}, {Wardle}, {Weintroub},
  {Wex}, {Wharton}, {Wielgus}, {Wong}, {Wu}, {Young}, {Young}, {Younsi},
  {Yuan}, {Yuan}, {Zensus}, {Zhao}, {Zhao}, {Zhu}, {Anczarski}, {Baganoff},
  {Eckart}, {Farah}, {Haggard}, {Meyer-Zhao}, {Michalik}, {Nadolski},
  {Neilsen}, {Nishioka}, {Nowak}, {Pradel}, {Primiani}, {Souccar},
  {Vertatschitsch}, {Yamaguchi}, \& {Zhang}}]{2019ApJ...875L...5E}
---. 2019{\natexlab{b}}, \apjl, 875, L5, \dodoi{10.3847/2041-8213/ab0f43}

\bibitem[{{Event Horizon Telescope Collaboration}
  {et~al.}(2019{\natexlab{c}}){Event Horizon Telescope Collaboration},
  {Akiyama}, {Alberdi}, {Alef}, {Asada}, {Azulay}, {Baczko}, {Ball},
  {Balokovi{\'c}}, {Barrett}, {Bintley}, {Blackburn}, {Boland}, {Bouman},
  {Bower}, {Bremer}, {Brinkerink}, {Brissenden}, {Britzen}, {Broderick},
  {Broguiere}, {Bronzwaer}, {Byun}, {Carlstrom}, {Chael}, {Chan}, {Chatterjee},
  {Chatterjee}, {Chen}, {Chen}, {Cho}, {Christian}, {Conway}, {Cordes}, {Crew},
  {Cui}, {Davelaar}, {De Laurentis}, {Deane}, {Dempsey}, {Desvignes}, {Dexter},
  {Doeleman}, {Eatough}, {Falcke}, {Fish}, {Fomalont}, {Fraga-Encinas},
  {Freeman}, {Friberg}, {Fromm}, {G{\'o}mez}, {Galison}, {Gammie},
  {Garc{\'\i}a}, {Gentaz}, {Georgiev}, {Goddi}, {Gold}, {Gu}, {Gurwell},
  {Hada}, {Hecht}, {Hesper}, {Ho}, {Ho}, {Honma}, {Huang}, {Huang}, {Hughes},
  {Ikeda}, {Inoue}, {Issaoun}, {James}, {Jannuzi}, {Janssen}, {Jeter}, {Jiang},
  {Johnson}, {Jorstad}, {Jung}, {Karami}, {Karuppusamy}, {Kawashima},
  {Keating}, {Kettenis}, {Kim}, {Kim}, {Kim}, {Kino}, {Koay}, {Koch}, {Koyama},
  {Kramer}, {Kramer}, {Krichbaum}, {Kuo}, {Lauer}, {Lee}, {Li}, {Li},
  {Lindqvist}, {Liu}, {Liuzzo}, {Lo}, {Lobanov}, {Loinard}, {Lonsdale}, {Lu},
  {MacDonald}, {Mao}, {Markoff}, {Marrone}, {Marscher}, {Mart{\'\i}-Vidal},
  {Matsushita}, {Matthews}, {Medeiros}, {Menten}, {Mizuno}, {Mizuno}, {Moran},
  {Moriyama}, {Moscibrodzka}, {M{\"u}ller}, {Nagai}, {Nagar}, {Nakamura},
  {Narayan}, {Narayanan}, {Natarajan}, {Neri}, {Ni}, {Noutsos}, {Okino},
  {Olivares}, {Oyama}, {{\"O}zel}, {Palumbo}, {Patel}, {Pen}, {Pesce},
  {Pi{\'e}tu}, {Plambeck}, {PopStefanija}, {Porth}, {Prather},
  {Preciado-L{\'o}pez}, {Psaltis}, {Pu}, {Ramakrishnan}, {Rao}, {Rawlings},
  {Raymond}, {Rezzolla}, {Ripperda}, {Roelofs}, {Rogers}, {Ros}, {Rose},
  {Roshanineshat}, {Rottmann}, {Roy}, {Ruszczyk}, {Ryan}, {Rygl},
  {S{\'a}nchez}, {S{\'a}nchez-Arguelles}, {Sasada}, {Savolainen}, {Schloerb},
  {Schuster}, {Shao}, {Shen}, {Small}, {Sohn}, {SooHoo}, {Tazaki}, {Tiede},
  {Tilanus}, {Titus}, {Toma}, {Torne}, {Trent}, {Trippe}, {Tsuda}, {van
  Bemmel}, {van Langevelde}, {van Rossum}, {Wagner}, {Wardle}, {Weintroub},
  {Wex}, {Wharton}, {Wielgus}, {Wong}, {Wu}, {Young}, {Young}, {Younsi},
  {Yuan}, {Yuan}, {Zensus}, {Zhao}, {Zhao}, {Zhu}, {Farah}, {Meyer-Zhao},
  {Michalik}, {Nadolski}, {Nishioka}, {Pradel}, {Primiani}, {Souccar},
  {Vertatschitsch}, \& {Yamaguchi}}]{2019ApJ...875L...4E}
---. 2019{\natexlab{c}}, \apjl, 875, L4, \dodoi{10.3847/2041-8213/ab0e85}

\bibitem[{{Event Horizon Telescope Collaboration}
  {et~al.}(2021{\natexlab{a}}){Event Horizon Telescope Collaboration},
  {Akiyama}, {Algaba}, {Alberdi}, {Alef}, {Anantua}, {Asada}, {Azulay},
  {Baczko}, {Ball}, {Balokovi{\'c}}, {Barrett}, {Benson}, {Bintley},
  {Blackburn}, {Blundell}, {Boland}, {Bouman}, {Bower}, {Boyce}, {Bremer},
  {Brinkerink}, {Brissenden}, {Britzen}, {Broderick}, {Broguiere}, {Bronzwaer},
  {Byun}, {Carlstrom}, {Chael}, {Chan}, {Chatterjee}, {Chatterjee}, {Chen},
  {Chen}, {Chesler}, {Cho}, {Christian}, {Conway}, {Cordes}, {Crawford},
  {Crew}, {Cruz-Osorio}, {Cui}, {Davelaar}, {De Laurentis}, {Deane}, {Dempsey},
  {Desvignes}, {Dexter}, {Doeleman}, {Eatough}, {Falcke}, {Farah}, {Fish},
  {Fomalont}, {Ford}, {Fraga-Encinas}, {Freeman}, {Friberg}, {Fromm},
  {Fuentes}, {Galison}, {Gammie}, {Garc{\'\i}a}, {Gentaz}, {Georgiev}, {Goddi},
  {Gold}, {G{\'o}mez}, {G{\'o}mez-Ruiz}, {Gu}, {Gurwell}, {Hada}, {Haggard},
  {Hecht}, {Hesper}, {Ho}, {Ho}, {Honma}, {Huang}, {Huang}, {Hughes}, {Ikeda},
  {Inoue}, {Issaoun}, {James}, {Jannuzi}, {Janssen}, {Jeter}, {Jiang},
  {Jimenez-Rosales}, {Johnson}, {Jorstad}, {Jung}, {Karami}, {Karuppusamy},
  {Kawashima}, {Keating}, {Kettenis}, {Kim}, {Kim}, {Kim}, {Kim}, {Kino},
  {Koay}, {Kofuji}, {Koch}, {Koyama}, {Kramer}, {Kramer}, {Krichbaum}, {Kuo},
  {Lauer}, {Lee}, {Levis}, {Li}, {Li}, {Lindqvist}, {Lico}, {Lindahl}, {Liu},
  {Liu}, {Liuzzo}, {Lo}, {Lobanov}, {Loinard}, {Lonsdale}, {Lu}, {MacDonald},
  {Mao}, {Marchili}, {Markoff}, {Marrone}, {Marscher}, {Mart{\'\i}-Vidal},
  {Matsushita}, {Matthews}, {Medeiros}, {Menten}, {Mizuno}, {Mizuno}, {Moran},
  {Moriyama}, {Moscibrodzka}, {M{\"u}ller}, {Musoke}, {Mej{\'\i}as},
  {Michalik}, {Nadolski}, {Nagai}, {Nagar}, {Nakamura}, {Narayan}, {Narayanan},
  {Natarajan}, {Nathanail}, {Neilsen}, {Neri}, {Ni}, {Noutsos}, {Nowak},
  {Okino}, {Olivares}, {Ortiz-Le{\'o}n}, {Oyama}, {{\"O}zel}, {Palumbo},
  {Park}, {Patel}, {Pen}, {Pesce}, {Pi{\'e}tu}, {Plambeck}, {PopStefanija},
  {Porth}, {P{\"o}tzl}, {Prather}, {Preciado-L{\'o}pez}, {Psaltis}, {Pu},
  {Ramakrishnan}, {Rao}, {Rawlings}, {Raymond}, {Rezzolla}, {Ricarte},
  {Ripperda}, {Roelofs}, {Rogers}, {Ros}, {Rose}, {Roshanineshat}, {Rottmann},
  {Roy}, {Ruszczyk}, {Rygl}, {S{\'a}nchez}, {S{\'a}nchez-Arguelles}, {Sasada},
  {Savolainen}, {Schloerb}, {Schuster}, {Shao}, {Shen}, {Small}, {Sohn},
  {SooHoo}, {Sun}, {Tazaki}, {Tetarenko}, {Tiede}, {Tilanus}, {Titus}, {Toma},
  {Torne}, {Trent}, {Traianou}, {Trippe}, {van Bemmel}, {van Langevelde}, {van
  Rossum}, {Wagner}, {Ward-Thompson}, {Wardle}, {Weintroub}, {Wex}, {Wharton},
  {Wielgus}, {Wong}, {Wu}, {Yoon}, {Young}, {Young}, {Younsi}, {Yuan}, {Yuan},
  {Zensus}, {Zhao}, \& {Zhao}}]{2021ApJ...910L..12E}
{Event Horizon Telescope Collaboration}, {Akiyama}, K., {Algaba}, J.~C.,
  {et~al.} 2021{\natexlab{a}}, \apjl, 910, L12,
  \dodoi{10.3847/2041-8213/abe71d}

\bibitem[{{Event Horizon Telescope Collaboration}
  {et~al.}(2021{\natexlab{b}}){Event Horizon Telescope Collaboration},
  {Akiyama}, {Algaba}, {Alberdi}, {Alef}, {Anantua}, {Asada}, {Azulay},
  {Baczko}, {Ball}, {Balokovi{\'c}}, {Barrett}, {Benson}, {Bintley},
  {Blackburn}, {Blundell}, {Boland}, {Bouman}, {Bower}, {Boyce}, {Bremer},
  {Brinkerink}, {Brissenden}, {Britzen}, {Broderick}, {Broguiere}, {Bronzwaer},
  {Byun}, {Carlstrom}, {Chael}, {Chan}, {Chatterjee}, {Chatterjee}, {Chen},
  {Chen}, {Chesler}, {Cho}, {Christian}, {Conway}, {Cordes}, {Crawford},
  {Crew}, {Cruz-Osorio}, {Cui}, {Davelaar}, {De Laurentis}, {Deane}, {Dempsey},
  {Desvignes}, {Dexter}, {Doeleman}, {Eatough}, {Falcke}, {Farah}, {Fish},
  {Fomalont}, {Ford}, {Fraga-Encinas}, {Friberg}, {Fromm}, {Fuentes},
  {Galison}, {Gammie}, {Garc{\'\i}a}, {Gelles}, {Gentaz}, {Georgiev}, {Goddi},
  {Gold}, {G{\'o}mez}, {G{\'o}mez-Ruiz}, {Gu}, {Gurwell}, {Hada}, {Haggard},
  {Hecht}, {Hesper}, {Himwich}, {Ho}, {Ho}, {Honma}, {Huang}, {Huang},
  {Hughes}, {Ikeda}, {Inoue}, {Issaoun}, {James}, {Jannuzi}, {Janssen},
  {Jeter}, {Jiang}, {Jimenez-Rosales}, {Johnson}, {Jorstad}, {Jung}, {Karami},
  {Karuppusamy}, {Kawashima}, {Keating}, {Kettenis}, {Kim}, {Kim}, {Kim},
  {Kim}, {Kino}, {Koay}, {Kofuji}, {Koch}, {Koyama}, {Kramer}, {Kramer},
  {Krichbaum}, {Kuo}, {Lauer}, {Lee}, {Levis}, {Li}, {Li}, {Lindqvist}, {Lico},
  {Lindahl}, {Liu}, {Liu}, {Liuzzo}, {Lo}, {Lobanov}, {Loinard}, {Lonsdale},
  {Lu}, {MacDonald}, {Mao}, {Marchili}, {Markoff}, {Marrone}, {Marscher},
  {Mart{\'\i}-Vidal}, {Matsushita}, {Matthews}, {Medeiros}, {Menten}, {Mizuno},
  {Mizuno}, {Moran}, {Moriyama}, {Moscibrodzka}, {M{\"u}ller}, {Musoke}, {Mus
  Mej{\'\i}as}, {Michalik}, {Nadolski}, {Nagai}, {Nagar}, {Nakamura},
  {Narayan}, {Narayanan}, {Natarajan}, {Nathanail}, {Neilsen}, {Neri}, {Ni},
  {Noutsos}, {Nowak}, {Okino}, {Olivares}, {Ortiz-Le{\'o}n}, {Oyama},
  {{\"O}zel}, {Palumbo}, {Park}, {Patel}, {Pen}, {Pesce}, {Pi{\'e}tu},
  {Plambeck}, {PopStefanija}, {Porth}, {P{\"o}tzl}, {Prather},
  {Preciado-L{\'o}pez}, {Psaltis}, {Pu}, {Ramakrishnan}, {Rao}, {Rawlings},
  {Raymond}, {Rezzolla}, {Ricarte}, {Ripperda}, {Roelofs}, {Rogers}, {Ros},
  {Rose}, {Roshanineshat}, {Rottmann}, {Roy}, {Ruszczyk}, {Rygl},
  {S{\'a}nchez}, {S{\'a}nchez-Arguelles}, {Sasada}, {Savolainen}, {Schloerb},
  {Schuster}, {Shao}, {Shen}, {Small}, {Sohn}, {SooHoo}, {Sun}, {Tazaki},
  {Tetarenko}, {Tiede}, {Tilanus}, {Titus}, {Toma}, {Torne}, {Trent},
  {Traianou}, {Trippe}, {van Bemmel}, {van Langevelde}, {van Rossum}, {Wagner},
  {Ward-Thompson}, {Wardle}, {Weintroub}, {Wex}, {Wharton}, {Wielgus}, {Wong},
  {Wu}, {Yoon}, {Young}, {Young}, {Younsi}, {Yuan}, {Yuan}, {Zensus}, {Zhao},
  \& {Zhao}}]{2021ApJ...910L..13E}
---. 2021{\natexlab{b}}, \apjl, 910, L13, \dodoi{10.3847/2041-8213/abe4de}

\bibitem[{{Fishbone} \& {Moncrief}(1976)}]{Fishbone_Moncrief_1976}
{Fishbone}, L.~G., \& {Moncrief}, V. 1976, \apj, 207, 962,
  \dodoi{10.1086/154565}

\bibitem[{{Ford} {et~al.}(1994){Ford}, {Harms}, {Tsvetanov}, {Hartig},
  {Dressel}, {Kriss}, {Bohlin}, {Davidsen}, {Margon}, \&
  {Kochhar}}]{1994ApJ...435L..27F}
{Ford}, H.~C., {Harms}, R.~J., {Tsvetanov}, Z.~I., {et~al.} 1994, \apjl, 435,
  L27, \dodoi{10.1086/187586}

\bibitem[{{Gabuzda} {et~al.}(2008){Gabuzda}, {Vitrishchak}, {Mahmud}, \&
  {O'Sullivan}}]{2008MNRAS.384.1003G}
{Gabuzda}, D.~C., {Vitrishchak}, V.~M., {Mahmud}, M., \& {O'Sullivan}, S.~P.
  2008, \mnras, 384, 1003, \dodoi{10.1111/j.1365-2966.2007.12773.x}

\bibitem[{{Gammie}(2004)}]{Gammie_2004}
{Gammie}, C.~F. 2004, \apj, 614, 309, \dodoi{10.1086/423443}

\bibitem[{{Gammie} {et~al.}(2003){Gammie}, {McKinney}, \&
  {T{\'o}th}}]{2003ApJ...589..444G}
{Gammie}, C.~F., {McKinney}, J.~C., \& {T{\'o}th}, G. 2003, \apj, 589, 444,
  \dodoi{10.1086/374594}

\bibitem[{{Gebhardt} {et~al.}(2011){Gebhardt}, {Adams}, {Richstone}, {Lauer},
  {Faber}, {G{\"u}ltekin}, {Murphy}, \& {Tremaine}}]{2011ApJ...729..119G}
{Gebhardt}, K., {Adams}, J., {Richstone}, D., {et~al.} 2011, \apj, 729, 119,
  \dodoi{10.1088/0004-637X/729/2/119}

\bibitem[{{Goddi} {et~al.}(2021){Goddi}, {Marti-Vidal}, {Messias}, {Bower},
  {Broderick}, {Dexter}, {Marrone}, {Moscibrodzka}, {Nagai}, {Algaba}, {Asada},
  {Crew}, {Gomez}, {Impellizzeri}, {Janssen}, {Kadler}, {Krichbaum}, {Lico},
  {Matthews}, {Nathanail}, {Ricarte}, {Ros}, {Younsi}, {The Event Horizon
  Telescope Collaboration}, {Bruni}, {Gopakumar}, {Hernandez-Gomez},
  {Herrero-Illana}, {Ingram}, {Komossa}, {Muders}, {Perucho}, {Rosch}, \&
  {Valtonen}}]{2021arXiv210502272G}
{Goddi}, C., {Marti-Vidal}, I., {Messias}, H., {et~al.} 2021, arXiv e-prints,
  arXiv:2105.02272.
\newblock \doarXiv{2105.02272}

\bibitem[{{Gold} {et~al.}(2017){Gold}, {McKinney}, {Johnson}, \&
  {Doeleman}}]{2017ApJ...837..180G}
{Gold}, R., {McKinney}, J.~C., {Johnson}, M.~D., \& {Doeleman}, S.~S. 2017,
  \apj, 837, 180, \dodoi{10.3847/1538-4357/aa6193}

\bibitem[{{Hada} {et~al.}(2011){Hada}, {Doi}, {Kino}, {Nagai}, {Hagiwara}, \&
  {Kawaguchi}}]{2011Natur.477..185H}
{Hada}, K., {Doi}, A., {Kino}, M., {et~al.} 2011, \nat, 477, 185,
  \dodoi{10.1038/nature10387}

\bibitem[{{Hada} {et~al.}(2013){Hada}, {Kino}, {Doi}, {Nagai}, {Honma},
  {Hagiwara}, {Giroletti}, {Giovannini}, \& {Kawaguchi}}]{2013ApJ...775...70H}
{Hada}, K., {Kino}, M., {Doi}, A., {et~al.} 2013, \apj, 775, 70,
  \dodoi{10.1088/0004-637X/775/1/70}

\bibitem[{{Hada} {et~al.}(2016){Hada}, {Kino}, {Doi}, {Nagai}, {Honma},
  {Akiyama}, {Tazaki}, {Lico}, {Giroletti}, {Giovannini}, {Orienti}, \&
  {Hagiwara}}]{2016ApJ...817..131H}
---. 2016, \apj, 817, 131, \dodoi{10.3847/0004-637X/817/2/131}

\bibitem[{{Hawley} {et~al.}(2011){Hawley}, {Guan}, \&
  {Krolik}}]{2011ApJ...738...84H}
{Hawley}, J.~F., {Guan}, X., \& {Krolik}, J.~H. 2011, \apj, 738, 84,
  \dodoi{10.1088/0004-637X/738/1/84}

\bibitem[{{Ho} {et~al.}(1997){Ho}, {Filippenko}, \&
  {Sargent}}]{1997ApJS..112..315H}
{Ho}, L.~C., {Filippenko}, A.~V., \& {Sargent}, W. L.~W. 1997, \apjs, 112, 315,
  \dodoi{10.1086/313041}

\bibitem[{{Hodge}(1982)}]{1982ApJ...263..595H}
{Hodge}, P.~E. 1982, \apj, 263, 595, \dodoi{10.1086/160530}

\bibitem[{{Homan} {et~al.}(2009){Homan}, {Lister}, {Aller}, {Aller}, \&
  {Wardle}}]{2009ApJ...696..328H}
{Homan}, D.~C., {Lister}, M.~L., {Aller}, H.~D., {Aller}, M.~F., \& {Wardle},
  J.~F.~C. 2009, \apj, 696, 328, \dodoi{10.1088/0004-637X/696/1/328}

\bibitem[{{Howes}(2010)}]{2010MNRAS.409L.104H}
{Howes}, G.~G. 2010, \mnras, 409, L104,
  \dodoi{10.1111/j.1745-3933.2010.00958.x}

\bibitem[{{Jeter} {et~al.}(2020){Jeter}, {Broderick}, \&
  {Gold}}]{2020MNRAS.493.5606J}
{Jeter}, B., {Broderick}, A.~E., \& {Gold}, R. 2020, \mnras, 493, 5606,
  \dodoi{10.1093/mnras/staa679}

\bibitem[{{Jim{\'e}nez-Rosales} \& {Dexter}(2018)}]{2018MNRAS.478.1875J}
{Jim{\'e}nez-Rosales}, A., \& {Dexter}, J. 2018, \mnras, 478, 1875,
  \dodoi{10.1093/mnras/sty1210}

\bibitem[{{Johnson} {et~al.}(2014){Johnson}, {Fish}, {Doeleman}, {Broderick},
  {Wardle}, \& {Marrone}}]{2014ApJ...794..150J}
{Johnson}, M.~D., {Fish}, V.~L., {Doeleman}, S.~S., {et~al.} 2014, \apj, 794,
  150, \dodoi{10.1088/0004-637X/794/2/150}

\bibitem[{{Johnson} {et~al.}(2015){Johnson}, {Fish}, {Doeleman}, {Marrone},
  {Plambeck}, {Wardle}, {Akiyama}, {Asada}, {Beaudoin}, {Blackburn},
  {Blundell}, {Bower}, {Brinkerink}, {Broderick}, {Cappallo}, {Chael}, {Crew},
  {Dexter}, {Dexter}, {Freund}, {Friberg}, {Gold}, {Gurwell}, {Ho}, {Honma},
  {Inoue}, {Kosowsky}, {Krichbaum}, {Lamb}, {Loeb}, {Lu}, {MacMahon},
  {McKinney}, {Moran}, {Narayan}, {Primiani}, {Psaltis}, {Rogers}, {Rosenfeld},
  {SooHoo}, {Tilanus}, {Titus}, {Vertatschitsch}, {Weintroub}, {Wright},
  {Young}, {Zensus}, \& {Ziurys}}]{2015Sci...350.1242J}
---. 2015, Science, 350, 1242, \dodoi{10.1126/science.aac7087}

\bibitem[{{Jones}(1988)}]{1988ApJ...332..678J}
{Jones}, T.~W. 1988, \apj, 332, 678, \dodoi{10.1086/166685}

\bibitem[{{Jones} \& {O'Dell}(1977)}]{1977ApJ...214..522J}
{Jones}, T.~W., \& {O'Dell}, S.~L. 1977, \apj, 214, 522, \dodoi{10.1086/155278}

\bibitem[{{Junor} {et~al.}(1999){Junor}, {Biretta}, \&
  {Livio}}]{1999Natur.401..891J}
{Junor}, W., {Biretta}, J.~A., \& {Livio}, M. 1999, \nat, 401, 891,
  \dodoi{10.1038/44780}

\bibitem[{{Kato} {et~al.}(2008){Kato}, {Fukue}, \&
  {Mineshige}}]{2008bhad.book.....K}
{Kato}, S., {Fukue}, J., \& {Mineshige}, S. 2008, {Black-Hole Accretion Disks
  --- Towards a New Paradigm ---}

\bibitem[{{Kawashima} {et~al.}(2019){Kawashima}, {Kino}, \&
  {Akiyama}}]{2019ApJ...878...27K}
{Kawashima}, T., {Kino}, M., \& {Akiyama}, K. 2019, \apj, 878, 27,
  \dodoi{10.3847/1538-4357/ab19c0}

\bibitem[{{Kawashima} {et~al.}(2021{\natexlab{a}}){Kawashima}, {Ohsuga}, \&
  {Takahashi}}]{Kawashima_2021}
{Kawashima}, T., {Ohsuga}, K., \& {Takahashi}, H.~R. 2021{\natexlab{a}}, arXiv
  e-prints, arXiv:2108.05131.
\newblock \doarXiv{2108.05131}

\bibitem[{{Kawashima} {et~al.}(2021{\natexlab{b}}){Kawashima}, {Toma}, {Kino},
  {Akiyama}, {Nakamura}, \& {Moriyama}}]{2021ApJ...909..168K}
{Kawashima}, T., {Toma}, K., {Kino}, M., {et~al.} 2021{\natexlab{b}}, \apj,
  909, 168, \dodoi{10.3847/1538-4357/abd5bb}

\bibitem[{{Kawazura} {et~al.}(2019){Kawazura}, {Barnes}, \&
  {Schekochihin}}]{2019PNAS..116..771K}
{Kawazura}, Y., {Barnes}, M., \& {Schekochihin}, A.~A. 2019, Proceedings of the
  National Academy of Science, 116, 771, \dodoi{10.1073/pnas.1812491116}

\bibitem[{{Kim} {et~al.}(2018){Kim}, {Krichbaum}, {Lu}, {Ros}, {Bach},
  {Bremer}, {de Vicente}, {Lindqvist}, \& {Zensus}}]{2018A&A...616A.188K}
{Kim}, J.~Y., {Krichbaum}, T.~P., {Lu}, R.~S., {et~al.} 2018, \aap, 616, A188,
  \dodoi{10.1051/0004-6361/201832921}

\bibitem[{{Kino} {et~al.}(2014){Kino}, {Takahara}, {Hada}, \&
  {Doi}}]{2014ApJ...786....5K}
{Kino}, M., {Takahara}, F., {Hada}, K., \& {Doi}, A. 2014, \apj, 786, 5,
  \dodoi{10.1088/0004-637X/786/1/5}

\bibitem[{{Koide} {et~al.}(1999){Koide}, {Shibata}, \&
  {Kudoh}}]{1999ApJ...522..727K}
{Koide}, S., {Shibata}, K., \& {Kudoh}, T. 1999, \apj, 522, 727,
  \dodoi{10.1086/307667}

\bibitem[{{Komissarov}(2005)}]{2005MNRAS.359..801K}
{Komissarov}, S.~S. 2005, \mnras, 359, 801,
  \dodoi{10.1111/j.1365-2966.2005.08974.x}

\bibitem[{{Kovalev} {et~al.}(2007){Kovalev}, {Lister}, {Homan}, \&
  {Kellermann}}]{2007ApJ...668L..27K}
{Kovalev}, Y.~Y., {Lister}, M.~L., {Homan}, D.~C., \& {Kellermann}, K.~I. 2007,
  \apjl, 668, L27, \dodoi{10.1086/522603}

\bibitem[{{Kravchenko} {et~al.}(2020){Kravchenko}, {Giroletti}, {Hada},
  {Meier}, {Nakamura}, {Park}, \& {Walker}}]{2020A&A...637L...6K}
{Kravchenko}, E., {Giroletti}, M., {Hada}, K., {et~al.} 2020, \aap, 637, L6,
  \dodoi{10.1051/0004-6361/201937315}

\bibitem[{{Kuo} {et~al.}(2014){Kuo}, {Asada}, {Rao}, {Nakamura}, {Algaba},
  {Liu}, {Inoue}, {Koch}, {Ho}, {Matsushita}, {Pu}, {Akiyama}, {Nishioka}, \&
  {Pradel}}]{2014ApJ...783L..33K}
{Kuo}, C.~Y., {Asada}, K., {Rao}, R., {et~al.} 2014, \apjl, 783, L33,
  \dodoi{10.1088/2041-8205/783/2/L33}

\bibitem[{{Legg} \& {Westfold}(1968)}]{1968ApJ...154..499L}
{Legg}, M.~P.~C., \& {Westfold}, K.~C. 1968, \apj, 154, 499,
  \dodoi{10.1086/149777}

\bibitem[{{Lu} {et~al.}(2014){Lu}, {Broderick}, {Baron}, {Monnier}, {Fish},
  {Doeleman}, \& {Pankratius}}]{2014ApJ...788..120L}
{Lu}, R.-S., {Broderick}, A.~E., {Baron}, F., {et~al.} 2014, \apj, 788, 120,
  \dodoi{10.1088/0004-637X/788/2/120}

\bibitem[{{Ly} {et~al.}(2007){Ly}, {Walker}, \& {Junor}}]{2007ApJ...660..200L}
{Ly}, C., {Walker}, R.~C., \& {Junor}, W. 2007, \apj, 660, 200,
  \dodoi{10.1086/512846}

\bibitem[{{Lynden-Bell}(1969)}]{1969Natur.223..690L}
{Lynden-Bell}, D. 1969, \nat, 223, 690, \dodoi{10.1038/223690a0}

\bibitem[{{Macchetto} {et~al.}(1997){Macchetto}, {Marconi}, {Axon}, {Capetti},
  {Sparks}, \& {Crane}}]{1997ApJ...489..579M}
{Macchetto}, F., {Marconi}, A., {Axon}, D.~J., {et~al.} 1997, \apj, 489, 579,
  \dodoi{10.1086/304823}

\bibitem[{{Mahadevan} {et~al.}(1996){Mahadevan}, {Narayan}, \&
  {Yi}}]{1996ApJ...465..327M}
{Mahadevan}, R., {Narayan}, R., \& {Yi}, I. 1996, \apj, 465, 327,
  \dodoi{10.1086/177422}

\bibitem[{{Marshall} {et~al.}(2002){Marshall}, {Miller}, {Davis}, {Perlman},
  {Wise}, {Canizares}, \& {Harris}}]{2002ApJ...564..683M}
{Marshall}, H.~L., {Miller}, B.~P., {Davis}, D.~S., {et~al.} 2002, \apj, 564,
  683, \dodoi{10.1086/324396}

\bibitem[{{McKinney} {et~al.}(2012){McKinney}, {Tchekhovskoy}, \&
  {Blandford}}]{2012MNRAS.423.3083M}
{McKinney}, J.~C., {Tchekhovskoy}, A., \& {Blandford}, R.~D. 2012, \mnras, 423,
  3083, \dodoi{10.1111/j.1365-2966.2012.21074.x}

\bibitem[{{Mei} {et~al.}(2007){Mei}, {Blakeslee}, {C{\^o}t{\'e}}, {Tonry},
  {West}, {Ferrarese}, {Jord{\'a}n}, {Peng}, {Anthony}, \&
  {Merritt}}]{2007ApJ...655..144M}
{Mei}, S., {Blakeslee}, J.~P., {C{\^o}t{\'e}}, P., {et~al.} 2007, \apj, 655,
  144, \dodoi{10.1086/509598}

\bibitem[{{Mertens} {et~al.}(2016){Mertens}, {Lobanov}, {Walker}, \&
  {Hardee}}]{2016A&A...595A..54M}
{Mertens}, F., {Lobanov}, A.~P., {Walker}, R.~C., \& {Hardee}, P.~E. 2016,
  \aap, 595, A54, \dodoi{10.1051/0004-6361/201628829}

\bibitem[{{Mizuno} {et~al.}(2021){Mizuno}, {Fromm}, {Younsi}, {Porth},
  {Olivares}, \& {Rezzolla}}]{2021arXiv210609272M}
{Mizuno}, Y., {Fromm}, C.~M., {Younsi}, Z., {et~al.} 2021, arXiv e-prints,
  arXiv:2106.09272.
\newblock \doarXiv{2106.09272}

\bibitem[{{Mo{\'s}cibrodzka} {et~al.}(2017){Mo{\'s}cibrodzka}, {Dexter},
  {Davelaar}, \& {Falcke}}]{2017MNRAS.468.2214M}
{Mo{\'s}cibrodzka}, M., {Dexter}, J., {Davelaar}, J., \& {Falcke}, H. 2017,
  \mnras, 468, 2214, \dodoi{10.1093/mnras/stx587}

\bibitem[{{Mo{\'s}cibrodzka} \& {Falcke}(2013)}]{2013A&A...559L...3M}
{Mo{\'s}cibrodzka}, M., \& {Falcke}, H. 2013, \aap, 559, L3,
  \dodoi{10.1051/0004-6361/201322692}

\bibitem[{{Mo{\'s}cibrodzka} {et~al.}(2016){Mo{\'s}cibrodzka}, {Falcke}, \&
  {Shiokawa}}]{2016A&A...586A..38M}
{Mo{\'s}cibrodzka}, M., {Falcke}, H., \& {Shiokawa}, H. 2016, \aap, 586, A38,
  \dodoi{10.1051/0004-6361/201526630}

\bibitem[{{Moscibrodzka} {et~al.}(2021){Moscibrodzka}, {Janiuk}, \& {De
  Laurentis}}]{2021arXiv210300267M}
{Moscibrodzka}, M., {Janiuk}, A., \& {De Laurentis}, M. 2021, arXiv e-prints,
  arXiv:2103.00267.
\newblock \doarXiv{2103.00267}

\bibitem[{{Nakamura} \& {Asada}(2013)}]{2013ApJ...775..118N}
{Nakamura}, M., \& {Asada}, K. 2013, \apj, 775, 118,
  \dodoi{10.1088/0004-637X/775/2/118}

\bibitem[{{Nakamura} {et~al.}(2018){Nakamura}, {Asada}, {Hada}, {Pu}, {Noble},
  {Tseng}, {Toma}, {Kino}, {Nagai}, {Takahashi}, {Algaba}, {Orienti},
  {Akiyama}, {Doi}, {Giovannini}, {Giroletti}, {Honma}, {Koyama}, {Lico},
  {Niinuma}, \& {Tazaki}}]{2018ApJ...868..146N}
{Nakamura}, M., {Asada}, K., {Hada}, K., {et~al.} 2018, \apj, 868, 146,
  \dodoi{10.3847/1538-4357/aaeb2d}

\bibitem[{{Narayan} {et~al.}(2003){Narayan}, {Igumenshchev}, \&
  {Abramowicz}}]{2003PASJ...55L..69N}
{Narayan}, R., {Igumenshchev}, I.~V., \& {Abramowicz}, M.~A. 2003, \pasj, 55,
  L69, \dodoi{10.1093/pasj/55.6.L69}

\bibitem[{{Narayan} {et~al.}(2012){Narayan}, {S{\"A} dowski}, {Penna}, \&
  {Kulkarni}}]{2012MNRAS.426.3241N}
{Narayan}, R., {S{\"A} dowski}, A., {Penna}, R.~F., \& {Kulkarni}, A.~K. 2012,
  \mnras, 426, 3241, \dodoi{10.1111/j.1365-2966.2012.22002.x}

\bibitem[{{Narayan} \& {Yi}(1995)}]{1995ApJ...452..710N}
{Narayan}, R., \& {Yi}, I. 1995, \apj, 452, 710, \dodoi{10.1086/176343}

\bibitem[{{Noble} {et~al.}(2006){Noble}, {Gammie}, {McKinney}, \& {Del
  Zanna}}]{2006ApJ...641..626N}
{Noble}, S.~C., {Gammie}, C.~F., {McKinney}, J.~C., \& {Del Zanna}, L. 2006,
  \apj, 641, 626, \dodoi{10.1086/500349}

\bibitem[{{Owen} {et~al.}(1990){Owen}, {Eilek}, \&
  {Keel}}]{1990ApJ...362..449O}
{Owen}, F.~N., {Eilek}, J.~A., \& {Keel}, W.~C. 1990, \apj, 362, 449,
  \dodoi{10.1086/169282}

\bibitem[{{Owen} {et~al.}(1989){Owen}, {Hardee}, \&
  {Cornwell}}]{1989ApJ...340..698O}
{Owen}, F.~N., {Hardee}, P.~E., \& {Cornwell}, T.~J. 1989, \apj, 340, 698,
  \dodoi{10.1086/167430}

\bibitem[{{Park} {et~al.}(2019{\natexlab{a}}){Park}, {Hada}, {Kino},
  {Nakamura}, {Ro}, \& {Trippe}}]{2019ApJ...871..257P}
{Park}, J., {Hada}, K., {Kino}, M., {et~al.} 2019{\natexlab{a}}, \apj, 871,
  257, \dodoi{10.3847/1538-4357/aaf9a9}

\bibitem[{{Park} {et~al.}(2019{\natexlab{b}}){Park}, {Hada}, {Kino},
  {Nakamura}, {Hodgson}, {Ro}, {Cui}, {Asada}, {Algaba}, {Sawada-Satoh}, {Lee},
  {Cho}, {Shen}, {Jiang}, {Trippe}, {Niinuma}, {Sohn}, {Jung}, {Zhao},
  {Wajima}, {Tazaki}, {Honma}, {An}, {Akiyama}, {Byun}, {Kim}, {Zhang},
  {Cheng}, {Kobayashi}, {Shibata}, {Lee}, {Roh}, {Oh}, {Yeom}, {Jung}, {Oh},
  {Kim}, {Hwang}, \& {Hagiwara}}]{2019ApJ...887..147P}
---. 2019{\natexlab{b}}, \apj, 887, 147, \dodoi{10.3847/1538-4357/ab5584}

\bibitem[{{Rees}(1984)}]{1984ARA&A..22..471R}
{Rees}, M.~J. 1984, \araa, 22, 471, \dodoi{10.1146/annurev.aa.22.090184.002351}

\bibitem[{{Ricarte} {et~al.}(2020){Ricarte}, {Prather}, {Wong}, {Narayan},
  {Gammie}, \& {Johnson}}]{2020MNRAS.498.5468R}
{Ricarte}, A., {Prather}, B.~S., {Wong}, G.~N., {et~al.} 2020, \mnras, 498,
  5468, \dodoi{10.1093/mnras/staa2692}

\bibitem[{{Ricarte} {et~al.}(2021){Ricarte}, {Qiu}, \&
  {Narayan}}]{2021MNRAS.tmp.1263R}
{Ricarte}, A., {Qiu}, R., \& {Narayan}, R. 2021, \mnras,
  \dodoi{10.1093/mnras/stab1289}

\bibitem[{{Ryan} {et~al.}(2018){Ryan}, {Ressler}, {Dolence}, {Gammie}, \&
  {Quataert}}]{2018ApJ...864..126R}
{Ryan}, B.~R., {Ressler}, S.~M., {Dolence}, J.~C., {Gammie}, C., \& {Quataert},
  E. 2018, \apj, 864, 126, \dodoi{10.3847/1538-4357/aad73a}

\bibitem[{{S{\"{a}}dowski} {et~al.}(2013){S{\"{a}}dowski}, {Narayan}, {Penna},
  \& {Zhu}}]{2013MNRAS.436.3856S}
{S{\"{a}}dowski}, A., {Narayan}, R., {Penna}, R., \& {Zhu}, Y. 2013, \mnras,
  436, 3856, \dodoi{10.1093/mnras/stt1881}

\bibitem[{{Sanders} {et~al.}(1989){Sanders}, {Phinney}, {Neugebauer}, {Soifer},
  \& {Matthews}}]{1989ApJ...347...29S}
{Sanders}, D.~B., {Phinney}, E.~S., {Neugebauer}, G., {Soifer}, B.~T., \&
  {Matthews}, K. 1989, \apj, 347, 29, \dodoi{10.1086/168094}

\bibitem[{{Sano} {et~al.}(2004){Sano}, {Inutsuka}, {Turner}, \&
  {Stone}}]{2004ApJ...605..321S}
{Sano}, T., {Inutsuka}, S.-i., {Turner}, N.~J., \& {Stone}, J.~M. 2004, \apj,
  605, 321, \dodoi{10.1086/382184}

\bibitem[{{Shcherbakov}(2008)}]{2008ApJ...688..695S}
{Shcherbakov}, R.~V. 2008, \apj, 688, 695, \dodoi{10.1086/592326}

\bibitem[{{Takahashi} {et~al.}(2018){Takahashi}, {Mineshige}, \&
  {Ohsuga}}]{Takahashi_2018}
{Takahashi}, H.~R., {Mineshige}, S., \& {Ohsuga}, K. 2018, \apj, 853, 45,
  \dodoi{10.3847/1538-4357/aaa082}

\bibitem[{{Takahashi} {et~al.}(2016){Takahashi}, {Ohsuga}, {Kawashima}, \&
  {Sekiguchi}}]{Takahashi_2016}
{Takahashi}, H.~R., {Ohsuga}, K., {Kawashima}, T., \& {Sekiguchi}, Y. 2016,
  \apj, 826, 23, \dodoi{10.3847/0004-637X/826/1/23}

\bibitem[{{Tchekhovskoy} {et~al.}(2011){Tchekhovskoy}, {Narayan}, \&
  {McKinney}}]{2011MNRAS.418L..79T}
{Tchekhovskoy}, A., {Narayan}, R., \& {McKinney}, J.~C. 2011, \mnras, 418, L79,
  \dodoi{10.1111/j.1745-3933.2011.01147.x}

\bibitem[{{Tsunetoe} {et~al.}(2020{\natexlab{a}}){Tsunetoe}, {Mineshige},
  {Ohsuga}, {Kawashima}, \& {Akiyama}}]{2020PASJ...72...32T}
{Tsunetoe}, Y., {Mineshige}, S., {Ohsuga}, K., {Kawashima}, T., \& {Akiyama},
  K. 2020{\natexlab{a}}, \pasj, 72, 32, \dodoi{10.1093/pasj/psaa008}

\bibitem[{{Tsunetoe} {et~al.}(2020{\natexlab{b}}){Tsunetoe}, {Mineshige},
  {Ohsuga}, {Kawashima}, \& {Akiyama}}]{2020arXiv201205243T}
---. 2020{\natexlab{b}}, arXiv e-prints, arXiv:2012.05243.
\newblock \doarXiv{2012.05243}

\bibitem[{{Walker} {et~al.}(2018){Walker}, {Hardee}, {Davies}, {Ly}, \&
  {Junor}}]{2018ApJ...855..128W}
{Walker}, R.~C., {Hardee}, P.~E., {Davies}, F.~B., {Ly}, C., \& {Junor}, W.
  2018, \apj, 855, 128, \dodoi{10.3847/1538-4357/aaafcc}

\bibitem[{{Wardle} \& {Homan}(2003)}]{2003Ap&SS.288..143W}
{Wardle}, J. F.~C., \& {Homan}, D.~C. 2003, \apss, 288, 143,
  \dodoi{10.1023/B:ASTR.0000005001.80514.0c}

\bibitem[{{Yuan} \& {Narayan}(2014)}]{2014ARA&A..52..529Y}
{Yuan}, F., \& {Narayan}, R. 2014, \araa, 52, 529,
  \dodoi{10.1146/annurev-astro-082812-141003}

\bibitem[{{Zavala} \& {Taylor}(2002)}]{2002ApJ...566L...9Z}
{Zavala}, R.~T., \& {Taylor}, G.~B. 2002, \apjl, 566, L9,
  \dodoi{10.1086/339441}

\bibitem[{{Zavala} \& {Taylor}(2003)}]{2003ApJ...589..126Z}
---. 2003, \apj, 589, 126, \dodoi{10.1086/374619}

\bibitem[{{Zavala} \& {Taylor}(2004)}]{2004ApJ...612..749Z}
---. 2004, \apj, 612, 749, \dodoi{10.1086/422741}

\bibitem[{{Zensus}(1997)}]{1997ARA&A..35..607Z}
{Zensus}, J.~A. 1997, \araa, 35, 607, \dodoi{10.1146/annurev.astro.35.1.607}

\end{thebibliography}
\bibliographystyle{aasjournal}



\end{document}